%% file: babar-pub-2104.tex
\documentclass[twocolumn,showpacs,aps,floatfix,prd,nofootinbib]{revtex4}
\usepackage[colorlinks=true]{hyperref}
\usepackage{graphics}
\usepackage{graphicx}
\usepackage{amssymb,amsmath}
\usepackage{tabularx}
\usepackage{relsize}
\def\babar{\mbox{\slshape B\kern-0.1em{\smaller A}\kern-0.1em
    B\kern-0.1em{\smaller A\kern-0.2em R}}}

\newcommand{\BABARPubNumber}{21/004}
\newcommand{\SLACPubNumber}{17620}
\newcolumntype{b}{>{\hsize=1.5\hsize}X}
\newcolumntype{s}{>{\hsize=0.5\hsize}X}
\newcolumntype{m}{>{\hsize=0.7\hsize}X}
\newcolumntype{Y}{>{\centering\arraybackslash}X}

\begin{document}
    
\begin{minipage}{.45\textwidth}
\begin{flushleft}
\babar\--PUB-\BABARPubNumber \\
SLAC-PUB-\SLACPubNumber \\
\end{flushleft}
\end{minipage}
\vspace*{5mm}

\title{\large \bf
\boldmath
Study of the process $e^+e^-\to \pi^+\pi^-\pi^0$ using 
initial state radiation with \babar\
}

\input{authors_sep2021_frozen.tex}
\begin{abstract}
The process $e^+e^-\to \pi^+\pi^-\pi^0\gamma$ 
is studied at a center-of-mass energy near the $\Upsilon(4S)$ resonance 
using a data sample of 469 fb$^{-1}$ collected with the \babar\ detector
at the PEP-II collider. We have performed a precise measurement of the 
$e^+e^-\to \pi^+\pi^-\pi^0$ cross section in the center-of-mass energy
range from 0.62 to 3.5 GeV.
In the energy regions of the $\omega$ and $\phi$ resonances, the cross
section is measured with a systematic uncertainty of 1.3\%.
The leading-order hadronic contribution to the muon
magnetic anomaly calculated using the measured $e^+e^-\to \pi^+\pi^-\pi^0$ 
cross section from threshold to 2.0 GeV is
$(45.86 \pm 0.14 \pm 0.58)\times 10^{-10}$.
From the fit to the measured $3\pi$ mass spectrum we have determined the
resonance parameters 
$\Gamma(\omega\to e^+e^-){\cal B}(\omega\to \pi^+\pi^-\pi^0)=
(0.5698\pm0.0031\pm0.0082) \mbox{ keV}$,
$\Gamma(\phi\to e^+e^-){\cal B}(\phi\to \pi^+\pi^-\pi^0)=
(0.1841\pm0.0021\pm0.0080) \mbox{ keV}$, and
${\cal B}(\rho\to 3\pi)=(0.88\pm 0.23\pm 0.30)\times 10^{-4}$.
The significance of the $\rho\to 3\pi$ signal is greater than $6\sigma$. 
For the $J/\psi$ resonance we have measured the product
${\Gamma(J/\psi\to e^+e^-){\cal B}(J/\psi\to 3\pi)}=
(0.1248\pm0.0019\pm0.0026)\mbox{ keV}$.
\end{abstract}
\maketitle

\setcounter{footnote}{0}

\section{Introduction
\label{intro}}
The process $e^+e^-\to\pi^+\pi^-\pi^0$
\footnote{Throughout this paper, $2\pi$, $3\pi$, and $4\pi$
mean  $\pi^+\pi^-$, $\pi^+\pi^-\pi^0$, and $\pi^+\pi^-\pi^0\pi^0$,
respectively. We also use the notation $\rho$, $\omega$, and $\phi$
for $\rho(770)$, $\omega(782)$, and $\phi(1020)$.}
has the second largest hadronic cross
section after $e^+e^-\to\pi^+\pi^-$ in the energy region below 1 GeV
and is therefore very important for the Standard Model calculation of
the anomalous magnetic moment of the muon $a_\mu\equiv (g-2)_\mu/2$.
Currently, the accuracy of the  $e^+e^-\to\pi^+\pi^-\pi^0$ contribution to 
the muon magnetic anomaly ($a_\mu^{3\pi}$) is about 3\%~\cite{dhmz} and
needs to be improved.

The most precise measurements of the $e^+e^-\to\pi^+\pi^-\pi^0$ cross section
in the energy region of the $\omega$ and $\phi$ resonances were performed by
the SND and CMD-2 Collaborations at the VEPP-2M $e^+e^-$ 
collider~\cite{snd1,snd2,cmd1,cmd2}.
Above the $\phi$ meson resonance the latest measurements come from the
\babar\ experiment~\cite{babar}, which used the initial-state radiation (ISR)
technique, and the SND experiment at the VEPP-2000 $e^+e^-$
collider~\cite{snd3}. There is also a preliminary result from the BESIII
experiment~\cite{bes3}, which measured the $e^+e^-\to\pi^+\pi^-\pi^0$ cross 
section in the energy range between 0.7 and 3.0 GeV using the ISR technique.

One of the reasons for the relatively low accuracy of $a_\mu^{3\pi}$ is the
difference between the cross section measurements in different experiments.
For example, the SND cross section near the $\omega$~\cite{snd2} is
about 8\% ($1.8\sigma$) larger than the cross section measured by
CMD-2~\cite{cmd1}. \babar\ did not measure the cross section in this region,
but fitted to the $3\pi$ mass spectrum in the $e^+e^-\to\pi^+\pi^-\pi^0\gamma$
reaction with the vector-meson-dominance (VMD) model~\cite{babar} and
determined the $\omega$ parameters. The \babar\ value for the $\omega$ peak
cross section as well as the BESIII preliminary result~\cite{bes3} support
 a larger cross section value, as obtained by SND~\cite{snd2}.

It is generally accepted that the process $e^+e^-\to\pi^+\pi^-\pi^0$ proceeds
mainly through the $\rho(770)\pi$ 
($\rho^+\pi^- + \rho^-\pi^+ + \rho^0\pi^0$) intermediate state. 
This assumption has been
well tested at the $\omega$ and $\phi$ resonances~\cite{bes-om,kloe-phi}.
The dynamics of $e^+e^-\to\pi^+\pi^-\pi^0$ in the energy range between 1.1 and
2 GeV were recently studied in Ref.~\cite{snd3}. This study confirms the
dominance of the $\rho(770)\pi$ channel below 1.5 GeV. However, in this
region there is a 10--20\% contribution from the isovector $\omega\pi^0$
mechanism and its interference with the dominant $\rho(770)\pi$ amplitude.
In the region of the $\omega(1650)$ resonance (1.55--1.75 GeV), a large 
contribution of the $\rho(1450)\pi$ intermediate
state was observed, which is comparable with that of the $\rho(770)\pi$.  
A relatively large fraction ($\sim$10\%) of the $\rho(1450)\pi$ channel
was also observed in the $J/\psi\to 3\pi$ decay~\cite{babar-psi}.

In this article we update the \babar\ $e^+e^-\to\pi^+\pi^-\pi^0$ 
measurement~\cite{babar} using a data set that is 5 times larger. We study
the process $e^+e^-\to\pi^+\pi^-\pi^0\gamma$, where the photon emission is
caused by initial-state radiation. The Born cross section for this process
integrated over the momenta of the hadrons is given by 
\begin{equation}
\frac{d\sigma(s,x,\theta_\gamma)}{dx\,d\cos{\theta_\gamma}} =
W(s,x,\theta_\gamma)\, \sigma_0(s(1-x)),
\label{eq1}
\end{equation}
where $\sqrt{s}$ is the $e^+e^-$ center-of-mass (c.m.) energy,
$x\equiv{2E_{\gamma}}/{\sqrt{s}}$,
$E_{\gamma}$ and $\theta_\gamma$ are the photon energy and polar angle in 
the c.m.~frame, and $\sigma_0$ is 
the Born cross section for $e^+e^-\to \pi^+\pi^-\pi^0$.
The so-called radiator function (see, for example, Ref.~\cite{ivanch})
\begin{equation}
W(s,x,\theta_\gamma)=\frac{\alpha}{\pi x}
\left(\frac{2-2x+x^2}{\sin^2\theta_\gamma}-
\frac{x^2}{2}\right)
\label{eq2} 
\end{equation}
describes the probability of ISR photon emission for 
$\theta_\gamma\gg m_e/\sqrt{s}$.
Here, $\alpha$ is the fine structure constant and $m_e$ is the electron mass. 
The ISR photons are emitted predominantly at small angles relative to
the initial electron or positron directions; however, about 10\% of the photons
have c.m.~polar angles in the range $30^\circ<\theta_\gamma<150^\circ$.
In the present analysis, we require that the ISR photon be detected.

The goal of this analysis is to improve the accuracy of the 
$e^+e^-\to \pi^+\pi^-\pi^0$ cross section measurement and the contribution
of this process to $a_\mu$.

\section{The \babar\ detector and data samples
\label{detector}}
In this article a data sample of 469 fb$^{-1}$, collected with
the \babar\ detector~\cite{babar-nim} at the PEP-II asymmetric-energy storage
ring at the SLAC National Accelerator Laboratory, is analyzed. At PEP-II, 
9 GeV electrons collide with 3.1 GeV positrons
at a center-of-mass energy of 10.58 GeV ($\Upsilon$(4S) resonance).
About 91\% of the integrated luminosity was recorded at 10.58 GeV,
while 9\% was recorded at 10.54 GeV.  

Charged-particle tracking for the \babar\ detector is
provided by a five-layer silicon vertex tracker (SVT) and
a 40-layer drift chamber (DCH), operating in a 1.5 T axial
magnetic field. The transverse momentum resolution
is 0.47\% at 1 GeV/$c$. Energies of photons and electrons
are measured with a CsI(Tl) electromagnetic calorimeter
(EMC) with a resolution of 3\% at 1 GeV. Charged-particle identification is
provided by measurements of ionization losses, $dE/dx$, in the SVT and DCH,
and by an internally reflecting ring-imaging Cherenkov detector. Muons are
identified in the solenoid's instrumented flux return.

Signal and background ISR processes are simulated by a Monte
Carlo (MC) event generator based on the approach suggested in
Ref.~\cite{ckhhad}. A model of the $\rho(770)\pi$ intermediate state
is used to simulate the signal process $e^+e^-\to 3\pi\gamma$.
The extra-photon radiation from the initial state 
is implemented with the structure function technique~\cite{strfun},
while the final-state radiation is simulated using the PHOTOS
package~\cite{PHOTOS}.
Since the ISR photon is emitted predominantly at small angles relative to the
beam directions, the events are generated with the restriction 
$20^\circ<\theta_\gamma<160^\circ$, where $\theta_\gamma$ is
the ISR photon polar angle in the c.m.~frame. We also require that
the invariant mass of the hadron system and ISR photon together be 
greater than 8 GeV/$c^2$. This condition restricts the maximum energy 
of extra photons emitted by the initial particles.

The following background ISR processes are simulated:
$e^+e^-\to \pi^+\pi^-\gamma$, $\mu^+\mu^-\gamma$, $K^+K^-\gamma$,
$K_SK_L\gamma$, $K^+K^-\pi^0\gamma$, $K_SK^-\pi^+\gamma$,
$\pi^+\pi^-\pi^0\pi^0\gamma$, $\pi^+\pi^-\eta\gamma$, 
$\omega\eta\gamma$, and $\omega\pi^0\pi^0\gamma$. The backgrounds from non-ISR
hadronic processes $e^+e^-\to q\bar{q}$, where $q=u,\,d,\,s$, and from 
$e^+e^-\to \tau^+\tau^-$ are simulated with the JETSET~\cite{Jetset} and
KK2f~\cite{KK2f} packages, respectively.
The interaction of the generated particles with the \babar\ detector 
and the detector response are simulated using the GEANT4~\cite{ref:geant4} 
package. The simulation takes into account the variation of the detector and 
accelerator conditions, and in particular describes the beam-induced 
background, which leads to the appearance of spurious photons and
tracks in the events of interest.

\section{Event selection\label{evsel}}
The selection of $e^+e^-\to \pi^+\pi^-\pi^0\gamma$
candidates is based on the requirement that all the final particles be detected
and well reconstructed. We select events with exactly two good quality 
opposite-sign charged tracks, which are considered as $\pi^+$ and $\pi^-$ 
candidates, and at least three photons. The ``good'' tracks are required to have
a transverse momentum above 100 MeV/$c$, originate from the interaction region,
and to be not identified as an electron. Their laboratory polar angle must be
between $23^\circ$ and $140^\circ$. An event
can contain any number of extra tracks not satisfying the above criteria. 

The photons must have energies above
100 MeV and be in the well-understood region of the calorimeter
$23^\circ < \theta < 137.5^\circ$. One of the photons (the ISR candidate 
photon) is required to have a c.m.~energy larger than 3 GeV.
The remaining photons must form at least one $\pi^0$ candidate, a pair of 
photons with invariant mass in the range 0.1--0.17 GeV/$c^2$.

\begin{figure}
\centering
\includegraphics[width=0.95\linewidth]{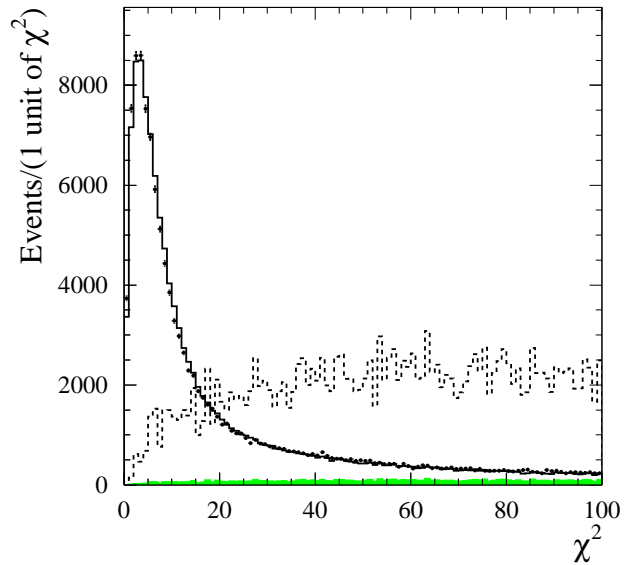}
\caption{The $\chi^2_{3\pi\gamma}$ distributions for data (points with error
bars) and simulated (histogram) signal plus background events from the 
$\omega$ mass region. The shaded (green) histogram shows the distribution for 
simulated background events.
The dashed histogram is the background distribution multiplied by a factor
of 25.}
\label{chi_omega}
\end{figure}
\begin{figure}
\centering
\includegraphics[width=0.95\linewidth]{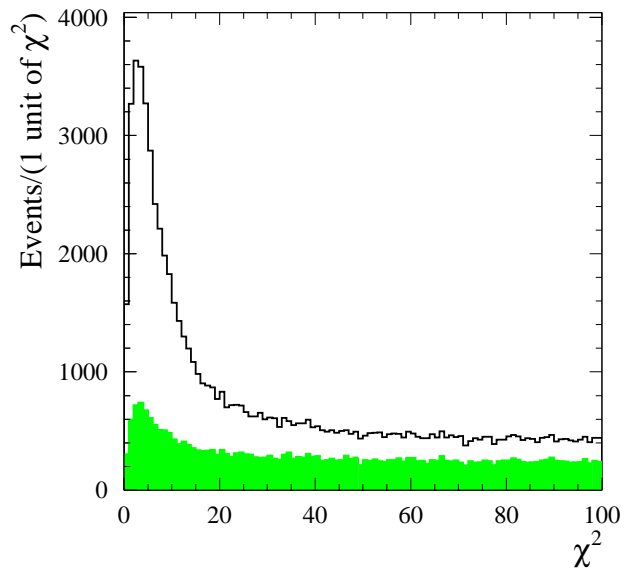}
\caption{The $\chi^2_{3\pi\gamma}$ distributions for data
from the mass range $1.05<M_{3\pi}<3.00$ GeV/$c^2$.
The shaded (green) histogram shows events rejected
by background suppression requirements, as explained in the text.}
\label{chi_13}
\end{figure}
For events satisfying the selection criteria described above,
a kinematic fit is performed with requirements of energy and momentum
conservation, and the $\pi^0$ mass constraint for the candidate $\pi^0$.
The MC simulation does not accurately reproduce the shape of the resolution
function for the photon energy. To reduce the effect of 
the data-MC simulation difference in the energy resolution, the fit uses only
the measured direction for the ISR photon candidate; its energy is a free fit
parameter.
For events with two or more candidate $\pi^0$s, all possible
$\pi^+\pi^-\pi^0\gamma$ combinations are tested and the one with the minimum
$\chi^2$ of the kinematic fit ($\chi^2_{3\pi\gamma}$) is used.
As a result of the kinematic fit we obtain the corrected three-pion
invariant mass ($M_{3\pi}$).

The $\chi^2_{3\pi\gamma}$ distribution for events from the $3\pi$ mass region
near the $\omega$ mass (0.67--0.87 GeV/$c^2$), where the contribution of 
background processes is small, is shown in Fig.~\ref{chi_omega}. 
In further analysis we use two conditions on this parameter: the standard 
$\chi^2_{3\pi\gamma} < 40$ and the tighter $\chi^2_{3\pi\gamma} < 20$.
The latter is applied for the $e^+e^-\to \pi^+\pi^-\pi^0$ cross section 
measurement. 
The $\chi^2_{3\pi\gamma}$ distribution
for the mass range $1.05<M_{3\pi}<3.00$ GeV/$c^2$ is shown in 
Fig.~\ref{chi_13}. In this region the background contribution is
significantly larger.

The main sources of background for the process under study are the ISR 
processes $e^+e^-\to\pi^+\pi^-\pi^0\pi^0\gamma$, $\pi^+\pi^-\gamma$,
$K^+K^-\pi^0\gamma$, etc., and non-ISR processes $e^+e^-\to q\bar{q}$ and 
$e^+e^-\to \tau^+\tau^-$. Additional conditions are 
applied to suppress background.

Events with charged kaons in the final state
($e^+e^-\to K^+K^-\pi^0\gamma$, $e^+e^-\to K^+K^-\gamma$, etc.)
are suppressed by the requirement that none of the ``good'' charged tracks be
identified as a kaon.

Two-body ISR events from the processes $e^+e^-\to\pi^+\pi^-\gamma$ and
$e^+e^-\to\mu^+\mu^-\gamma$ with extra spurious photons are suppressed by
the two conditions: $E_{\pi^0}>0.4$ GeV, where $E_{\pi^0}$ is the energy of
the candidate $\pi^0$, and $M^2_{\rm rec}>5$ $\mbox{GeV}^2/c^4$, where
$M_{\rm rec}$ is the mass recoiling against the $\pi^+\pi^-$ pair.

Some fraction of background $e^+e^-\to\pi^+\pi^-\pi^0\pi^0\gamma$ events
contain additional $\pi^0$ candidates. For these events we perform a kinematic
fit to the $4\pi\gamma$ hypothesis and apply the condition
$\chi^2_{4\pi\gamma}>30$, which reduces the $4\pi\gamma$ background by a factor
of 2.

Another important background source is  $e^+e^-\to q\bar{q}$ events
containing a very energetic $\pi^0$ in the final state.
A fraction of these events is seen as a peak at the $\pi^0$ mass in 
the $M_{\gamma\gamma}^\ast$ distribution, where
$M_{\gamma\gamma}^\ast$ is the invariant mass 
of two photons, one of which is the most energetic in an event.
The second photon is required to have an energy above 100 MeV.
Once all possible photon pair combinations are checked, the one with closest
invariant mass to the $\pi^0$ mass is chosen.
Events with $0.10<M_{\gamma\gamma}^\ast<0.17$ GeV/$c^2$ are rejected.

The $e^+e^-\to q\bar{q}$ background is dominated by 
$e^+e^-\to \pi^+\pi^-\pi^0\pi^0$ events. Events of this process
passing the $\pi^+\pi^-\pi^0\gamma$ selection criteria 
have a $\chi^2_{3\pi\gamma}$ distribution peaked at low values, similar to the
signal. A fraction of these events proceeding via $\rho^+\rho^-$ intermediate
state is rejected by the condition $M_{\pi\gamma}>1.5$ GeV/$c^2$, where
$M_{\pi\gamma}$ is the invariant mass of the most energetic
photon and one of the charged pions. This condition also rejects
$e^+e^-\to\tau^+\tau^-$ events, which imitate $\pi^+\pi^-\pi^0\gamma$ events
when both $\tau$'s decay into $\rho\nu$. 

In the $3\pi$ mass region below 1.1 GeV/$c^2$, which is the most important for 
the calculation of $a_\mu^{3\pi}$, the background suppression requirements
decrease the fraction of 
background events from 5\% to 2\%, with loss of signal events of 15\%. The 
$\chi^2_{3\pi\gamma}$ distribution for data events rejected by the background
suppression requirements in the mass region $1.05<M_{3\pi}<3$ GeV/$c^2$ is 
shown as 
the shaded histogram in Fig.~\ref{chi_13}. In this region, the background is
suppressed by a factor of 2.6 with a loss of signal events of 17\%.

\section{Background estimation and subtraction}
\label{subtraction}
To estimate background, the samples of simulated events listed in
Sec.~\ref{detector} are normalized to the collected integrated luminosity.
Before normalization, the hadron mass spectrum for a particular
simulated process is reweighted using Eq.~(\ref{eq1}) and the existing data on
its Born cross section.
For the most important background ISR processes $e^+e^-\to K^+K^-\pi^0\gamma$,
$e^+e^-\to \pi^+\pi^-\gamma$, and $e^+e^-\to \pi^+\pi^-\pi^0\pi^0\gamma$, data
samples selected with special criteria are used to determine additional
scale factors.

The mass distribution for events with two charged kaons surviving our selection
($dN_{0K}/dm$) is obtained from the distribution of events with two identified 
kaons: $N_{0K} = R_K(dN_{2K}/dm)$. The coefficient $R_K$ is determined from
$e^+e^-\to K^+K^-\pi^0\gamma$ simulation corrected for 
data-simulation differences in the charged-kaon identification
efficiency. The observed spectrum of two-kaon background events is almost
completely saturated by the $e^+e^-\to K^+K^-\pi^0\gamma$ process.

The scale factor for the $e^+e^-\to\pi^+\pi^-\gamma$ process is estimated
using events with $40<\chi^2_{3\pi\gamma}<250$ and 
$M^2_{\rm rec}<10$ $\mbox{GeV}^2/c^4$. The latter condition suppresses
contributions of all processes except $e^+e^-\to\pi^+\pi^-\gamma$.
The scale factor is found to be $1.6\pm0.2$.
The quoted systematic uncertainty is estimated by variation of the
conditions on $\chi^2_{3\pi\gamma}$ and $M^2_{\rm rec}$. The large difference
between the fitted and expected numbers of $e^+e^-\to\pi^+\pi^-\gamma$
background events may be the result of an inaccurate simulation of the nuclear 
interactions of charged $\pi$ mesons in the calorimeter. In particular,
the number of fake photons due to nuclear interactions may be different in data
and simulation.

\begin{figure}
\centering
\includegraphics[width=0.95\linewidth]{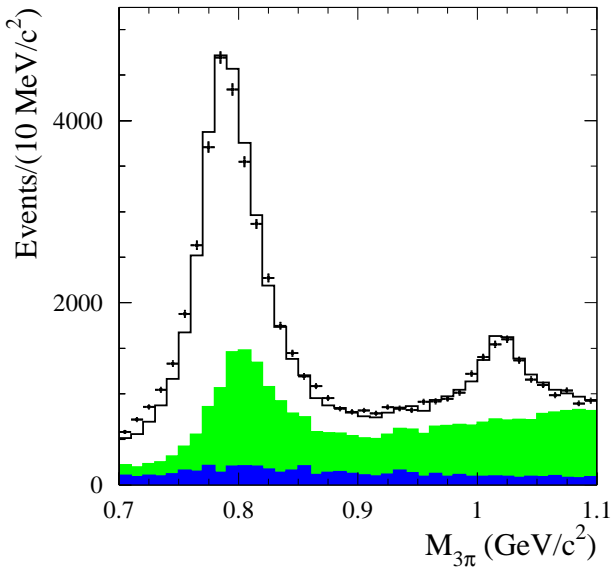}
\caption{The $3\pi$ invariant mass spectrum for data events
with $50<\chi^2_{3\pi\gamma}<500$ and $M^2_{\rm rec}>10$ $\mbox{GeV}^2/c^4$.
The solid histogram represents the result of the fit with a sum
of signal and background distributions. The light-shaded (green) area
represents the fitted $e^+e^-\to\pi^+\pi^-\pi^0\pi^0\gamma$ contribution,
while the dark-shaded (blue) histograms is the spectrum for all
other background processes.
\label{4pig_norm}}
\end{figure}

The process $e^+e^-\to\pi^+\pi^-\pi^0\pi^0\gamma$ is the main source
of background for the process under study. Several intermediate states 
($\omega\pi^0$, $a_1\pi$, $\rho^+\rho^-$, etc.) contribute to this process.
Our MC event generator incorrectly reproduces both the $4\pi$ mass spectrum
for $e^+e^-\to\pi^+\pi^-\pi^0\pi^0\gamma$ events and the relation between
intermediate states, in particular, the fraction of $\omega\pi^0$ events.
Therefore, the normalization for this process is performed
in two stages. In the first stage, we select events with two charged particles
and at least five photons, perform a kinematic fit to the 
$e^+e^-\to\pi^+\pi^-\pi^0\pi^0\gamma$ hypothesis, and
select events with $\chi^2_{4\pi\gamma}<30$.  We measure the $4\pi$ mass
spectrum and reweight the $e^+e^-\to\pi^+\pi^-\pi^0\pi^0\gamma$ simulation
using the ratio of the data and simulated spectra as a weight function. The
reweighting is performed separately for $\omega\pi^0$ and non-$\omega\pi^0$ 
events. In the second stage, we analyze the $3\pi$ mass spectrum below
1.1 GeV/$c^2$ for events with $50<\chi^2_{3\pi\gamma}<500$ and
$M^2_{\rm rec}>10$ $\mbox{GeV}^2/c^4$. The latter condition is applied to
suppress the $e^+e^-\to\pi^+\pi^-\gamma$ background. The spectrum shown in
Fig.~\ref{4pig_norm} is fitted with a sum of simulated signal and background
distributions.  The fitted parameters are scale factors for the
$e^+e^-\to\pi^+\pi^-\pi^0\gamma$ and $e^+e^-\to\pi^+\pi^-\pi^0\pi^0\gamma$
distributions. The difference in the line shape of the $\omega$ peak between
data and simulation seen in Fig.~\ref{4pig_norm} is attributed to inaccurate
simulation of the tails of the $M_{3\pi}$ resolution function at large 
$\chi^2_{3\pi\gamma}$ values. The $\pi^+\pi^-\pi^0\pi^0\gamma$ scale factor 
is found to be $1.30\pm0.15$. The quoted uncertainty is systematic. It is 
estimated by variation of the conditions on $\chi^2_{3\pi\gamma}$ and
$M^2_{\rm rec}$.
The total contribution to the background from other ISR processes at 
$M_{3\pi}<1.1$ GeV is calculated to be less than 1/50 of the
$e^+e^-\to\pi^+\pi^-\pi^0\pi^0\gamma$ background.

The $e^+e^-\to q\bar{q}$ background events can be divided into two classes.
The first class ($4\pi$) contains events from the 
$e^+e^-\to\pi^+\pi^-\pi^0\pi^0$ process. The second (non-$4\pi$) contains
events from all other processes. The $4\pi$ events has a $\chi^2_{3\pi\gamma}$
distribution peaked at small values similar to the signal process 
$e^+e^-\to\pi^+\pi^-\pi^0\gamma$. For the second class, the
$\chi^2_{3\pi\gamma}$ distribution has a wide maximum near
$\chi^2_{3\pi\gamma}=300$. The ISR photon in $4\pi$ and most non-$4\pi$
$q\bar{q}$ events is imitated by a photon from the $\pi^0$ decay. Therefore,
to estimate these backgrounds we study the $M_{\gamma\gamma}^\ast$
distribution.

\begin{figure}
\centering
\includegraphics[width=0.95\linewidth]{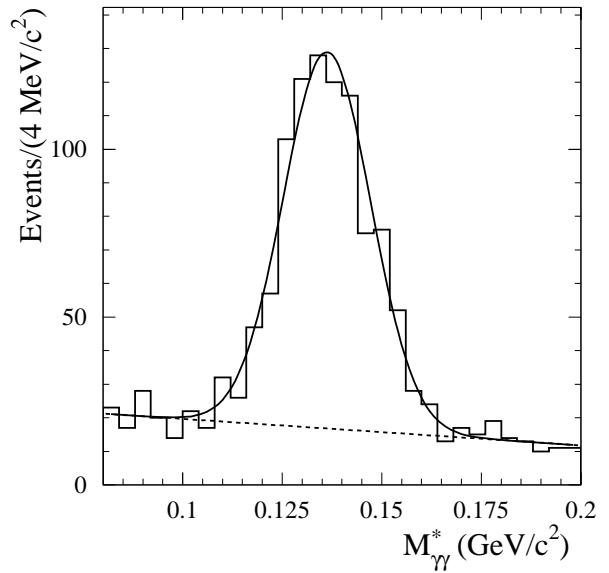}
\caption{The $M_{\gamma\gamma}^\ast$ distribution for data events from the
region $0.6<M_{3\pi}<3.5$ GeV/$c^2$ selected with our standard selection
criteria except for the condition that $M_{\gamma\gamma}^\ast$ be outside 
the window 0.10--0.17 GeV/$c^2$. The curve is the result of the fit
by the sum of a Gaussian and a linear function. The linear function is 
shown separately by the dashed line. 
\label{pi0}}
\end{figure}

The $M_{\gamma\gamma}^\ast$ distribution for data events with 
$0.6<M_{3\pi}<3.5$ GeV/$c^2$ selected using our standard selection criteria 
except for the condition on $M_{\gamma\gamma}^\ast$ is shown
in Fig.~\ref{pi0}. 
The events in the $\pi^0$ peak originate mainly from
the $4\pi$ class, while the nearly flat distribution is dominated by 
$e^+e^-\to\pi^+\pi^-\pi^0\gamma$ events. The distribution is fitted with 
the sum of a Gaussian and a linear function. 
The non-$4\pi$ background is also estimated from the number of events in 
the $\pi^0$ peak in the $M_{\gamma\gamma}^\ast$ distribution, but for events
with $40<\chi^2_{3\pi\gamma}<200$.

\begin{figure}
\centering
\includegraphics[width=0.95\linewidth]{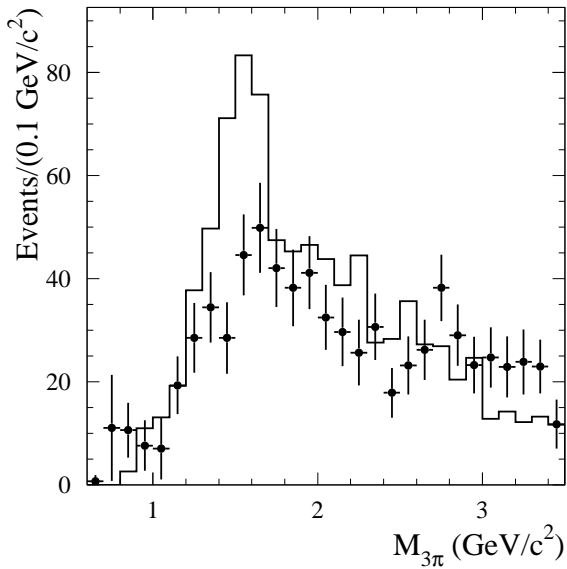}
\caption{The $M_{3\pi}$ spectra for data (points with error bars) and
simulated (histogram)  $e^+e^-\to\pi^+\pi^-\pi^0\pi^0$ events obtained from
the fits to the $M_{\gamma\gamma}^\ast$ distributions as described in the text.
\label{qqbarsp}}
\end{figure}

The $3\pi$ mass region 0.6--3.5 GeV/$c^2$ is divided into 29 intervals
with 0.1 GeV/$c^2$ width. For each $M_{3\pi}$ interval, we determine the
numbers of $4\pi$ and non-$4\pi$ events in data and $e^+e^-\to q\bar{q}$
simulation from the fit to the $M_{\gamma\gamma}^\ast$ distribution. 
The obtained data spectrum for $4\pi$ events is compared with the same 
spectrum for simulated events in Fig.~\ref{qqbarsp}.
It is seen that the $e^+e^-\to q\bar{q}$ simulation reproduces reasonably well
the total number of selected  $e^+e^-\to\pi^+\pi^-\pi^0\pi^0$ events. The 
overall scale factor for the simulation is $0.83\pm0.05$. However the shapes 
of the $M_{3\pi}$ spectra for data and simulation are different, especially in
the region 1.3--1.8 GeV/$c^2$. At $M_{3\pi}>0.9$ GeV/$c^2$ the ratio of the 
data and simulated spectra shown in Fig.~\ref{qqbarsp} is used to 
reweight the yield of simulated $e^+e^-\to\pi^+\pi^-\pi^0\pi^0$ events. It 
should be noted that the ratio of the number of $4\pi$ events selected
with our standard criteria to the number of events shown in Fig.~\ref{qqbarsp}
is about five. The uncertainty in the number of $4\pi$ background events
obtained using the reweighted simulation is dominated by the uncertainty in
the number of events in each mass bin in Fig.~\ref{qqbarsp}. 

\begin{figure}
\centering
\includegraphics[width=0.95\linewidth]{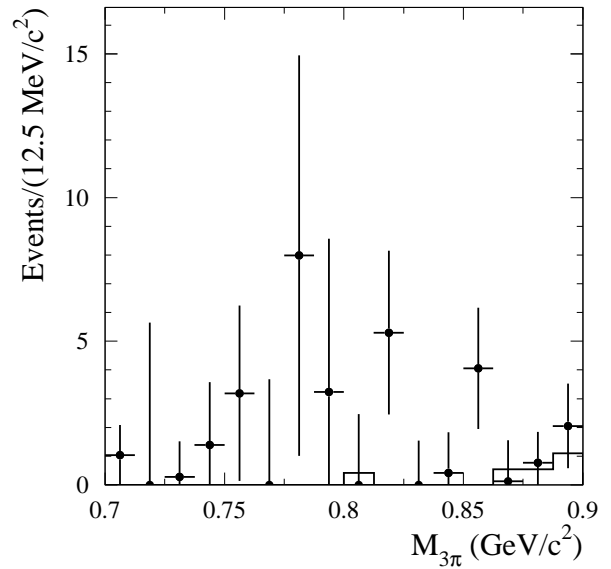}
\caption{The $M_{3\pi}$ spectrum for data (points with error bars) and 
simulated (histogram) $e^+e^-\to\pi^+\pi^-\pi^0\pi^0$ events with the
finer binning.
\label{4pispfb}}
\end{figure}
An excess of data over simulation is seen in Fig.~\ref{qqbarsp} in
the mass region 0.7--0.9 GeV/$c^2$. This excess may be an indication of a
contribution from the $e^+e^-\to\omega\pi^0\to\pi^+\pi^-\pi^0\pi^0$ process, 
which is absent in our $e^+e^-\to q\bar{q}$ simulation. This process produces
events peaked at the $\omega$ mass. We repeat the fitting procedure described 
above with finer binning. The result is shown in Fig.~\ref{4pispfb}. This
spectrum is used to estimate the $e^+e^-\to\pi^+\pi^-\pi^0\pi^0$ background
in the $\pi^+\pi^-\pi^0$ mass region from 0.7 to 0.9 GeV/$c^2$. To do this,
the data spectrum in Fig.~\ref{4pispfb} is multiplied by a scale factor of 
five, obtained in the region $M_{3\pi}>0.9$ GeV. The systematic
uncertainty in this estimation is taken to be 100\%. The same scale factor
is used for the interval $0.6<M_{3\pi}<0.7$ GeV/$c^2$, where the number
of fitted $4\pi$ events in Fig.~\ref{qqbarsp} is $0.7\pm1.2$.

\begin{figure}
\centering
\includegraphics[width=0.95\linewidth]{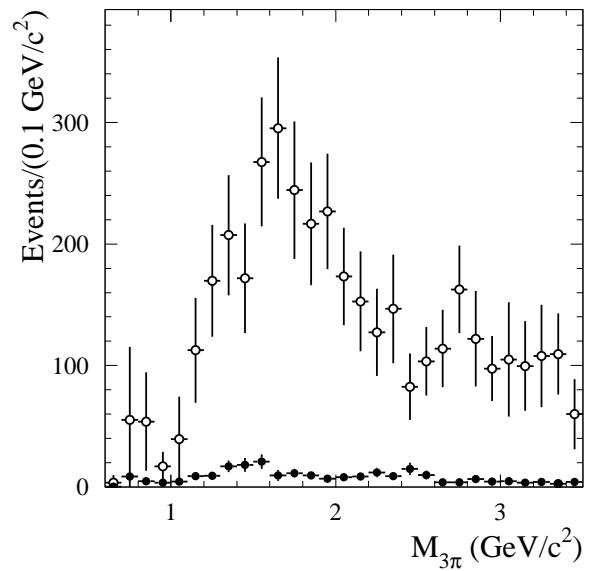}
\caption{The $M_{3\pi}$ spectrum for the $4\pi$ (open circles) and
non-$4\pi$ $q\bar{q}$ (filled circles) background events selected
with the standard criteria. The spectra are obtained using 
reweighted $q\bar{q}$ simulation.
\label{totqqbar}}
\end{figure}
A similar procedure is used to reweight the non-$4\pi$ $q\bar{q}$ simulation.
For this background we also analyze events with $M_{\gamma\gamma}^\ast$ near
the $\eta$ meson mass.
The $M_{3\pi}$ spectra for the $4\pi$ and non-$4\pi$ $q\bar{q}$ background
events selected with the standard criteria are shown in Fig.~\ref{totqqbar}. It
is seen that the fraction of non-$4\pi$ $q\bar{q}$ events is relatively small.

\begin{figure*}[p]
\centering
\includegraphics[width=0.75\textwidth]{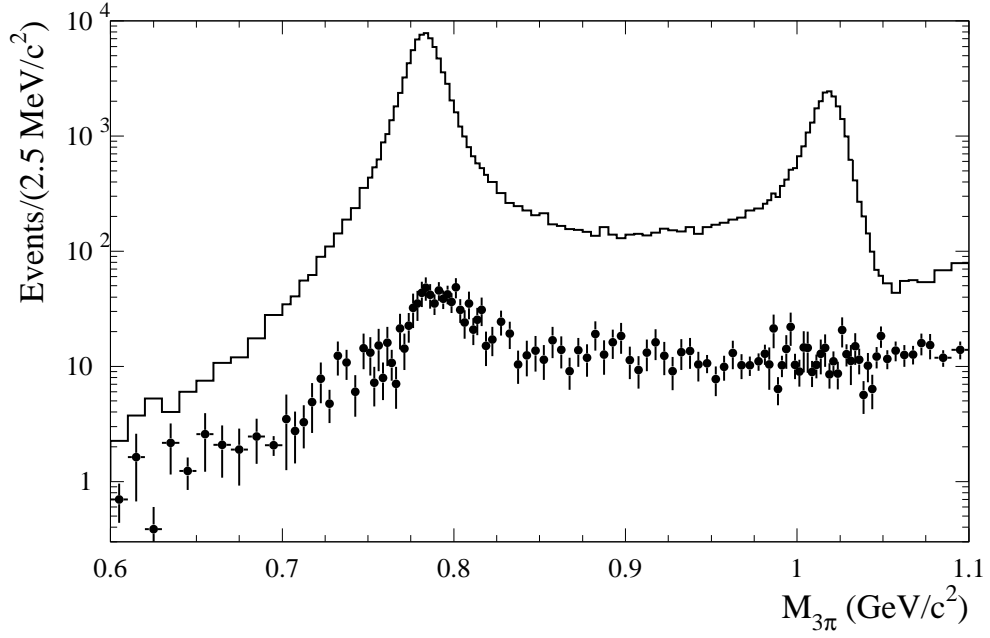}
\caption{The $M_{3\pi}$ spectrum for selected data events in the range
from 0.6 to 1.1 GeV/$c^2$ (histogram). The points with error bars represent
the estimated background contribution.
\label{bkg1}}
\end{figure*}
\begin{figure*}[p]
\centering
\includegraphics[width=0.47\textwidth]{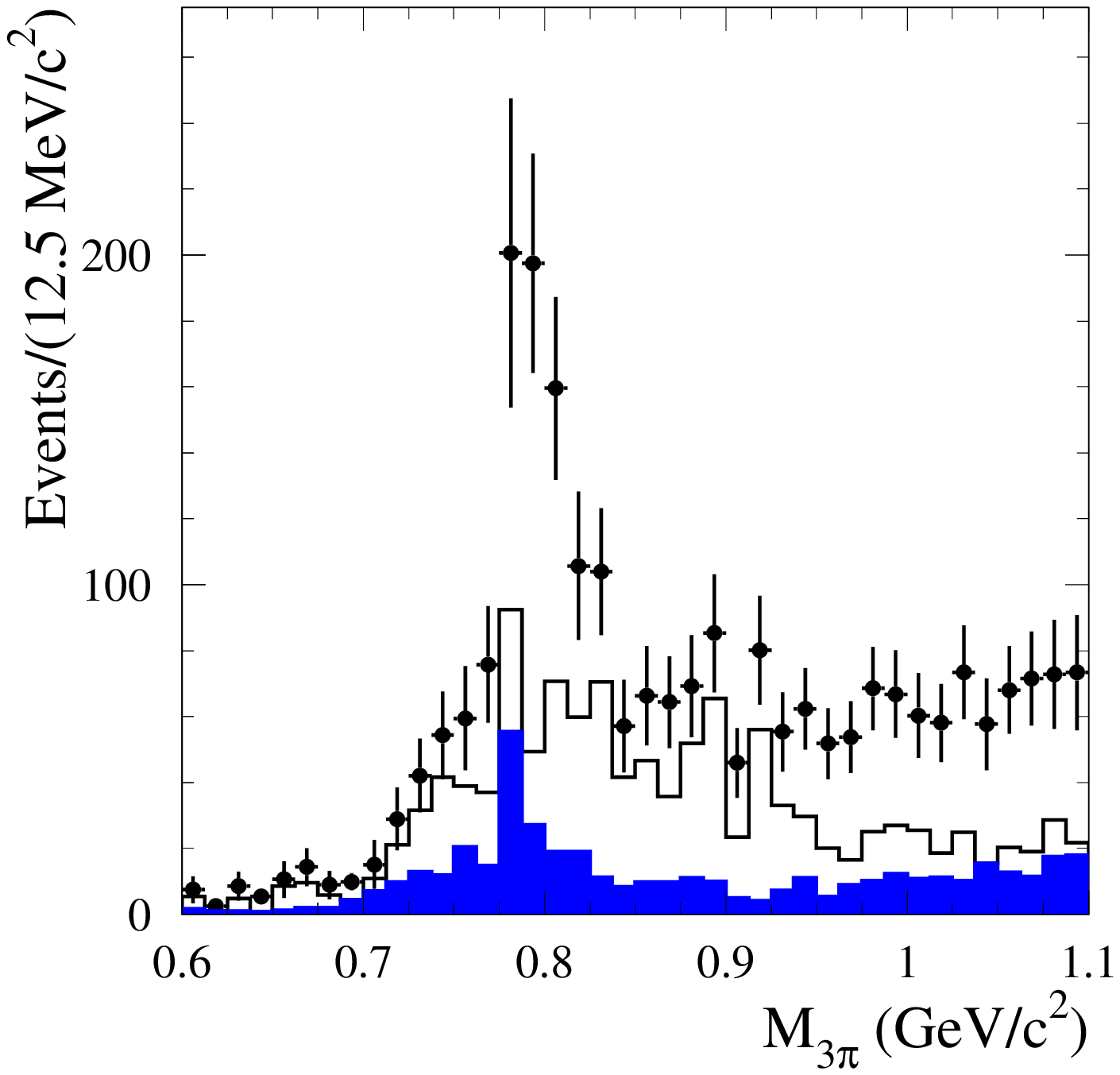}\hfill
\includegraphics[width=0.47\textwidth]{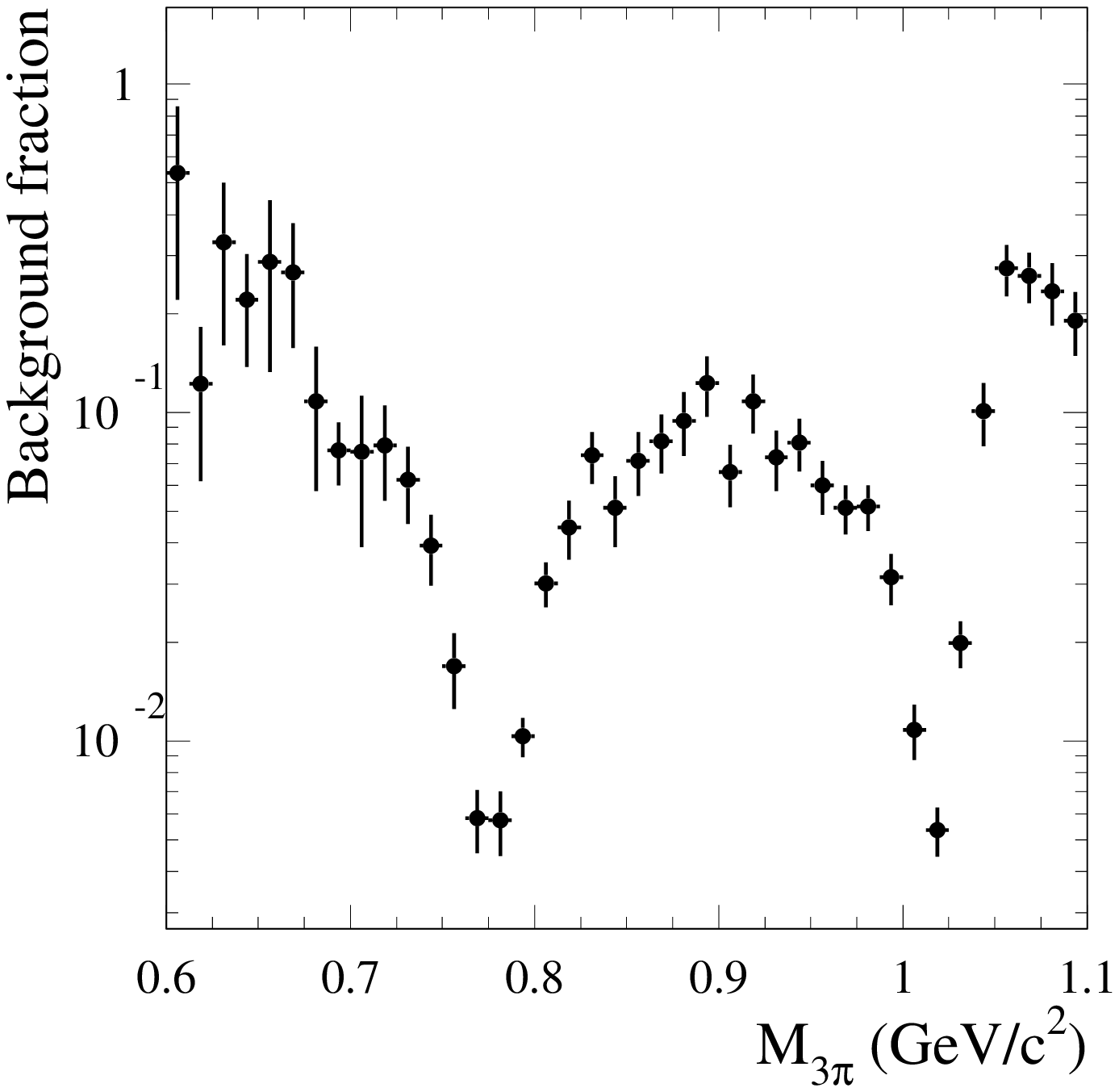}
\caption{Left panel: The $M_{3\pi}$ spectrum for background events (points with
error bars). The filled (blue) histogram represents the contribution of all 
background sources except $e^+e^-\to\pi^+\pi^-\pi^0\pi^0\gamma$ and
$e^+e^-\to\pi^+\pi^-\gamma$. The open histogram is a sum of the filled 
histogram and the spectrum for $e^+e^-\to\pi^+\pi^-\gamma$ events.
Right panel: The ratio of the background spectrum to the spectrum for
selected data events.
\label{bkg2}}
\end{figure*}

The mass region $0.6<M_{3\pi}<1.1$ GeV/$c^2$ is divided into 116 bins.
The bin width varies from 2.5 MeV/$c^2$ near the peaks of the $\omega$ and
$\phi$ resonances to 5 MeV/$c^2$ between the resonances and 10 MeV/$c^2$
near 0.6 and 1.1 GeV/$c^2$. The $M_{3\pi}$ spectrum for data events selected
with the standard criteria is shown in Fig.~\ref{bkg1}. The points with
error bars in Fig.~\ref{bkg1} represent the estimated total background 
contribution from the sources described above. The background $M_{3\pi}$
spectrum on a linear scale is displayed in Fig.~\ref{bkg2} (left). The filled 
histogram represents the contribution of all background sources except 
$2\pi\gamma$ and $4\pi\gamma$. About two-thirds of events in this histogram
come from the $e^+e^-\to\pi^+\pi^-\pi^0\pi^0$ process. The open histogram 
is a sum of the filled histogram and the $e^+e^-\to\pi^+\pi^-\gamma$ background
spectrum. It is seen from Fig.~\ref{bkg2} (left) that the background in this
$M_{3\pi}$ region is dominated by the processes
$e^+e^-\to\pi^+\pi^-\pi^0\pi^0\gamma$ and $e^+e^-\to\pi^+\pi^-\gamma$.
The ratio of the background spectrum to the data spectrum is shown in 
Fig.~\ref{bkg2} (right).
The background fraction decreases from $(25\pm 15)\%$ at 
0.65 GeV/$c^2$ to $(7\pm 3)\%$ at 0.7 GeV/$c^2$ and to $(0.5\pm 0.1)\%$ 
in the $\omega$ region, then increases to $(9\pm 2)\%$ at 0.9 GeV/$c^2$ 
and decreases again to $(0.5\pm 0.1)\%$ at the $\phi$. Near 1.05 GeV/$c^2$,
where the $e^+e^-\to \pi^+\pi^-\pi^0$ cross section has a minimum, the
background fraction is $(27\pm 5)\%$. With the tighter selection
$\chi^2_{3\pi\gamma} < 20$ the background fraction decreases by a factor
of about two.

In the region $0.6<M_{3\pi}<1.1$ GeV/$c^2$ the estimated background is 
subtracted from the number of selected data events in each $M_{3\pi}$ bin.
It should be noted that the numbers of background events in different mass 
bins are correlated. This correlation
arises from the uncertainties in the scale factors for 
$e^+e^-\to\pi^+\pi^-\pi^0\pi^0\gamma$ and $e^+e^-\to\pi^+\pi^-\gamma$
events, which are equal to 10.5\% and 12.5\%, respectively. 

\begin{figure*}[p]
\centering
\includegraphics[width=0.47\textwidth]{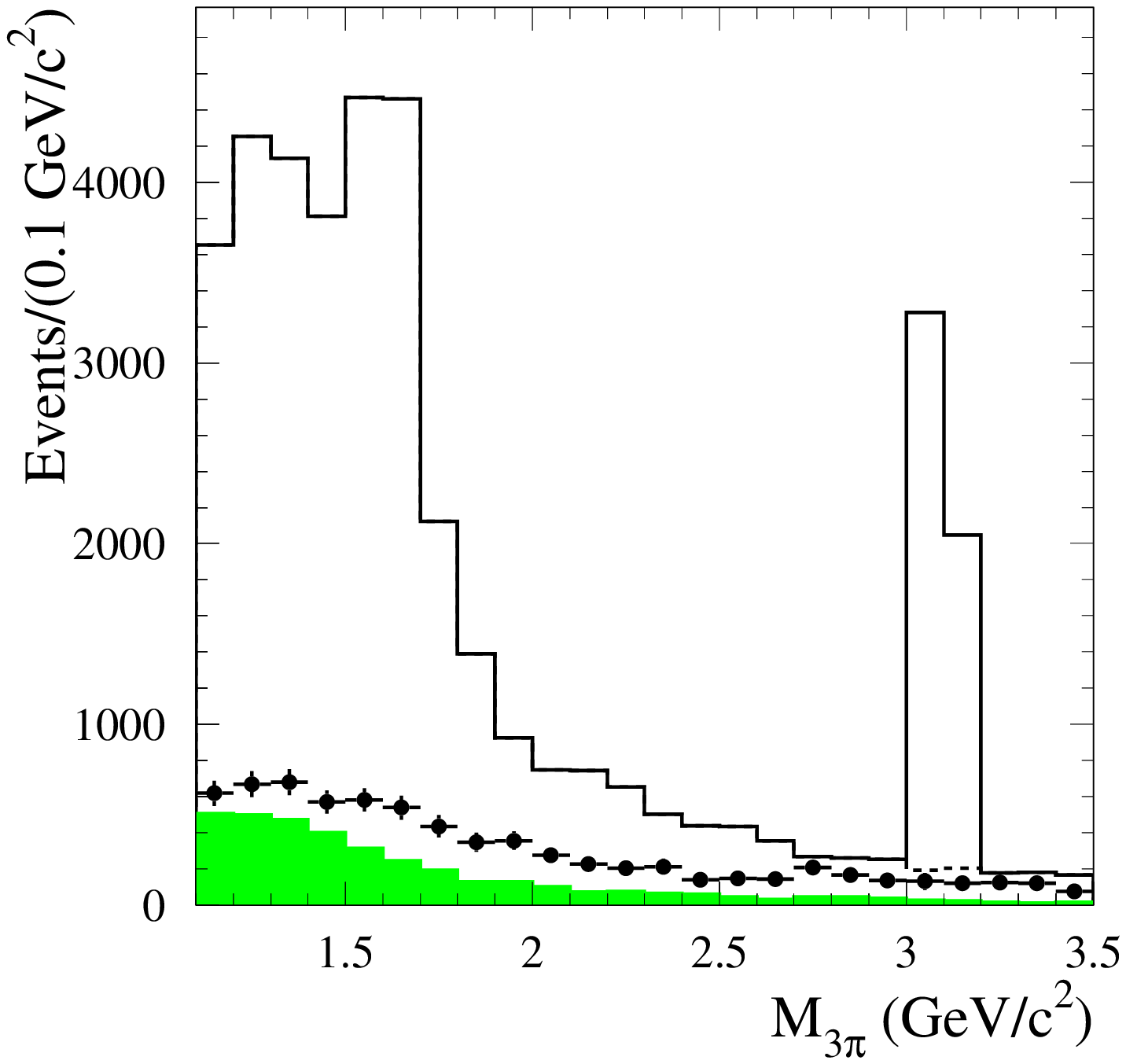}\hfill
\includegraphics[width=0.47\textwidth]{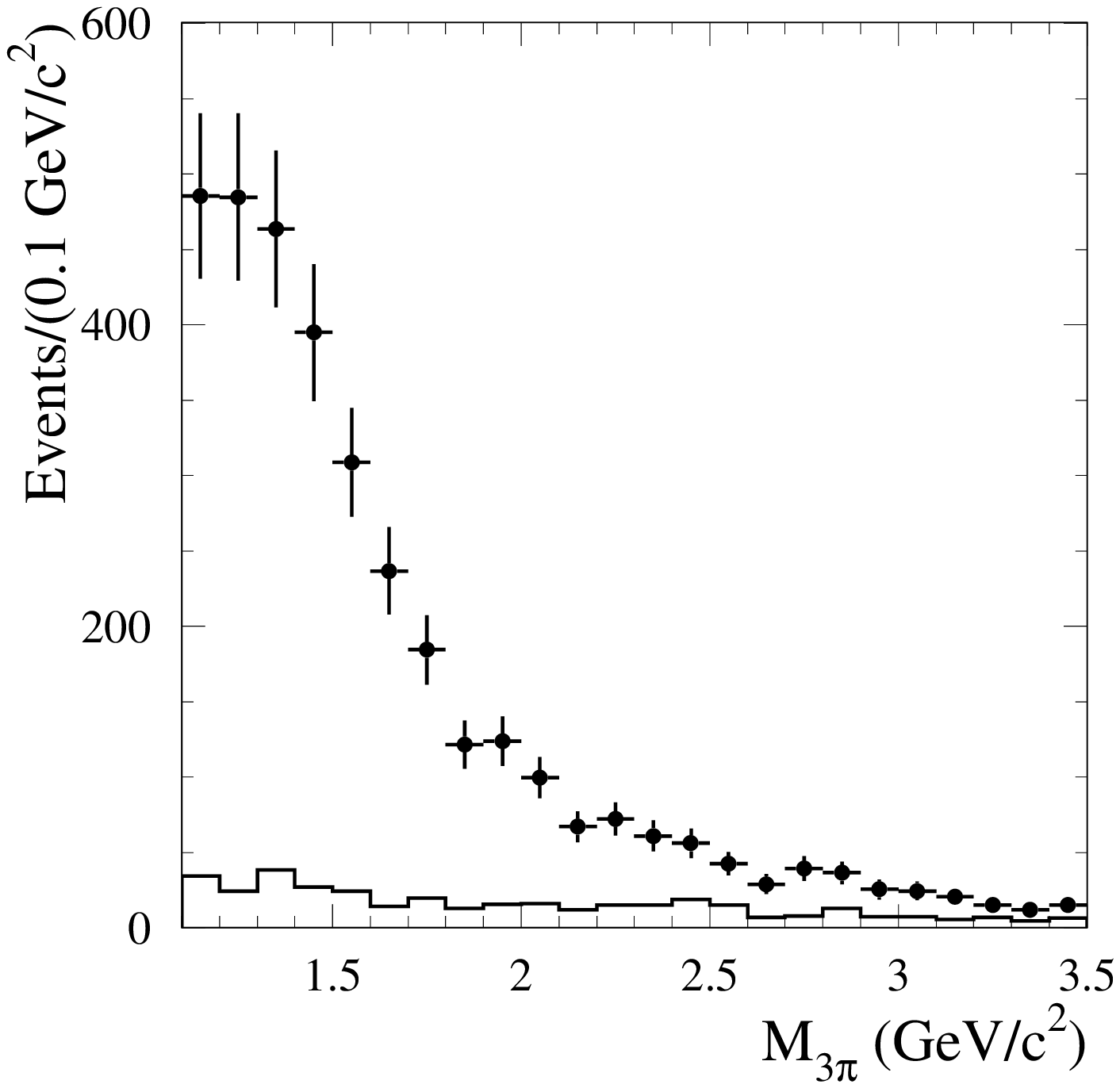}
\caption{Left panel: The $M_{3\pi}$ spectrum for selected data events with
$1.1<M_{3\pi}<3.5$ GeV/$c^2$ (open histogram). The dashed histogram at
$3.0<M_{3\pi}<3.2$ GeV/$c^2$ represents the spectrum after subtraction
of the $J/\psi$ resonance contribution (see Sec.~\ref{jpsi}).
The points with error bars show the calculated spectrum for background events.
The filled histogram represents the background spectrum with the
$e^+e^-\to\pi^+\pi^-\pi^0\pi^0$ contribution subtracted.
Right panel: The $M_{3\pi}$ spectrum for background events from all sources
except $e^+e^-\to\pi^+\pi^-\pi^0\pi^0$ and $e^+e^-\to K^+K^-\pi^0\gamma $ 
(points with error bars). The histogram represents the same spectrum 
with the $e^+e^-\to\pi^+\pi^-\pi^0\pi^0\gamma$ contribution subtracted.
\label{bkghigh}}
\end{figure*}
The $M_{3\pi}$ spectrum for selected data events with
$1.1<M_{3\pi}<3.5$ GeV/$c^2$ is shown in Fig.~\ref{bkghigh} (left).
The points with error bars in Fig.~\ref{bkghigh} (left) represent the 
calculated spectrum for background events, while the filled histogram shows
the background spectrum with the $e^+e^-\to\pi^+\pi^-\pi^0\pi^0$ contribution
subtracted. It is seen that
the process $e^+e^-\to\pi^+\pi^-\pi^0\pi^0$ becomes the dominant background
source above 1.5 GeV/$c^2$. This background has a $\chi^2_{3\pi\gamma}$ 
distribution similar to that for signal events. It is estimated as described
above and subtracted from the data $M_{3\pi}$ spectrum.
The $e^+e^-\to K^+K^-\pi^0\gamma$ 
background is also estimated from data. It is found
to be relatively small, about 4\% of the $e^+e^-\to\pi^+\pi^-\pi^0\pi^0$ 
contribution. Figure~\ref{bkghigh} (right) displays the calculated background
from all other sources. Here, the dominant contribution arises from the
$e^+e^-\to\pi^+\pi^-\pi^0\pi^0\gamma$ process. The next largest contribution
comes from non-$4\pi$ $q\bar{q}$ events. 

The mass region $1.1<M_{3\pi}<3.5$ GeV/$c^2$ is divided into 72 bins.
The bin width is 25 MeV/$c^2$ below 2.7 GeV/$c^2$ and 100 MeV/$c^2$ above.
In this region, the background from ISR processes, even 
from $e^+e^-\to\pi^+\pi^-\pi^0\pi^0\gamma$,
cannot be estimated with the same precision as at low masses, because
the MC event generator does not include many intermediate states
contributing to the ISR processes. Therefore, a procedure
of background subtraction based on the difference in $\chi^2_{3\pi\gamma}$ 
distributions for signal and background events is used.
In each mass bin, we subtract background events of the
$e^+e^-\to K^+K^-\pi^0\gamma$ and $e^+e^-\to\pi^+\pi^-\pi^0\pi^0$ processes
and determine the numbers of events with $\chi^2_{3\pi\gamma}\leq 20$
($N_1$) and $20<\chi^2_{3\pi\gamma}<40$ ($N_2$). 
The numbers of signal ($N_{\rm sig}$) and  remaining background
($N_{\rm bkg}$) events 
are then determined from the system of linear equations: 
\begin{eqnarray}
N_1&=&\alpha_{\rm sig} N_{\rm sig}+\alpha_{\rm bkg} N_{\rm bkg},\nonumber\\
N_2&=&(1-\alpha_{\rm sig}) N_{\rm sig}+(1-\alpha_{\rm bkg}) N_{\rm bkg}.
\label{bkg_sub}
\end{eqnarray}
The coefficients 
$\alpha=N_1/(N_1+N_2)$ for pure signal and background events are determined
from simulation. 
\begin{figure}
\centering
\includegraphics[width=0.95\linewidth]{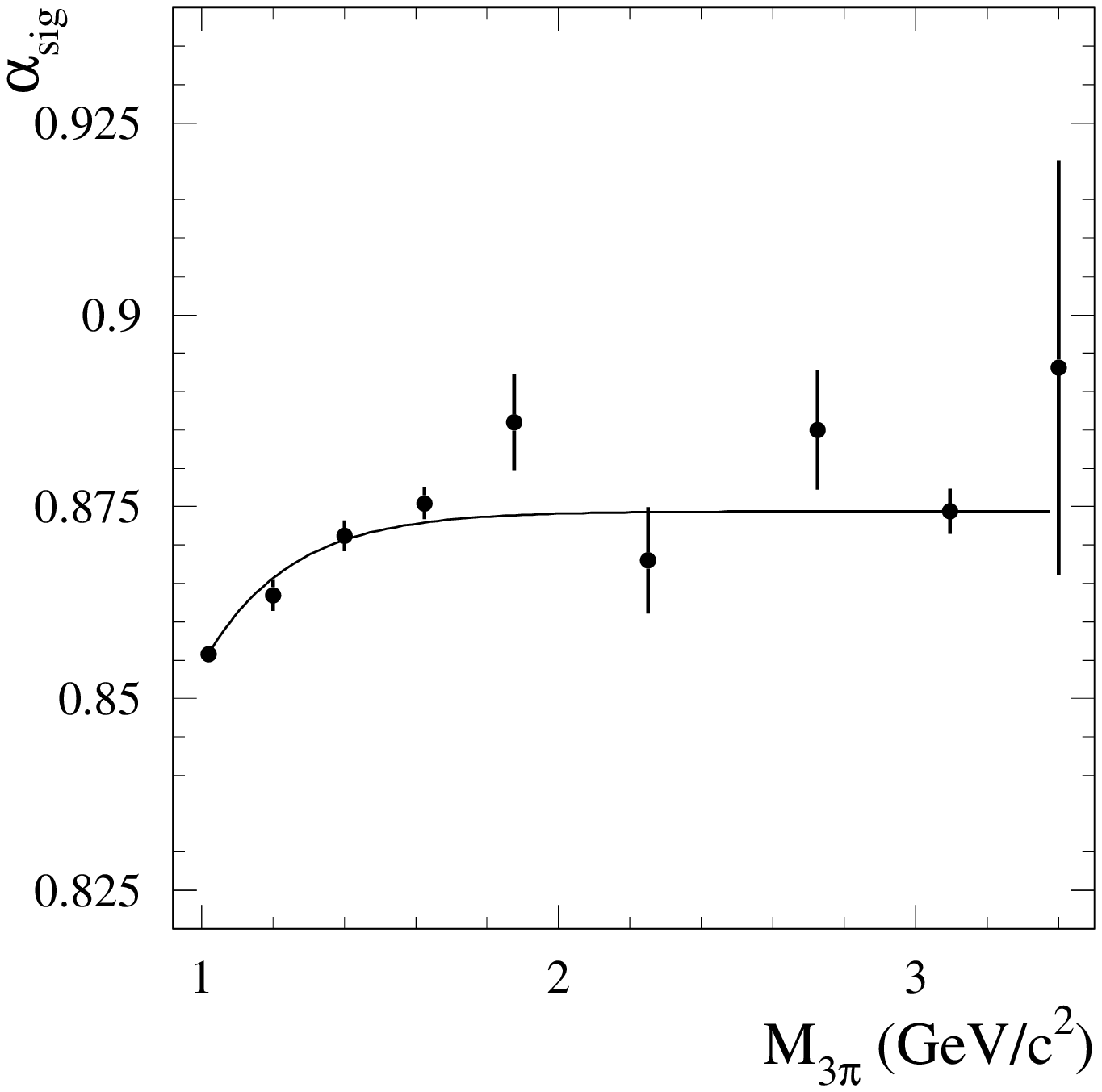}
\caption{
The $M_{3\pi}$ dependence of the $\alpha_{\rm sig}$ coefficient.
The dependence is fitted by the function $y=p_1[1-p_2\exp(-p_3M_{3\pi})]$,
in which two parameters $p_i$ are determined from the relations
$y(m_\phi)=\alpha_{\rm sig}(m_\phi)$ and 
$y(m_{J/\psi})=\alpha_{\rm sig}(m_{J/\psi})$.
\label{alpha1}}
\end{figure}

The mass dependence of the coefficient $\alpha_{\rm sig}$ is shown
in Fig.~\ref{alpha1}. The values of $\alpha_{\rm sig}$ at the $\phi$ and 
$J/\psi$ masses can be extracted from data. In the $\phi$ mass region, we 
determine $N_1$ and $N_2$ for pure signal events by subtracting
the calculated background. In the $J/\psi$ mass region, the same numbers
are obtained using a fit to the $M_{3\pi}$ spectrum by a sum of a $J/\psi$ 
line shape and a linear function (see Sec.~\ref{jpsi}). The resulting values 
of $\alpha_{\rm sig}$ are $0.859\pm 0.003$ at the $\phi$ mass and $0.890\pm 0.005$
at the $J/\psi$ mass. Their ratios to the corresponding values obtained from 
simulation are $R_\phi=1.004\pm 0.004$ and $R_{J/\psi}=1.018\pm 0.007$, 
respectively. In Eqs.~(\ref{bkg_sub}), we use for $\alpha_{\rm sig}$ the fitting
function shown in Fig.~\ref{alpha1} multiplied by a linear function 
interpolating between $R_\phi$ and $R_{J/\psi}$.

\begin{figure}
\centering
\includegraphics[width=0.95\linewidth]{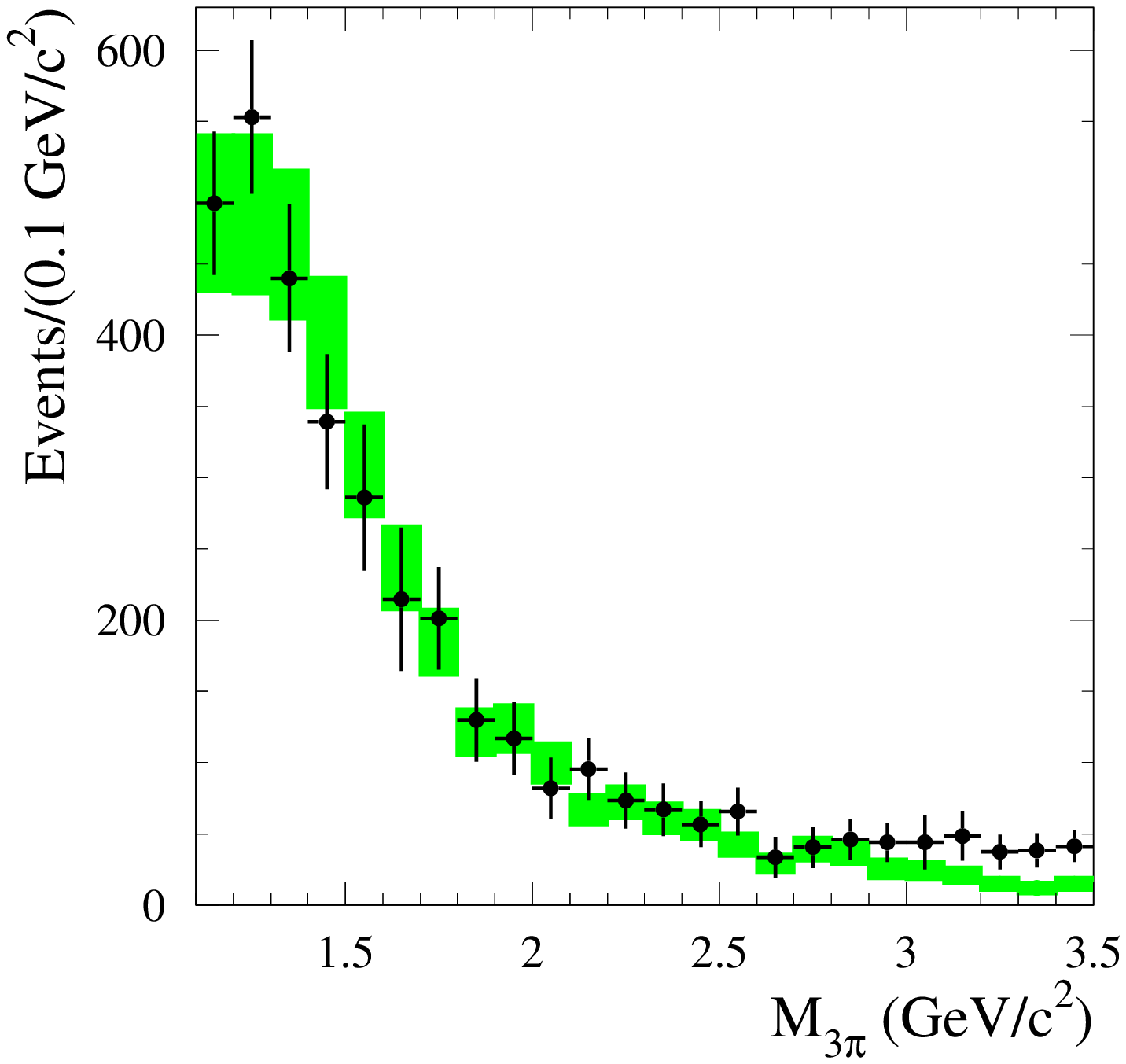}
\caption{The $M_{3\pi}$ spectrum for background events extracted from data in
each mass bin by solving the system of equations~(\ref{bkg_sub}) (point with 
error bars) compared with the spectrum obtained using simulation (filled
rectangles).
\label{alpha2}}
\end{figure}
The $\alpha_{\rm bkg}$ coefficient is determined using a mixture of background 
simulated events shown in Fig.~\ref{bkghigh} (right). The coefficient
is practically independent of mass and equal to
$0.316\pm0.007$. To estimate the systematic uncertainty in 
$\alpha_{\rm bkg}$, we vary the fraction of non-$\pi^+\pi^-\pi^0\pi^0\gamma$
events in the mixture of simulated background events by a factor of two. 
The variation in the $\alpha_{\rm bkg}$ value is taken as a measure of the 
systematic uncertainty. It is less than 5\% below 2 GeV/$c^2$, 
8\% between 2 and 3 GeV/$c^2$, and 15\% above 3 GeV/$c^2$.

\begin{figure*}
\centering
\includegraphics[width=0.7\textwidth]{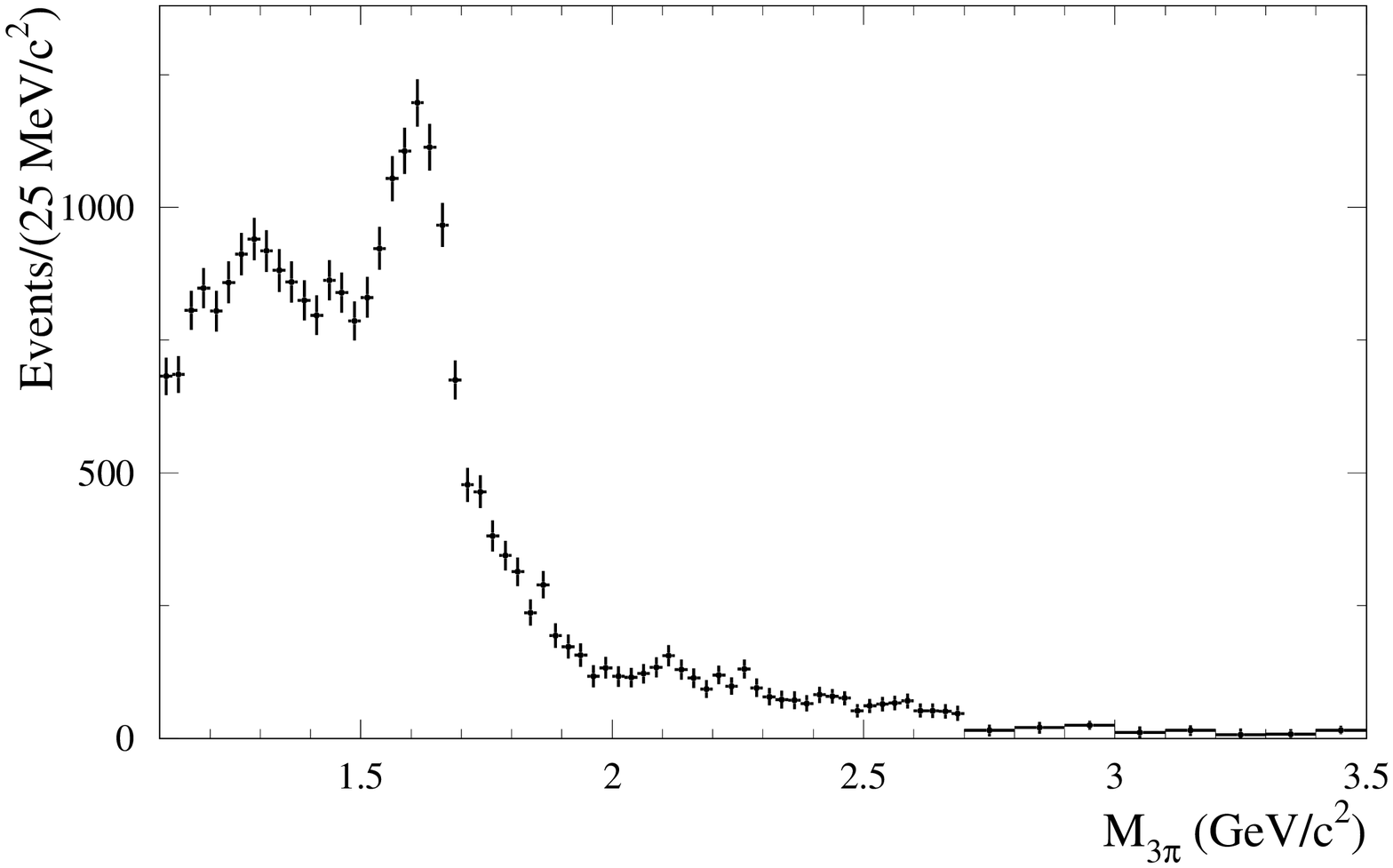}
\caption{The $M_{3\pi}$ spectrum for signal events extracted from data in 
each mass bin by solving the system of equations~(\ref{bkg_sub}).
\label{spec_high}}
\end{figure*}
The $M_{3\pi}$ spectrum for background events obtained by the solution of the
system of equations~(\ref{bkg_sub}) is shown in Fig.~\ref{alpha2}
in comparison with the same spectrum obtained using simulation. It is seen
that the simulation reproduces the data spectrum reasonably well up to 3 GeV.
The $M_{3\pi}$ spectrum for signal events is shown in Fig.~\ref{spec_high}.

\section{Final-state radiation}
A high-energy photon can be also emitted from the final state. Since
the $3\pi$ system in the ISR and final-state radiation (FSR) processes
has different $C$-parity, the contribution of the interference between them
to the total cross section vanishes when integrating over the final hadron
momenta. 

We analyze two FSR mechanisms. 
The first is emission of the photon by charged pions. Its
cross section is calculated as $\sigma_{3\pi}(10.58\mbox{ GeV})f_{\rm FSR}$,
where $f_{\rm FSR}$ is the FSR probability. The $e^+e^-\to\pi^+\pi^-\pi^0$
cross section at 10.58 GeV can be estimated from the CLEO
measurement at 3.67 GeV $\sigma_{3\pi}(3.67\mbox{ GeV})=
(13.1^{+1.0}_{-1.7}\pm 2.1)$~pb~\cite{cleo}.
Perturbative QCD (pQCD) predicts the same asymptotic energy dependence
$1/E^8$, where $E$ is the c.m. energy,
for all vector-pseudoscalar ($e^+e^-\to VP$) cross sections~\cite{qcd1,qcd2}.
This prediction can be tested experimentally using the CLEO~\cite{cleo} and
Belle~\cite{belle1,belle2} results for $e^+e^-\to VP$ cross sections at 
3.67~GeV
and 10.58~GeV, respectively. For the most accurately measured cross sections
for $e^+e^-\to \rho\eta$, $\omega\pi^0$, and $K^\ast K$, the ratio
$\sigma(3.67\mbox{ GeV})/\sigma(10.58\mbox{ GeV})\approx 3000$,
which corresponds to the dependence $1/E^{7.6}$. With this dependence,
$\sigma_{3\pi}(10.58\mbox{ GeV})$ is expected to be about 4.4 fb.

The mass region under study $M_{3\pi}<3.5$ GeV/$c^2$ corresponds to
the FSR photon c.m.~energy $E_\gamma^\ast > 4.7$ GeV. Such a photon can be 
radiated only by the most energetic pion in the $e^+e^-\to\pi^+\pi^-\pi^0$
process. For the dominant mechanism 
$e^+e^-\to \rho(770)\pi \to \pi^+\pi^-\pi^0$, the c.m.~energy of the
most energetic pion is 5.26 GeV. To estimate the FSR probability, we
use the formula for the FSR $e^+e^-\to\pi^+\pi^-\gamma$ cross 
section from Ref.~\cite{achasov} obtained for point-like pions. The FSR 
probability for the $\pi^+\pi^-$ final state at 10.52 GeV 
($f_{2\pi}(E_\gamma^\ast > 4.7\mbox{ GeV})=0.26\alpha/\pi$) must be multiplied
by a factor of $1/3$ (only the most energetic pion in the $3\pi$ final state
can emit such a photon and this pion must be charged).
Thus, the FSR contribution to the $e^+e^-\to\pi^+\pi^-\pi^0\gamma$ cross 
section under the assumption that the photon is emitted by charged pions
is estimated to be about 0.001 fb and is negligible.

The second FSR mechanism is photon emission from the quarks, which then 
hadronize into $\pi^+\pi^-\pi^0$. In the $3\pi$ mass region under study, this
process is expected to be dominated by production of $C=+1$ resonances
decaying to $\pi^+\pi^-\pi^0$, e.g., the processes $e^+e^- \to
\eta\gamma$, $a_1(1260)\gamma$, $a_2(1320)\gamma$, $\pi(1300)\gamma$.

The process $e^+e^- \to \eta\gamma$ has a $3\pi$ invariant mass well below 
the mass range under study. This process and the process $e^+e^- \to
\eta^\prime\gamma$ were studied by \babar\ in Ref.~\cite{etag}. The measured 
$e^+e^- \to \eta\gamma$ and $e^+e^- \to \eta^\prime\gamma$ cross sections
are $4.5^{+1.2}_{-1.1}\pm0.3$~fb and $5.4\pm 0.8\pm 0.3$~fb, respectively.
In Ref.~\cite{etag} they are compared with the pQCD prediction obtained
with asymptotic $\eta$ and $\eta^\prime$ distribution amplitudes,
2.2~fb and 5.5 fb, respectively.

The cross section for the processes $e^+e^- \to a_1(1260)\gamma$,
$a_2(1320)\gamma$ at large c.m.~energy is given by~\cite{chernyak}
\begin{equation}
\frac{d\sigma(e^+e^-\to M\gamma)}{d\cos{\theta_\gamma}}=
\frac{\pi^2\alpha^3}{4}|F_{M\gamma\gamma}|^2(1+\cos^2{\theta_\gamma}),
\label{fsrcs}
\end{equation}
where $F_{M\gamma\gamma}$ is a meson-photon transition form factor for the
helicity-zero state, which dominates at large momentum transfers,
\begin{equation}
q^2|F_{M\gamma\gamma}|=\frac{1}{3}\frac{|f_M|}{\sqrt{2}}I_M,
\end{equation}
and where $I_M$ is an integral depending on the shape of the meson 
distribution amplitude. For the asymptotic distribution amplitude,
$I_{a_1}=6$ and $I_{a_2}=10$. With the meson decay constants,
$f_{a_1}\approx 200$ MeV~\cite{fa1}, and $f_{a_2}\approx f_{f_2}\approx 110$ 
MeV~\cite{fa2}, the cross sections for the processes
$e^+e^- \to  a_1(1260)\gamma$ and $a_2(1320)\gamma$ are estimated to be
6.4 fb and 5.4 fb, respectively. There are no experimental data for these
cross sections. There is, however, a measurement of the 
$e^+e^- \to  f_2(1260)\gamma$ cross section at 10.58 GeV performed
by \babar~\cite{f2cs}: $(37^{+24}_{-18})$ fb, which is in reasonable
agreement with the prediction $\sigma_{f_2\gamma}\approx
(25/9)\sigma_{a_2\gamma}\approx 15$ fb~\cite{chernyak}.
The radiative process with an excited pion $e^+e^- \to \pi(1300)\gamma$ is 
expected to be small because of the suppression of the $\pi(1300)$ leptonic
decay constant~\cite{pi1300}.

The next group of $C$-even resonances decaying to $\pi^+\pi^-\pi^0$ is located
near 1.7 GeV. It consists of the radial excitations of $a_1$ and $a_2$ mesons,
$a_1(1640)$ and $a_2(1700)$, and the $D$-wave $q\bar{q}$ state
$\pi_2(1670)$.
We do not expect a significant decrease of the leptonic decay constants for
radially excited $P$-wave $q\bar{q}$ states compared with the ground states.
However, because of their larger masses their branching fractions to
$\pi^+\pi^-\pi^0$ must be lower. The theoretical predictions for them are 
about 30--50\%~\cite{barnes,pang}. For the $a_2$ family, we can assume that 
$f_{a_2(1700)}^2/f_{a_2(1320)}^2\sim
\Gamma(a_2(1700)\to \gamma\gamma)/\Gamma(a_2(1320)\to \gamma\gamma)$ and
use the measurements of the products 
$\Gamma(a_2(1320)\to \gamma\gamma){\cal B}(a_2(1320)\to\pi^+\pi^-\pi^0)=
0.65\pm 0.02\pm 0.02$ keV and
$\Gamma(a_2(1700)\to \gamma\gamma){\cal B}(a_2(1700)\to\pi^+\pi^-\pi^0)=
0.37\pm 0.10\pm 0.10$ keV~\cite{l3} to obtain 
$f_{a_2(1700)}^2{\cal B}(a_2(1700)\to\pi^+\pi^-\pi^0)\approx 0.4f_{a_2(1320)}^2$.
The same relation is used to estimate the 
$e^+e^- \to  a_1(1640)\gamma$ cross section.
The $\pi_2(1670)$ two-photon width is found to be low compared with 
$\Gamma(a_2(1320)\to \gamma\gamma)$~\cite{l3}. As a consequence, we neglect
the contribution of the $e^+e^- \to  \pi_2(1670)\gamma$ process.
\begin{figure}
\centering
\includegraphics[width=0.95\linewidth]{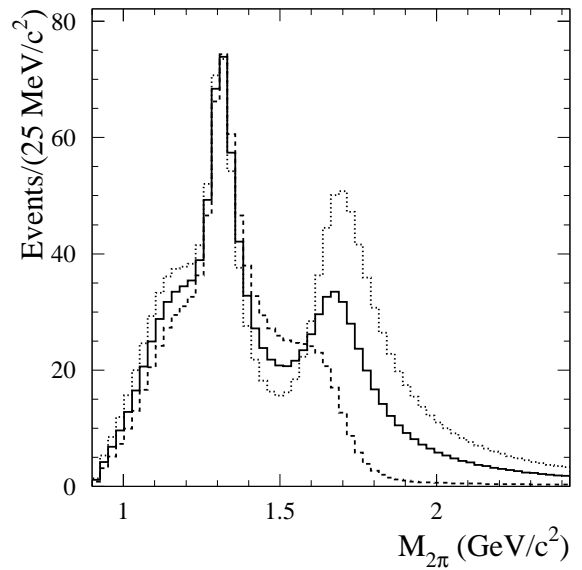}
\caption{The expected $M_{3\pi}$ spectrum from the FSR processes
$e^+e^- \to M\gamma\to \pi^+\pi^-\pi^0\gamma$, where $M=a_1(1260)\gamma$,
$a_2(1320)\gamma$, $a_1(1640)$, and $a_2(1700)$. The solid histogram
represents the incoherent sum of the processes. The dotted (dashed) histogram
demonstrates the effect of constructive (destructive) interference between
the $a_1(1260)$ and $a_1(1640)$ amplitudes, and the $a_2(1320)$ and 
$a_2(1700)$ amplitudes.
\label{fsr}}
\end{figure}

To estimate the detection efficiency for the FSR processes we assume that the
efficiency is weakly dependent on the internal structure of the $3\pi$ state
and reweight 
simulated ISR $e^+e^- \to 3\pi\gamma$ events to reproduce the photon angular
distribution given by Eq.~(\ref{fsrcs}). The obtained detection efficiency
at the $a_2(1320)$ mass is 17.9\% for the standard selection criteria.
The $3\pi$ mass distribution for the $e^+e^- \to M\gamma$ process has
a resonance shape. The expected mass spectrum for the FSR processes, calculated
as a sum of the $ a_1(1260)$, $a_2(1320)$, $a_1(1640)$, and $a_2(1700)$ 
Breit-Wigner functions,
is shown in Fig.~\ref{fsr} by the solid histogram. Interference
between amplitudes of different resonances may strongly modify
this spectrum. The effect of interference is demonstrated in Fig.~\ref{fsr}.
We take into account the interference between the $ a_1(1260)$ and
$a_1(1640)$ amplitudes, and $a_2(1320)$ and $a_2(1700)$ amplitudes, but
neglect the interference between $a_1$ and $a_2$ states. The dotted
(dashed) histogram represents the result for relative phases between 
resonances equal to 0 ($\pi$). 
We subtract the spectrum without interference from the spectra for the selected
data events shown in Figs.~\ref{bkg1} and \ref{spec_high}. The systematic 
uncertainty in the FSR contribution, which takes into account the uncertainty
in the theoretical prediction and the effect of interference, is estimated to
be 100\%. The fraction of the FSR background is maximal (7--8\%) in the region
1.05--1.08 GeV/$c^2$, where the measured $M_{3\pi}$ spectrum has a minimum,
and near $M_{3\pi}=1.32$ GeV/$c^2$. Near 1.7 GeV/$c^2$, the background fraction
is about 6\%. 

In the mass region near 2 GeV/$c^2$, there are several poorly
established excited $a_1$ and $a_2$ states~\cite{pdg}. We model their 
contribution by a sum of the $a_1(1930)$ and $a_2(2030)$ resonances 
assuming that 
$f_{a_1(1930)}{\cal B}(a_1(1930)\to\pi^+\pi^-\pi^0)\approx 0.2 f_{a_1(1260)}^2$ and
$f_{a_2(2030)}{\cal B}(a_2(2030)\to\pi^+\pi^-\pi^0)\approx 0.2 f_{a_2(1320)}^2$.
The latter relation is based on the results of the measurement of
the $\gamma\gamma\to \pi^+\pi^-\pi^0$ cross section in Ref.~\cite{l3}.
We find that the radiative production of the excited $a_1$ and $a_2$ 
states with mass near 2 GeV/$c^2$ may give a 10\% contribution to the
measured $M_{3\pi}$ spectrum above 1.8 GeV. This value is taken as an estimate
of the systematic uncertainty associated with FSR at $M_{3\pi}>1.8$
GeV/$c^2$.

\section{Detection efficiency}\label{sdetef}
\begin{figure*}
\centering
\includegraphics[width=0.9\textwidth]{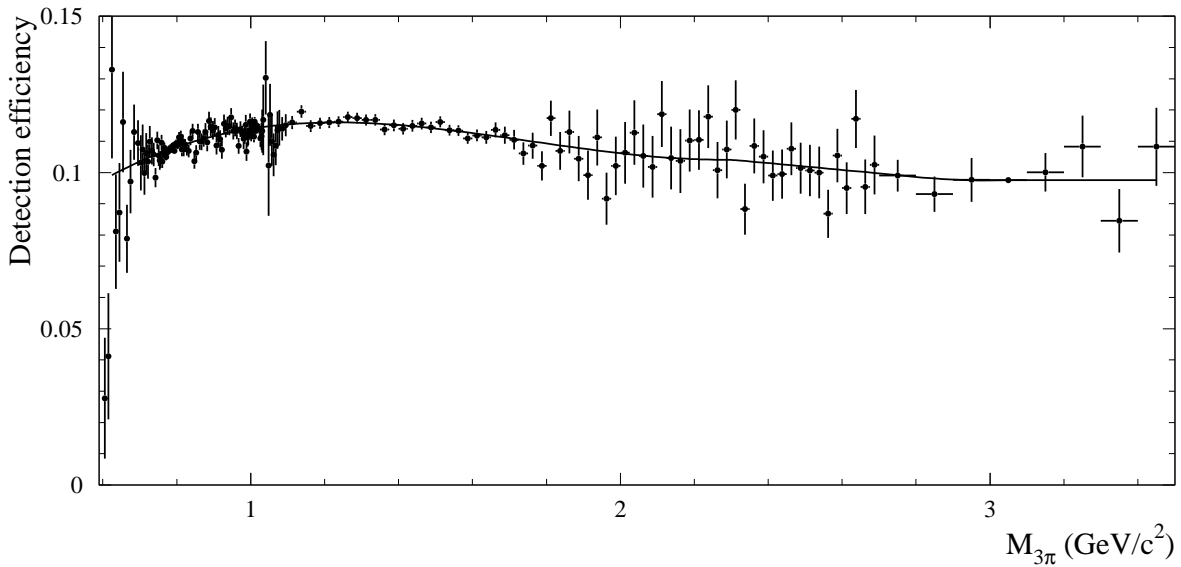}
\caption{The $3\pi$ mass dependence of the detection efficiency obtained
using MC simulation. The curve is the result of the fit described in the text. 
\label{deteff}}
\end{figure*}

The detection efficiency is determined using MC simulation as the ratio of 
the true $3\pi$ mass spectra computed after and before applying the selection
criteria. The detection efficiency calculated in this way is
shown in Fig.~\ref{deteff}. Its mass dependence is fitted by a combination of 
a third-order polynomial in the range 0.62--2.3 GeV/$c^2$, a linear function 
in the range 2.3--2.9 GeV/$c^2$, and a constant above 2.9 GeV/$c^2$. 
For the tighter requirement $\chi^2_{3\pi\gamma}<20$ the detection efficiency
is smaller by 12--17\%. The statistical uncertainty of the fitted detection
efficiency is about 0.1\% at the $\omega$, 0.2\% at the $\phi$, and then
increases to 1\% at 2 GeV/$c^2$ and up to 2.2\% at 2.5 GeV and above.

The decrease in efficiency below 0.62 GeV/$c^2$ is due to the merging of 
clusters from photons and charged pions in the calorimeter. This effect
leads to $\pi^0$ loss, which increases as the $3\pi$ mass decreases. To avoid 
a possible systematic uncertainty due to imperfect simulation of this 
effect, we perform the measurement of the $e^+e^-\to \pi^+\pi^-\pi^0$ cross 
section at masses above 0.62 GeV/$c^2$.

The efficiency ($\varepsilon_{MC}$) found using MC simulation must be 
corrected to account for data-MC simulation differences in detector response:
\begin{equation}
\varepsilon=\varepsilon_{MC}\Pi_i (1+\delta_i),
\label{eff}
\end{equation}
where $\delta_i$ are efficiency corrections for the different effects discussed
below.

\subsection{ISR photon inefficiency\label{phef}} 
A correction is applied to the ISR photon detection efficiency. There are two
sources of this correction: data-MC simulation differences in the probability
of photon conversion in the detector material before the DCH, and dead 
calorimeter channels.  A sample of $e^+e^-\to \mu^+\mu^-\gamma$ events is 
used to determine the calorimeter photon inefficiency in data. Events with
exactly two charged tracks identified as muons are selected, and a 
one-constraint kinematic fit is performed with the requirement that the recoil
mass against the muon pair be zero. A tight condition on the $\chi^2$ of the
kinematic fit selects events with only one photon in the final state. The 
photon direction is determined from the fit. The detection inefficiency is
calculated as the ratio of the number of events not satisfying the condition
$E_\gamma^\ast>3$ GeV, to the total number of selected events. The same 
procedure is applied to simulated $e^+e^-\to \mu^+\mu^-\gamma$ events. 
The efficiency correction is determined from the data-MC simulation ratio
as a function of the photon polar angle and the $\mu^+\mu^-$ invariant mass. 
The data-MC simulation difference in the probability of photon conversion is 
also studied using $e^+e^-\to \mu^+\mu^-\gamma$ events.  In
addition to two identified muons, we require that an event contain a
converted-photon candidate, i.e., a pair of oppositely charged tracks
with $e^+e^-$ invariant mass close to zero, momentum directed along the 
expected photon direction, and forming a secondary vertex well separated from
the interaction region. The data-MC difference in the probability 
of photon conversion is measured as a function of the photon polar angle.
Then we calculate the total correction to the ISR photon efficiency due to
calorimeter inefficiency and photon conversion.

The measured angular dependence of the correction is used to reweight the 
simulated $e^+e^-\to \pi^+\pi^-\pi^0\gamma$ events and calculate the efficiency
correction. It is found to be $-(1.0\pm0.2)\%$ for $3\pi$ masses below 
1.1 GeV/$c^2$, $-(1.2\pm0.2)\%$ in the mass range 1.1-2.0 GeV/$c^2$, 
and $-(1.4\pm0.2)\%$ in the range 2.0-3.5 GeV/$c^2$. The contribution to this
correction from photon conversion is about $-0.2\%$.

\subsection{
$\pi^0$ efficiency
and kinematic-fit $\chi^2$ distribution}
From the study of the ISR photon inefficiency it is expected that the 
difference between data and simulation in the $\pi^0$ detection efficiency
is at least $-2\%$. To study the $\pi^0$ losses more accurately, we perform
a kinematic fit for data and simulated events to the
$e^+e^- \to \pi^+\pi^-\pi^0\gamma$ hypothesis using the measured parameters
for only the two charged tracks and the ISR photon.  The $\pi^0$ energy and
angles are determined as a result of the fit.  
We apply a very tight condition on the fit quality and the background
suppression conditions described in Sec.~\ref{evsel}.
Because of the high level of remaining background, we restrict our study to the
$\omega$ mass region. 

The $\pi^0$ detection efficiency is determined as the fraction of selected 
signal events with a detected $\pi^0$. The result depends on the definition of 
the $\pi^0$ candidate. For the simple $\pi^0$ definition as a pair of photons
with invariant mass near the $\pi^0$ mass, for example, in the range 
0.1--0.17 GeV/$c^2$, there is a substantial probability to observe a false 
$\pi^0$ candidate due to a large number of spurious photons in an event. To
avoid difficulties with false $\pi^0$'s, we require that an event containing 
the $\pi^0$ candidate satisfy our standard kinematic-fit condition 
$\chi^2_{3\pi\gamma}<40$.

\begin{figure*}
\centering
\includegraphics[width=0.47\textwidth]{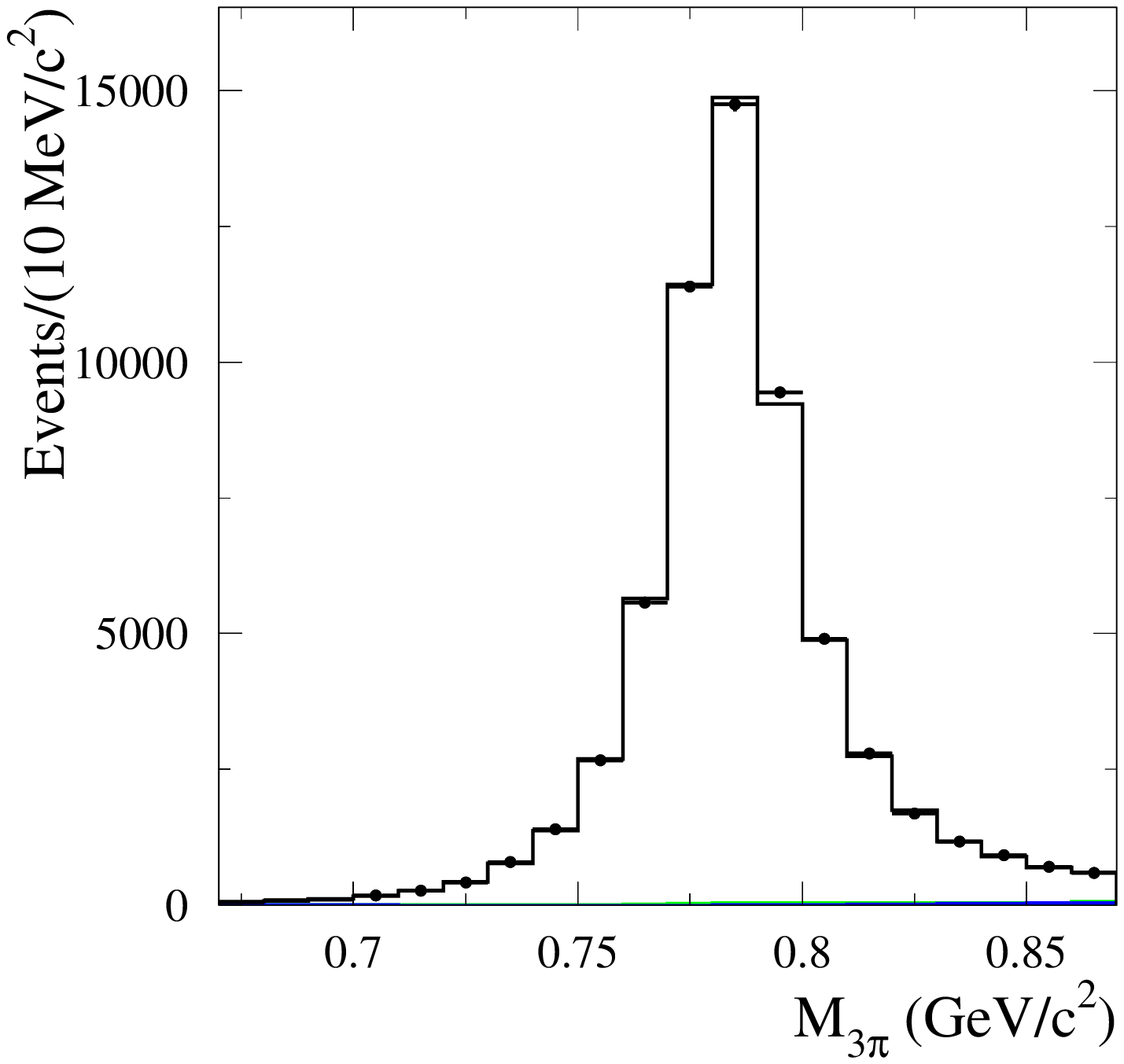}\hfill
\includegraphics[width=0.47\textwidth]{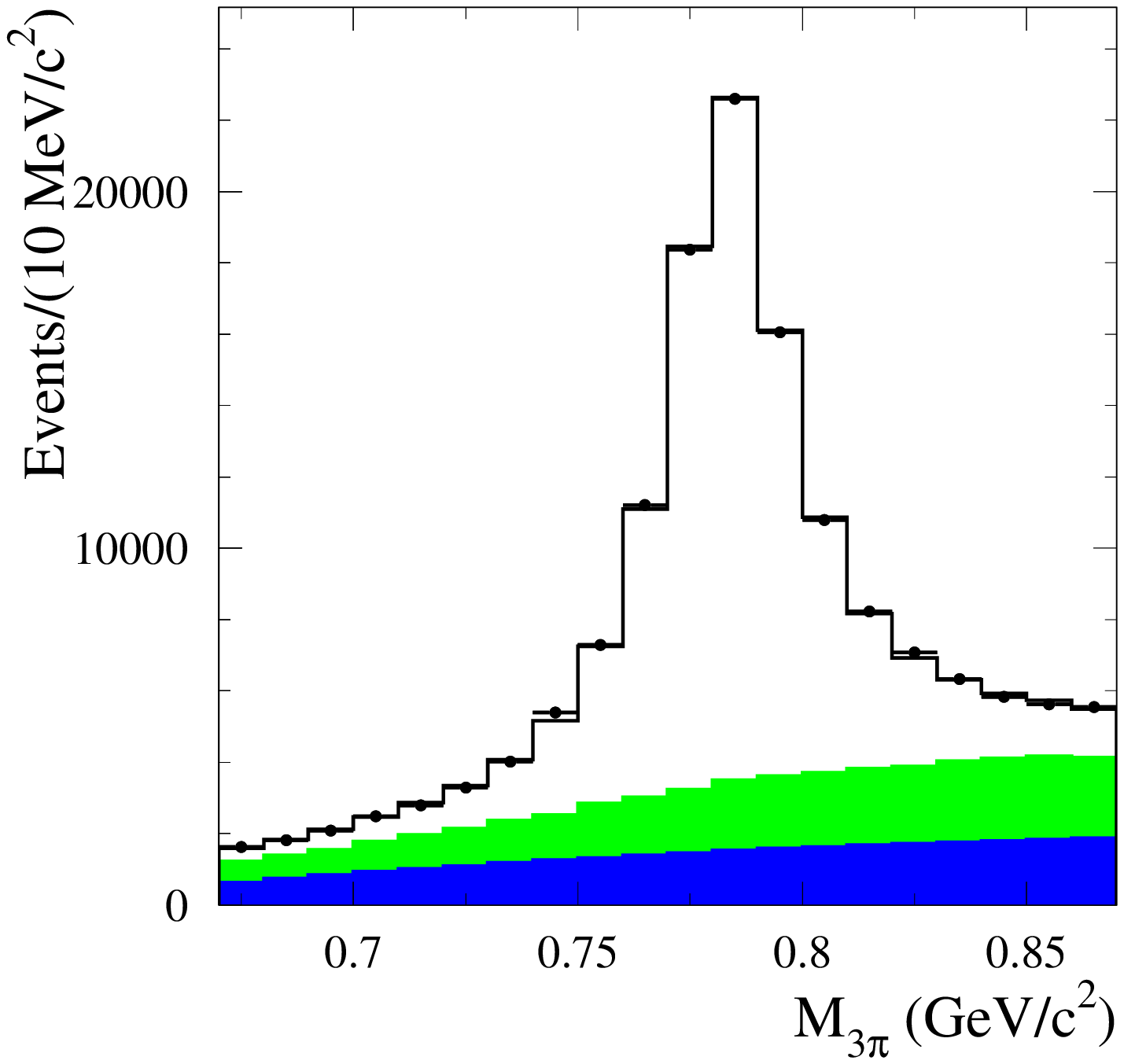}
\caption{The $3\pi$ mass distributions for events
selected without using the photons from the $\pi^0$ decay.
The left (right) plot is for events with $\chi^2_{3\pi\gamma} < 40$ ($\chi^2_{3\pi\gamma} > 40$).
The points with error bars show the data distribution. The solid histogram is
the result of the fit described in the text. The light shaded (green) region
represents the simulated $e^+e^- \to\pi^+\pi^-\pi^0\pi^0\gamma$ contribution. 
The dark shaded (blue) histogram is the fitted background contribution from other 
sources. 
\label{pi0loss}}
\end{figure*}

The $3\pi$ mass spectra for selected events with $\chi^2_{3\pi\gamma} < 40$ and 
$\chi^2_{3\pi\gamma} >40$ are shown in Fig.~\ref{pi0loss}.
The mass spectra are fitted with a sum of distributions
for signal and background events. The signal distribution is extracted from
the simulation. The background spectrum is a sum of the simulated distribution
for $e^+e^- \to \pi^+\pi^-\pi^0\pi^0\gamma$ events and a second-order
polynomial with free coefficients. The efficiency correction due to $\pi^0$
losses is determined to be 
$\delta_2=f_{\rm data}/f_{\rm MC}-1=-(3.4\pm 0.5)\%$.
Here $f$ is the fraction of selected events with $\chi^2_{3\pi\gamma} < 40$.

\begin{figure}
\centering
\includegraphics[width=0.95\linewidth]{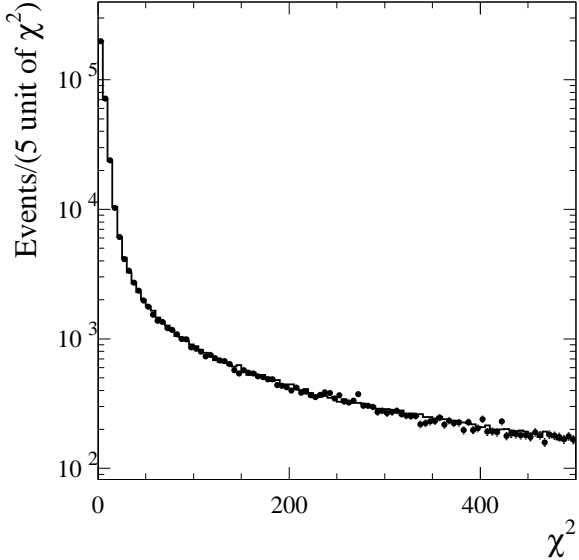}
\caption{ 
The distributions of the $\chi^2$ of the kinematic fit for selected data 
(histogram) and simulated (points with error bars) $e^+e^-\to \mu^+\mu^-\gamma$
events with the $\mu^+\mu^-$ invariant mass in the range 0.2-1.1 GeV/$c^2$. 
\label{pi0loss1}}
\end{figure}
In a similar way, we determine the efficiency correction in different
ranges of the $\pi^0$ energy. At the current level of statistical precision
the correction is found to be independent of
the $\pi^0$ energy. Therefore, the efficiency correction due to $\pi^0$
losses determined at the $\omega$ region is also used for higher $3\pi$
masses.

The $\pi^0$ correction includes a part of the efficiency correction due to the
$\chi^2_{3\pi\gamma}<40$ requirement related to the photons from the $\pi^0$
decay.
To understand the influence of the data-simulation difference in the parameters
of the charged tracks and the ISR photon, we study $e^+e^-\to \mu^+\mu^-\gamma$
events. We select events with two charged particles identified as muons and a 
photon with c.m.~energy larger than 3 GeV. As mentioned in
Sec.~\ref{detector},
the simulation uses the requirement that the invariant mass of the muon pair 
and ISR photon be greater than 8 GeV/$c^2$. To ensure compliance with this 
requirement in the data, we apply an additional condition that the invariant
mass of the muon and the ISR photon candidates be greater than 9 GeV/$c^2$. 

For such selected events, a kinematic fit is performed with the requirements 
of energy and momentum balance. The fit uses measured momenta and angles of the
muons and only angles of the ISR photon. The $\chi^2$ distributions for 
selected data and simulated $e^+e^-\to \mu^+\mu^-\gamma$ events with 
$\mu^+\mu^-$ invariant mass $M_{\mu\mu}<1.1$ GeV/$c^2$ are compared in 
Fig.~\ref{pi0loss1}. It is seen that the data and simulated
distributions are in agreement. To estimate the difference between them 
numerically, we calculate the double ratio 
$R_\chi^2=[N(\chi^2<c)/N_0]_{\rm data}/[N(\chi^2<c)/N_0]_{\rm MC}$,
where $N_0$ is the total number of selected $\mu^+\mu^-\gamma$ events, and
$N(\chi^2<c)$ is the number of events satisfying the condition $\chi^2<c$. This
ratio is practically independent of the $c$ value in the range $20<c<40$.
Its deviation from unity, $R_\chi^2-1$, in the invariant mass
range $0.6<M_{\mu\mu}<1.1$ GeV/$c^2$, equal to $-(0.4\pm 0.2)\%$, can be 
used as an estimation of the efficiency correction for $e^+e^-
\to\pi^+\pi^-\pi^0\gamma$ events.

We take into account the difference in the charged-particle 
momentum distributions for the processes $e^+e^-\to\mu^+\mu^-\gamma$ and 
$e^+e^- \to \pi^+\pi^-\pi^0\gamma$. To understand a possible effect of this 
difference, we study the dependence of $R_\chi^2$ on the minimum muon momentum
in an event and do not observe any statistically significant dependence.
However, since the phase space distribution of charged pions from the 
$e^+e^- \to \pi^+\pi^-\pi^0\gamma$ reaction cannot be fully reproduced using 
the $e^+e^-\to\mu^+\mu^-\gamma$ events, we assign a 100\% systematic 
uncertainty to this correction.

In summary, the efficiency correction associated with the difference in the $\chi^2$
distribution between data and simulation is estimated to be $-(0.4\pm 0.4)\%$
in the $3\pi$ mass region 0.6--1.1 GeV/$c^2$. For higher masses the
correction is larger. Its average value in the mass range 1.1-3.5 GeV/$c^2$ is
$(1\pm 1)\%$.

\subsection{Efficiency correction due to the selection criteria}
Our preliminary selection contains the requirement of exactly two good quality
charged tracks in an event. The definition of a good charged track is given
in Sec.~\ref{evsel}. To determine an efficiency correction due to this
requirement, we analyze events with three good tracks.
Two of them with opposite charge having closest distance to the 
beam axis are selected as candidates for charged pions from 
the $e^+e^- \to \pi^+\pi^-\pi^0\gamma$ reaction.
The fraction of three-track events determined in the $3\pi$ mass
regions near the $\omega$ and $\phi$ resonances is about 0.4\% both in
data and simulation. No efficiency correction due to the 
requirement of exactly two charged tracks is needed.

Radiative Bhabha events are rejected by the requirement that none
of the good charged tracks be identified as an electron. The rejected 
events are prescaled by a factor of 40. We study a sample of prescaled 
events passing our standard selection criteria, except for the electron
identification requirement, and find that the 
efficiency correction is $-(0.01\pm 0.12)\%$.

The efficiency correction for the background suppression requirements
described in Sec.~\ref{evsel} is determined near the $\omega$, $\phi$, and
$J/\psi$ resonances from ratios of the number of events selected with and
without these requirements, in data and MC simulation. The fraction of signal
events rejected by the background suppression requirements varies from
15\% in the $\omega$ and $\phi$ mass region to 25\% at the $J/\psi$. 
This dependence is reproduced by the simulation.  The efficiency correction is
$(0.4\pm 0.2)\%$ at the $\omega$ and $\phi$, and $(0.6\pm0.8)\%$ at 
the $J/\psi$. The latter correction is used in the energy region above
1.1 GeV.

\subsection{Efficiency correction due to track losses}
\begin{figure}
\centering
\includegraphics[width=0.95\linewidth]{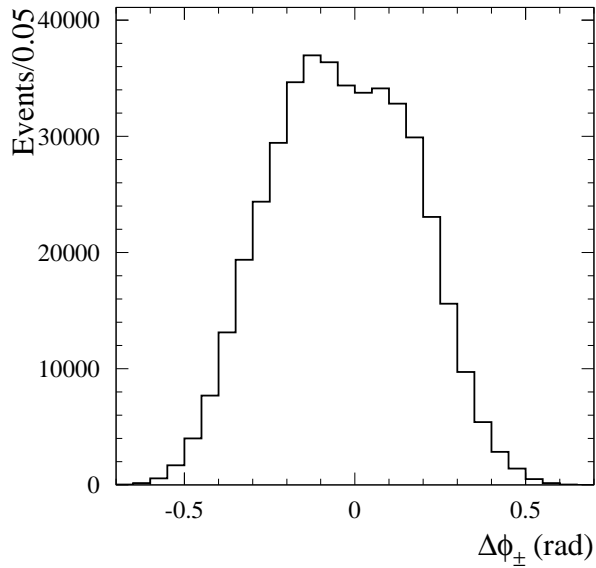}
\caption{The $\Delta \varphi_\pm$ distribution for simulated signal events
from the $\omega$ mass region.
\label{dphipm}}
\end{figure}
The data-MC simulation difference in track losses for isolated tracks is
studied using $e^+e^-\to \tau^+\tau^-$ events, with one $\tau$ decaying
leptonically and the other $\tau$ hadronically with three charged particles.
No difference between data and simulation in the tracking efficiency
is observed within an uncertainty of 0.24\% per track. In $e^+e^-\to 
\pi^+\pi^-\pi^0\gamma$ events, especially at small $M_{3\pi}$, the angle 
between charged tracks may be small, and the effect of track overlap in the
DCH should be taken into account. To study this effect we analyze the
distribution of the azimuthal angle difference between the positive and 
negative tracks $\Delta \varphi_\pm=\varphi^+-\varphi^-$. In the \babar\
magnetic field, events with $\Delta \varphi_\pm>0$ exhibit a ``fishtail'' 
two-track configuration in which the tracks tend to overlap. The
$\Delta \varphi_\pm$ distribution for simulated signal events from the
$\omega$ mass region is shown in Fig.~\ref{dphipm}. The track overlap
leads to an asymmetry in the distribution. It should be noted that 
larger $\Delta \varphi_\pm$ values correspond to larger differences between
charged pion momenta. Therefore, the asymmetry in the distribution is seen 
even at relatively large $\Delta \varphi_\pm\sim 0.5$. With larger values of
$M_{3\pi}$, the $\Delta \varphi_\pm$ distribution becomes wider and more
symmetric. We estimate the fraction of events lost because of track overlap as 
\begin{equation}
f_{\rm overlap}=\frac{N(\Delta \varphi_\pm<0)-N(\Delta \varphi_\pm>0)}
{2N(\Delta\varphi_\pm<0)}.
\end{equation}
This fraction is 11\% at the $\omega$, 8\% at the $\phi$, and about 1\%
at the $J/\psi$. We do not observe any significant difference in this
fraction between data and simulation. The difference calculated over the
$3\pi$ mass range 0.6--1.1 GeV/$c^2$ is $-(0.03\pm 0.23)\%$. The uncertainty
in this difference is used as an estimate of the systematic uncertainty
associated with track overlap in the DCH.
The total systematic uncertainty in the detector efficiency due to data-MC 
simulation differences in the tracking efficiency and track overlap is 
estimated to be 0.5\%. 

\subsection{Efficiency correction due to trigger and background
filters }
We also studied the quality of the simulation of the trigger and background
filters~\cite{bgf}
used in event reconstruction. In the analysis we use events 
passing through the two trigger lines L3OutDCH and L3OutEMC, which are
based on signals from the DCH and EMC, respectively.
The inefficiency of these lines in the simulation is 6.1\% for
L3OutDCH and $(2.3\pm 0.2)\times 10^{-4}$ for L3OutEMC. A logical OR of the 
L3OutDCH and L3OutEMC lines has a very small inefficiency,
$(1.2\pm 0.1)\times 10^{-4}$. 
The inefficiencies for the trigger lines in data can be estimated using the 
overlap of the samples of events passing through them. The simulation
shows that these estimates are very close to the true inefficiencies of the
trigger lines. This method applied to data results in an inefficiency of
$(6.6\pm 0.1)\%$ for L3OutDCH and $(3.8\pm 0.6)\times 10^{-4}$ for L3OutEMC. 
Although the efficiency in data is lower, the efficiency itself is very
close to 100\%. Therefore, no correction is applied for the trigger 
inefficiency.

The inefficiency in the background filters in simulation is about
1.8\% at the $\omega$ and $\phi$ mass regions and then decreases to 0.5\%
at 2 GeV/$c^2$ and to 0.3\% at the $J/\psi$. To measure this
inefficiency in data we use a subsample of prescaled events that
does not pass through the background filters. The prescale factor
is 200. The filter inefficiency for $3\pi$ masses below 1.1 GeV is
measured to be $(3.2\pm 0.7)\%$. The efficiency correction in this mass region
is $-(1.4\pm 0.7)\%$. For $M_{3\pi}>1.1$ GeV/$c^2$, insufficient statistical
precision and large background do not allow us to determine the
inefficiency with acceptable accuracy. Therefore, in this region we use the
correction $-(1\pm 1)\%$, which covers the range of its possible variations
as a function of mass.

The efficiency corrections $\delta_i$ are summarized in Table~\ref{effcorr}.
The total efficiency correction is about $-6\%$.
\begin{table*}
\caption{Efficiency corrections (in \%) for different effects in three
$M_{3\pi}$ regions.\label{effcorr}}
\begin{ruledtabular}
\begin{tabular}{lccc}
Effect & $M_{3\pi}<1.1$ GeV/$c^2$ & $1.1<M_{3\pi}<2$ GeV/$c^2$ & $M_{3\pi}>2$ GeV/$c^2$\\
\hline
Photon efficiency         & $-1.0\pm 0.2$ & $-1.2\pm 0.2$ & $-1.4\pm 0.2$ \\
$\pi^0$ loss              & $-3.4\pm 0.5$ & $-3.4\pm 0.5$ & $-3.4\pm 0.5$\\
$\chi^2_{3\pi\gamma}$ distribution     & $-0.4\pm 0.4$ & $-1\pm 1$     & $-1\pm 1$     \\
Rad. Bhabha suppression   & $0.0\pm 0.1$ & $0.0\pm 0.1$ & $0.0\pm 0.1$ \\
Background suppression    & $0.4\pm 0.2$ & $0.6\pm 0.5$ & $0.6\pm 0.8$ \\
Track loss                & $0.0\pm 0.5$ & $0.0\pm 0.5$ & $0.0\pm 0.5$ \\
Trigger and background filters       & $-1.4\pm 0.7$ & $-1\pm 1$ & $-1\pm 1$ \\
\hline
Total                  & $-5.8\pm1.1$  & $-6.0\pm1.7$ &$-6.2\pm1.8$      \\
\hline
$\chi^2_{3\pi\gamma}<20$ & $0.1\pm0.1$ at $\omega$ & 0.5--1.1  & 1.1--1.8 \\
                         & $0.4\pm0.4$ at $\phi$ \\
\end{tabular}
\end{ruledtabular}
\end{table*}

In Sec.~\ref{masssp}, we also analyze the mass spectrum for events with
$\chi^2_{3\pi\gamma}<20$. The additional correction related to this
requirement is $0.001\pm0.001$ at the $\omega$, $0.004\pm0.004$ at the $\phi$,
and $0.018\pm0.007$ at the $J/\psi$. A linear interpolation is used between
the resonances.

\subsection{Model uncertainty} 
The signal simulation uses the model of the $\rho(770)\pi$ intermediate state.
This model works reasonably well for the the $\omega$ and $\phi$
decays~\cite{bes-om,kloe-phi}.
A comparison of the data and simulated two-pion distributions in different
mass regions for \babar\ $3\pi$ data was performed in our previous
work~\cite{babar}. Data and simulation agree well below 1.1 GeV, in the
$\omega$ and $\phi$ regions. For higher masses, the difference was observed
associated with additional intermediate mechanisms $\omega\pi^0$ and 
$\rho(1450)\pi$.

To estimate the model dependence of
the detection efficiency in the mass range 1.1--2 GeV/$c^2$, the simulated
signal events are reweighted using the model with a sum of the $\rho(770)\pi$,
$\omega\pi^0$, and $\rho(1450)\pi$ mechanisms with coefficients and relative 
phases taken from the SND measurement~\cite{snd3}. The difference
in the detection efficiencies between the two models depends on energy 
but does not exceed 1.5\%. This number is taken as an estimate of the model
uncertainty in the detection efficiency in the region 1.1--3.5 GeV/$c^2$.

A similar procedure is used to find the correction to the detection
efficiency at the $J/\psi$. Here we use the result of the Dalitz plot analysis
of Ref.~\cite{babar-psi}. 
An $\sim$10\% contribution from the $\rho(1450)\pi$ channel leads
to a shift in the detection efficiency of $-(0.5\pm0.1)\%$.

\section{
Fit to the $\pi^+\pi^-\pi^0$ invariant mass 
distribution \label{masssp}}
\begin{figure*}
\centering
\includegraphics[width=0.47\textwidth]{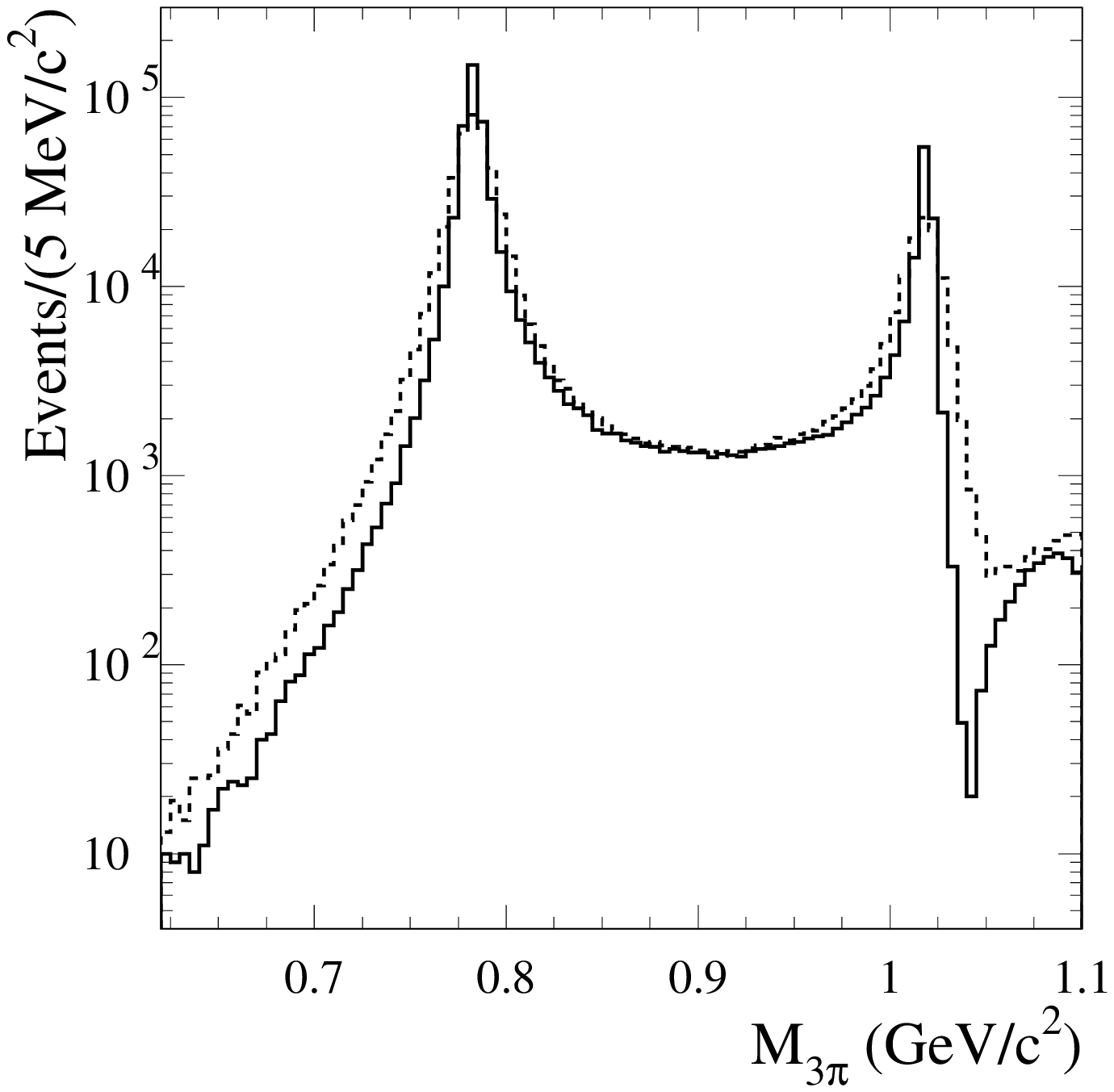}\hfill
\includegraphics[width=0.47\textwidth]{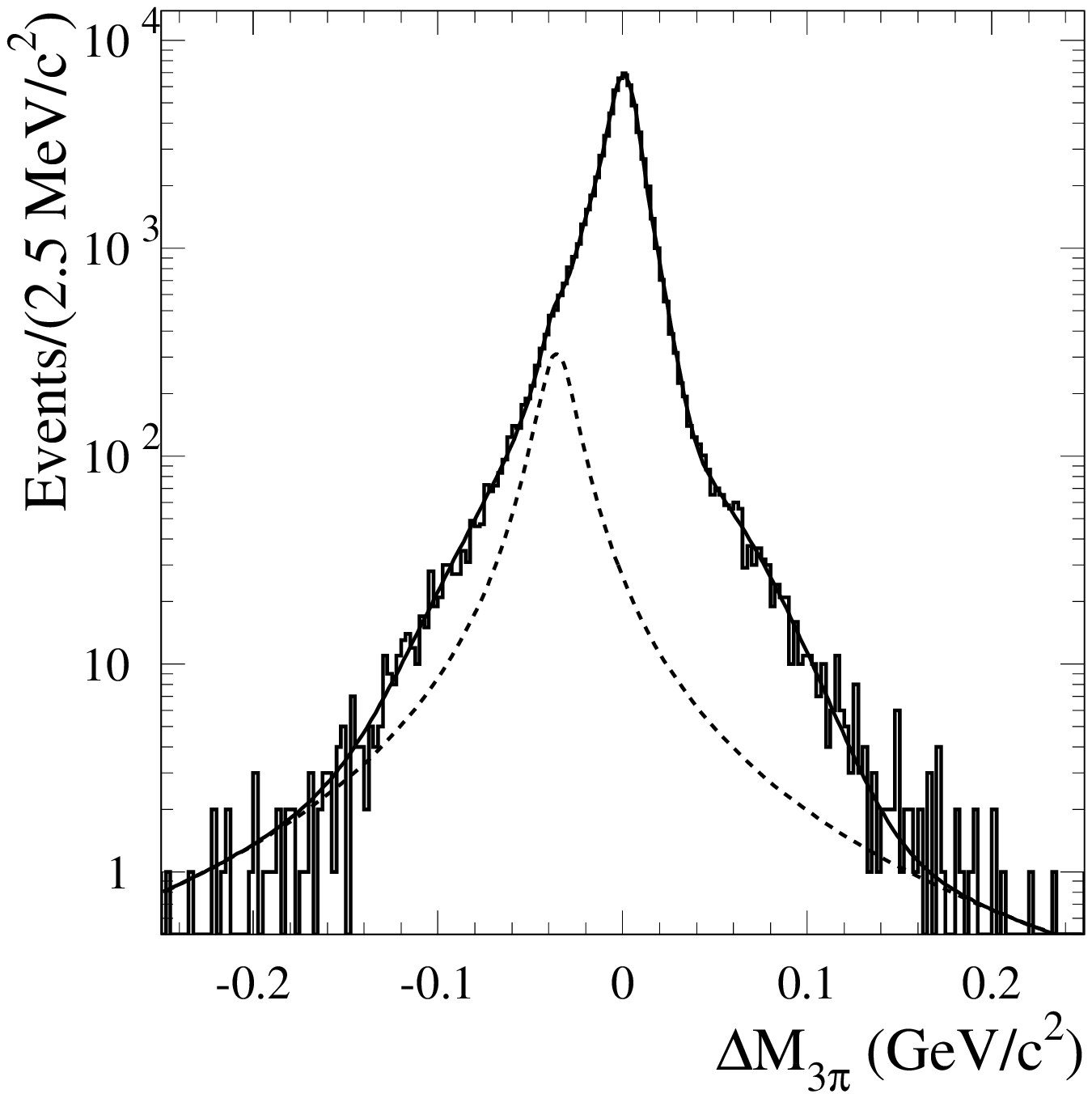}
\caption{Left panel: The distributions of the true (solid histogram) and 
reconstructed (dashed histogram) $3\pi$ mass for simulated signal events. 
Right panel: The distribution of the difference between
the reconstructed and true mass for simulated signal events with
$20<\chi^2_{3\pi\gamma}<40$ from the $\omega$ peak. The curve is the result
of the fit described in the text. The dashed histogram represents the
fitted Lorentzian contribution.
\label{mcspec}}
\end{figure*}

To measure the $e^+e^-\to \pi^+\pi^-\pi^0$ cross section, detector
resolution effects need to be unfolded from the measured $3\pi$
invariant-mass spectrum. In Fig.~\ref{mcspec} (left) the simulated
distribution of the true $3\pi$ mass in the energy regions of 
the $\omega$ and $\phi$ resonances is compared with the distribution
of the reconstructed $3\pi$ mass. The true spectrum varies by four orders of
magnitude and has two narrow peaks.  The reconstructed 
spectrum strongly differs from the true one. 
For such a spectrum, the result of the unfolding procedure is very sensitive
to the quality of simulation used to obtain the resolution function. 
To study the difference between data and simulation in resolution, we
fit the measured $3\pi$ mass spectrum with the vector-meson-dominance
model including several resonances. The $\omega$ and $\phi$ masses and widths
are known with relatively high accuracy. Therefore, from the fit we can extract
the mass shift and standard deviation of an additional smearing Gaussian
function needed
to describe the data-MC simulation difference in the mass resolution. These 
parameters are determined separately for the $\omega$ and $\phi$ resonances.

The detector resolution function has long non-Gaussian tails as seen
in Fig.~\ref{mcspec} (right), where the distribution of the difference between
the reconstructed and true mass ($\Delta M_{3\pi}$) is shown for events from
the $\omega$ peak. To increase the fraction of events in the non-Gaussian 
tails, 
events are selected with the condition $20<\chi^2_{3\pi\gamma}<40$. The 
distribution is fitted by a sum of three Gaussians and a Lorentzian function
$L(x)=(\gamma/\pi)/((x-x_0)^2+\gamma^2)$.
The latter is shown in Fig.~\ref{mcspec} (right) by the dashed histogram.
Because of the asymmetry in the $\Delta M_{3\pi}$ distribution, the maximum of 
the Lorentzian function is shifted from zero by about $-30$ MeV. The 
same shift is observed in the $\Delta M_{3\pi}$ distribution at 750, 900
GeV/$c^2$, and at the $\phi$ resonance. 

To describe a possible difference between data and simulation in the tails
of the resolution function, we introduce to the fit to the $M_{3\pi}$ data
spectrum a smearing Lorentzian function with $x_0=-30$ MeV.

The following probability density function is used in the fit to the
measured $M_{3\pi}$ spectrum 
\begin{eqnarray}
\left( \frac{dN}{dm}\right)^{\rm meas}_i&=&
(1-\epsilon)\sum_j P_{ij}\left[
\left(\frac{dN}{dm}\right)\ast G\right]_j\nonumber\\
&+&\epsilon\left[\left(\frac{dN}{dm}\right)\ast L\right]_i, 
\end{eqnarray}
where the theoretical spectrum of true $3\pi$ mass (${dN}/{dm}$) is convolved
with the smearing Gaussian ($G$) and Lorentzian ($L$) functions.
The spectra of true and measured masses are presented as histograms
with the same binning. The folding matrix $P_{ij}$ obtained using simulation
gives the probability that an event with true mass in bin $j$ is actually 
reconstructed in bin $i$. From the fit we determine the standard deviations of
the smearing Gaussian and mass shifts at the $\omega$ and $\phi$, $\epsilon$,
and $\gamma$ of the smearing Lorentzian.

The width of the mass bins near the $\omega$ and $\phi$ resonances
is chosen to be 2.5 MeV/$c^2$. This width is not much smaller than the
resonance widths. Therefore, the elements of the matrix $P_{ij}$ depend on the
values of the resonance parameters used in simulation. We correct the folding
matrix using an iterative procedure. The procedure uses results of the fit 
without Lorentzian smearing (Model 4 in Table~\ref{model40}) for events with
$\chi^2_{3\pi\gamma}<20$. Simulated events are reweighted by the
ratio of the fitted spectrum $(dN/dm)\ast G$ to the true simulated
spectrum. The reweighting is performed with a bin width of 0.5 MeV/$c^2$.
Then a new matrix $P_{ij}$ is obtained, and the fit is repeated. 
We iterate until the change in $(dN/dm)^{\rm true}\ast G$ between two successive
iterations is less than 0.1\%.

The true mass spectrum in the fit is described by the following function:
\begin{equation}
\frac{dN}{dm}=\sigma_{3\pi}(m)\frac{d{\cal L}}{dm}\,R\,\varepsilon,
\label{thmspt}
\end{equation}
where $\sigma_{3\pi}(m)$ is the Born cross section for $e^+e^-\to 3\pi$,
$d{\cal L}/dm$ is the so-called ISR differential
luminosity, $\varepsilon$ is the detection efficiency as a function of mass,
and $R$ is a radiative correction factor accounting for the Born mass
spectrum distortion due to emission of several photons by the initial
electron and positron. The ISR luminosity is calculated
using the total integrated luminosity ${\cal L}$ and the probability density 
function for ISR photon emission~(Eq.~(\ref{eq2})):
\begin{equation}
\frac{d{\cal L}}{dm}=\frac{\alpha}{\pi x}\left(
(2-2x+x^2)\log\frac{1+C}{1-C}-x^2 C\right)\frac{2m}{s}\,{\cal L}.
\label{ISRlum}
\end{equation}
Here, $x=1-m^2/s$, $\sqrt{s}$ is the $e^+e^-$ c.m.~energy, $C=\cos{\theta_0}$,
and $\theta_0$ determines the range of polar angles in the c.m.~frame:
$\theta_0<\theta_\gamma<180^\circ-\theta_0$ for the ISR photon.
In our case  $\theta_0$  is equal to 20$^\circ$,
since we determine the detector efficiency using the simulation with
$20^\circ<\theta_\gamma<160^\circ$.
The total integrated luminosity (${\cal L}=468.6$ fb$^{-1}$) is measured
with an accuracy of 0.43\%~\cite{babarlumi}.

The Born cross section for $e^+e^-\to 3\pi$ can be written as the sum
of the contributions of five resonances $\rho\equiv \rho(770)$,
$\omega\equiv \omega(782)$, $\phi\equiv \phi(1020)$, 
$\omega^\prime\equiv \omega(1420)$, 
and $\omega^{\prime\prime}\equiv \omega(1650)$:
\begin{widetext}
\begin{equation}
\sigma_{3\pi}(m)=\frac{12\pi}{m^3}F_{\rho\pi}(m)
\left|\sum_{V=\rho,\omega,\phi,\omega^\prime,\omega^{\prime\prime}}
\frac{\Gamma_V m_V^{3/2}\sqrt{{\cal B}(V\to e^+e^-){\cal B}(V\to 3\pi)}}{D_V(m)}
\frac{e^{i\varphi_V}}{\sqrt{F_{\rho\pi}(m_V)}}\right|^2,\label{born}
\end{equation}
\end{widetext}
where $m_V$ and $\Gamma_V$ are the mass and width of the resonance $V$,
$\varphi_V$ is its phase, and
${\cal B}(V\to e^+e^-)$ and ${\cal B}(V\to 3\pi)$ are the branching fractions
of $V$ into $e^+e^-$ and $3\pi$,
\begin{eqnarray}
D_V(m)&=&m_V^2 - m^2 - i m\Gamma_V(m),\nonumber\\
\Gamma_V(m)&=&\sum_{f}\Gamma_f(m).
\end{eqnarray}
Here $\Gamma_f(m)$ is the mass-dependent partial width of the resonance decay
into the final state $f$, and $\Gamma_f(m_V)=\Gamma_V {\cal B}(V\to f)$.
The mass-dependent width for the $\omega$ and $\phi$ mesons has been
calculated taking into account all significant decay modes.
The corresponding formulae can be found,
for example, in Ref.~\cite{snd2}. We assume that the $V\to 3\pi$ decay proceeds
via the $\rho\pi$ intermediate state, and $F_{\rho\pi}(m)$ is the $3\pi$
phase space volume calculated under this hypothesis.
The formula for the $F_{\rho\pi}$ calculation can be found in Ref.~\cite{snd2}.

The radiative correction factor $R$ is determined using Monte Carlo
simulation (at the generator level, with no detector simulation) 
with the PHOKHARA event generator~\cite{phokhara1}. This generator includes
the next-to-leading order (NLO) ISR contributions. The accuracy of the cross
section calculation for ISR processes with the ISR photon emitted at large 
angle is estimated to be 0.5\%~\cite{phokhara2}. Since the radiative
correction is independent of process, we generate $e^+ e^- \to \mu^+
\mu^-\gamma$ events with no FSR included in the simulation, and calculate
a ratio of the mass spectra obtained in the NLO and LO generator modes. With
the requirement on the invariant mass of the $\mu^+\mu^-\gamma$ system
$M_{\mu\mu\gamma} > 8$ GeV/$c^2$, this ratio is weakly dependent on mass and is
equal to $R=1.0077\pm0.0005$ below 1.1 GeV/$c^2$, $1.0086\pm0.0004$ between
1.1 and 2 GeV/$c^2$, and $1.0091\pm0.0004$ in the range 2--3.5 GeV/$c^2$. The
quoted uncertainty reflects the observed $R$ variation in the specified mass
range.
The radiative correction factor does not include 
the corrections due to leptonic and hadronic vacuum polarization. Here we 
follow the generally accepted practice~\cite{vacuum} of including the vacuum 
polarization correction to the resonance electronic width. 

The free parameters in the fit are the scale factors for products of the 
branching fractions
${\cal B}(V\to e^+e^-){\cal B}(V\to 3\pi)$, and the masses and widths of
the $\omega^\prime$ and $\omega^{\prime\prime}$. The masses and
widths of the $\omega$ and $\phi$ mesons are fixed at the Particle Data 
Group (PDG) values~\cite{pdg}.

The phase $\varphi_\omega$ is set to zero. The relative phase
between the $\omega$ and $\phi$ amplitudes, $\varphi_\phi=(163\pm7)^\circ$, is 
taken from Ref.~\cite{snd2}. The phases of the $\omega^\prime$ and 
$\omega^{\prime\prime}$ are fixed at values of $180^\circ$ and 
$0^\circ$~\cite{Clegg} with an uncertainty of $20^\circ$. This uncertainty is
estimated from the deviation of $\varphi_\phi$ from $180^\circ$.
Our fitting function does not take into account the isovector
$e^+e^-\to\omega\pi^0\gamma\to 3\pi\gamma$ contribution and the presence of the
$\omega^{\prime\prime}\to \rho(1450)\pi$ decay. Therefore, we do not expect
that the parameters of excited $\omega$ states are determined correctly. 
Their inclusion into the fit is needed to study the effect of interference
of the $\omega$ and $\phi$ amplitudes with the contributions of the excited
states. The fitted mass region is restricted to masses below 1.8 GeV.

The branching fraction of the $\rho\to\pi^+\pi^-\pi^0$ decay
can be estimated assuming that the dominant mechanism of the 
$\omega\to\pi^+\pi^-$ and  $\rho\to\pi^+\pi^-\pi^0$ decay is 
$\rho$-$\omega$ mixing. Under this assumption, the coupling 
$g_{\rho\to 3\pi}=\xi g_{\omega\to 3\pi}$,
where the mixing parameter 
$|\xi|^2\approx\Gamma(\omega\to 2\pi)/\Gamma(\rho\to 2\pi)$,
and ${\cal B}(\rho\to 3\pi)\approx 0.4\times 10^{-4}$. The phase 
$\varphi_\rho$ is expected to be close to $-90^\circ$. A significantly larger
value of $\xi$ is obtained in Ref.~\cite{maltman}, where
data on the pion electromagnetic form factor are analyzed in the model
including both $\rho$-$\omega$ mixing and direct isospin-breaking 
$\omega\to 2\pi$ decay. Using the result of Ref.~\cite{maltman}, we obtain 
${\cal B}(\rho\to 3\pi)\approx (2.5\pm 1.0)\times 10^{-4}$ and 
$\varphi_\rho=-(40\pm 13)^\circ$.
The values of the branching fraction and phase measured in the SND 
experiment~\cite{snd2}
are $(1.01^{+0.54}_{-0.36}\pm 0.34)\times 10^{-4}$ and $-(135^{+17}_{-13}\pm
0.9)^\circ$, respectively.
With the current experimental accuracy, $\sim$1\% at the $\omega$ resonance,
the contribution of the $\rho\to\pi^+\pi^-\pi^0$ decay with a branching
fraction of about $10^{-4}$ must be taken into account in the fit.

\begin{table*}
\caption{Models used to describe the $3\pi$ mass spectrum (142 bins) and
the $\chi^2/\nu$ values from the fits, where $\nu$ is the number of 
degrees of freedom. The first four rows show the results for the standard
selection criteria, while the last four are for the tighter 
cut on the $\chi^2$ of the kinematic fit ($\chi^2_{3\pi\gamma}<20$). 
\label{model40}}
\begin{ruledtabular}
\begin{tabular}{lccccc}
Model & Lorentzian smearing & ${\cal B}(\rho\to 3\pi)$ & $\chi^2/\nu$ \\
\hline
1 & yes & free       &  136/127 \\
2 & no  & $\equiv 0$ &  201/131 \\
3 & yes & $\equiv 0$ &  180/129 \\
4 & no  & free       &  147/129 \\
\hline
1 & yes & free       &  135/127 \\
2 & no  & $\equiv 0$ &  181/131 \\
3 & yes & $\equiv 0$ &  178/129 \\
4 & no  & free       &  136/129 \\
\end{tabular}
\end{ruledtabular}
\end{table*}

\begin{figure*}
\centering
\includegraphics[width=0.43\textwidth]{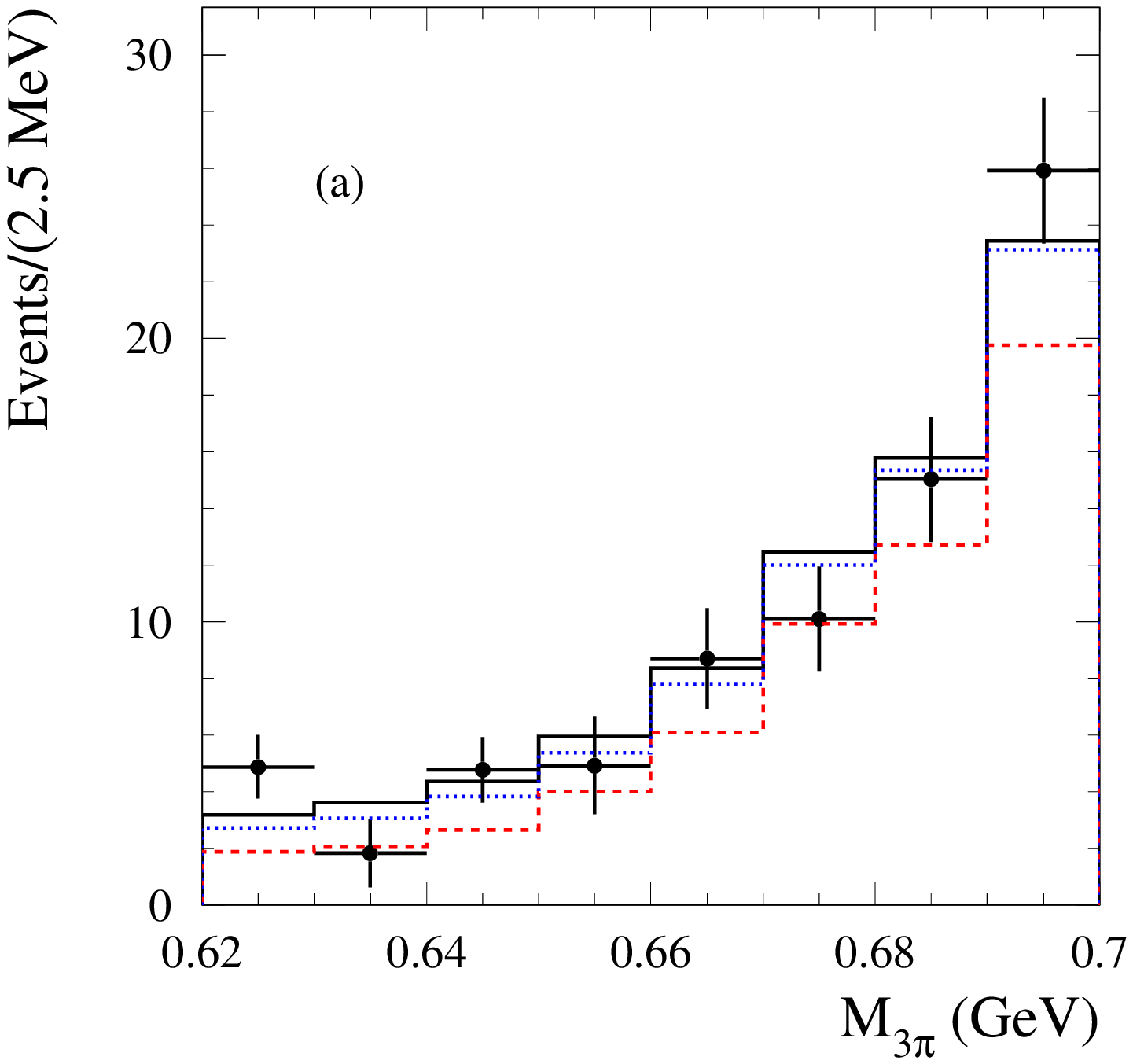}\hfill
\includegraphics[width=0.43\textwidth]{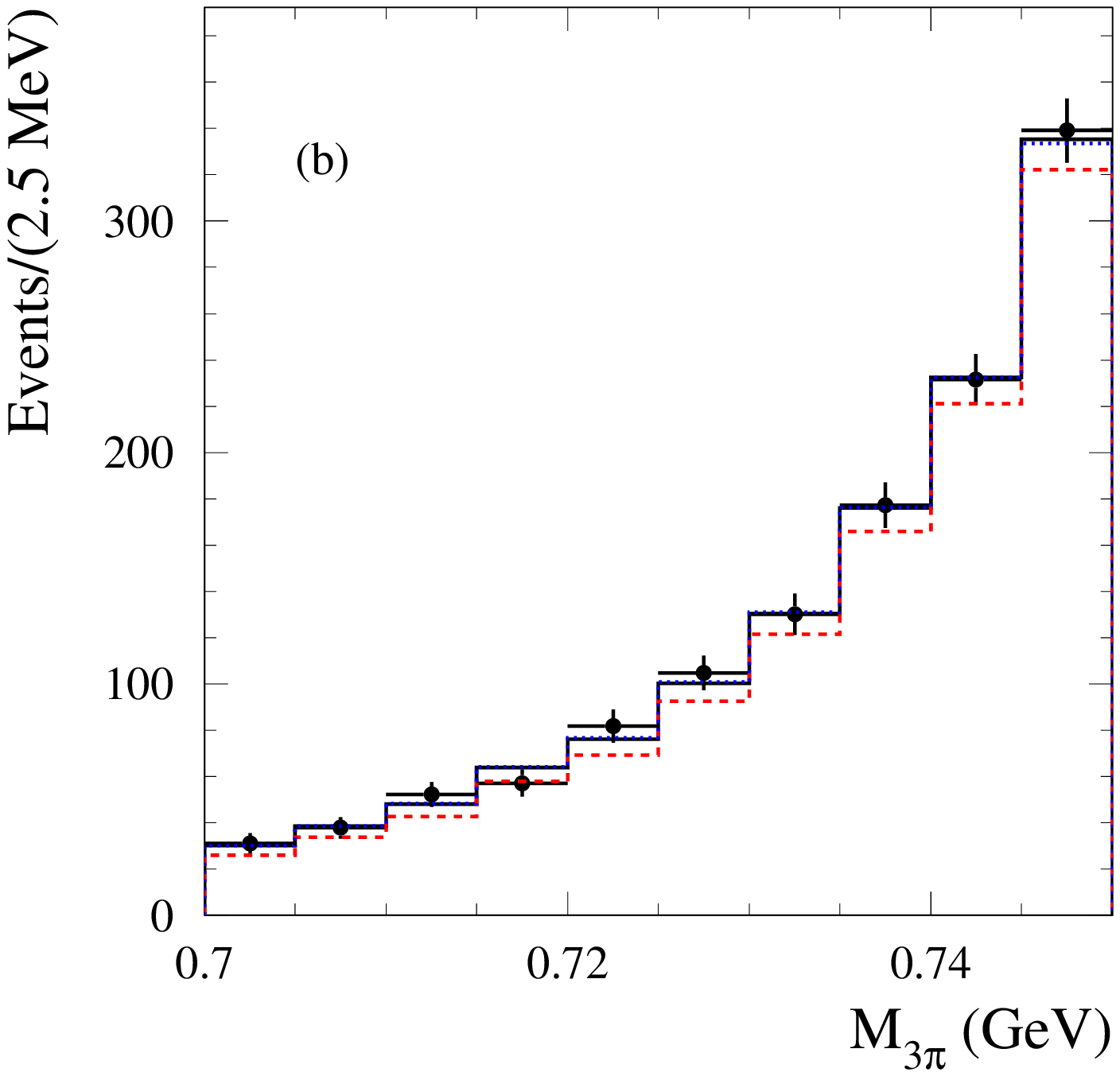}
\includegraphics[width=0.43\textwidth]{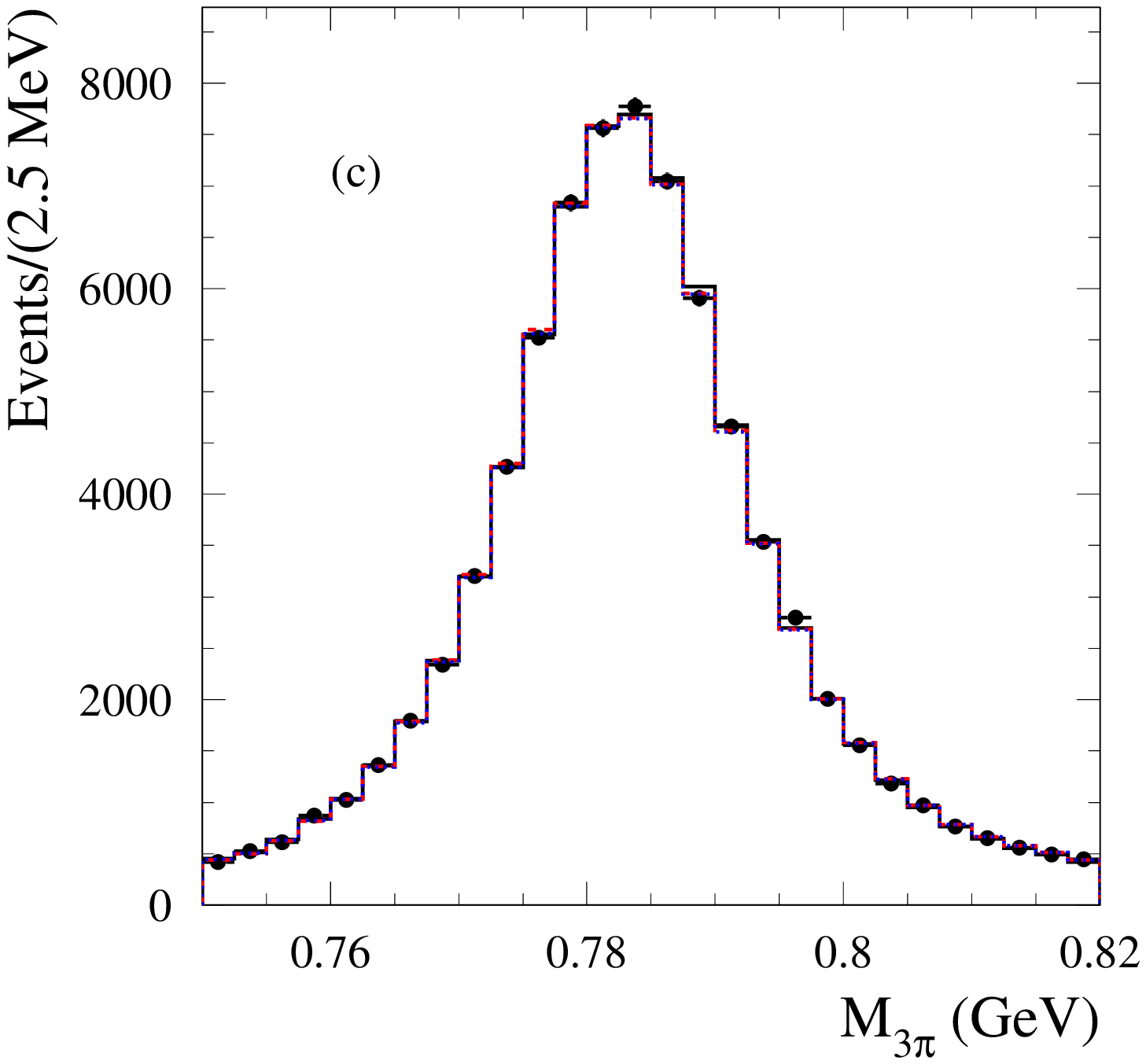}\hfill
\includegraphics[width=0.43\textwidth]{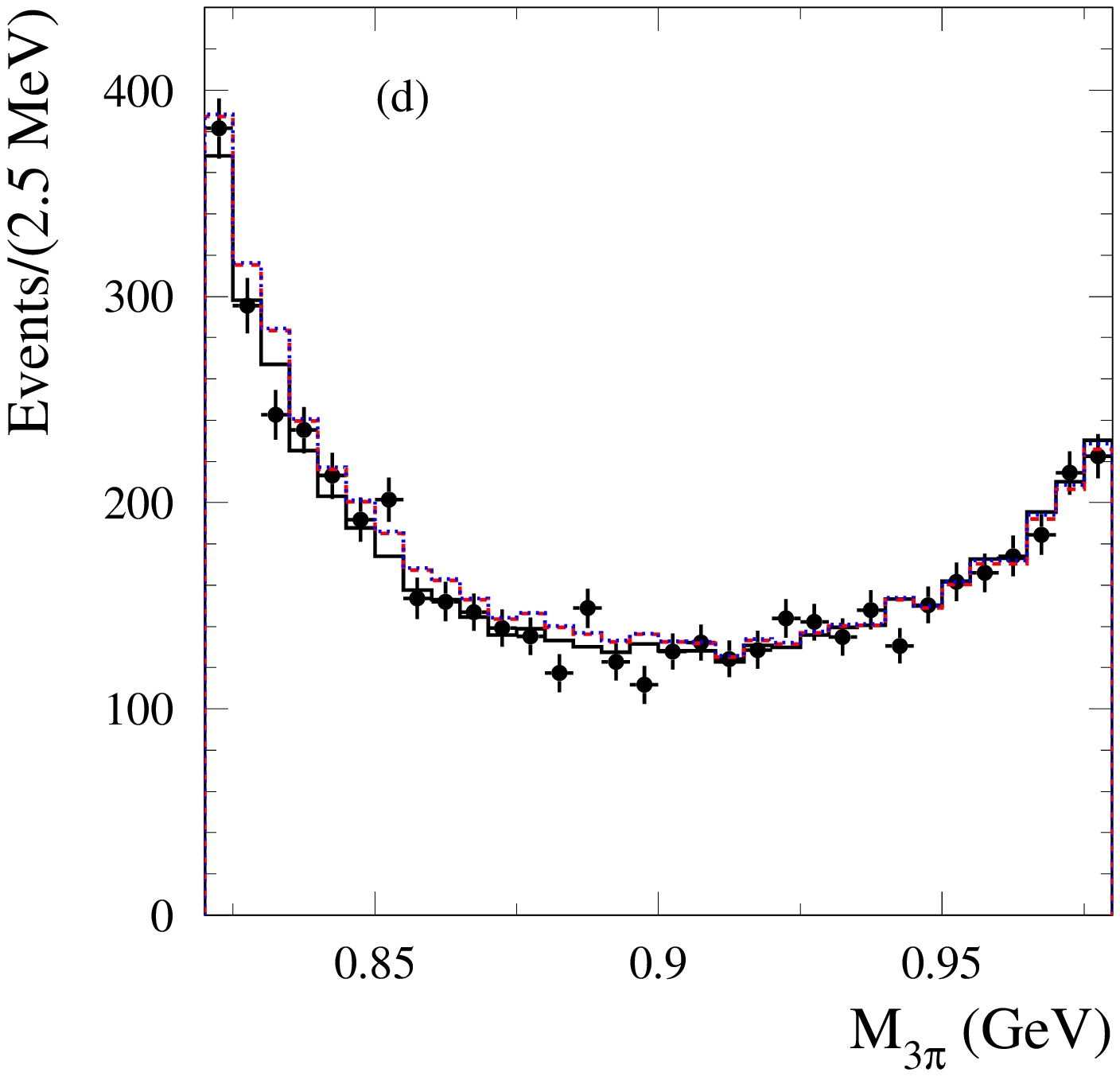}
\includegraphics[width=0.43\textwidth]{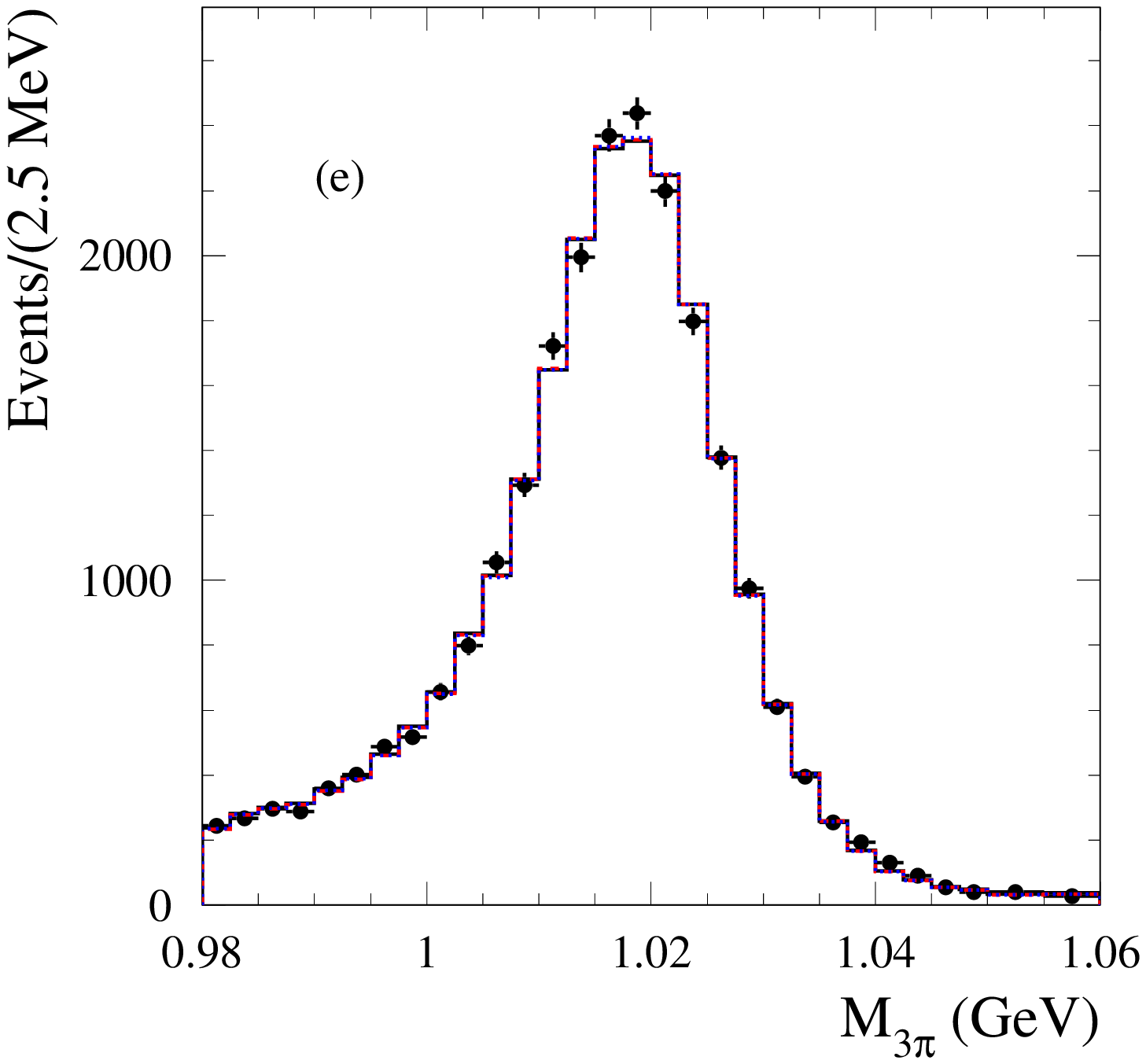}\hfill
\includegraphics[width=0.43\textwidth]{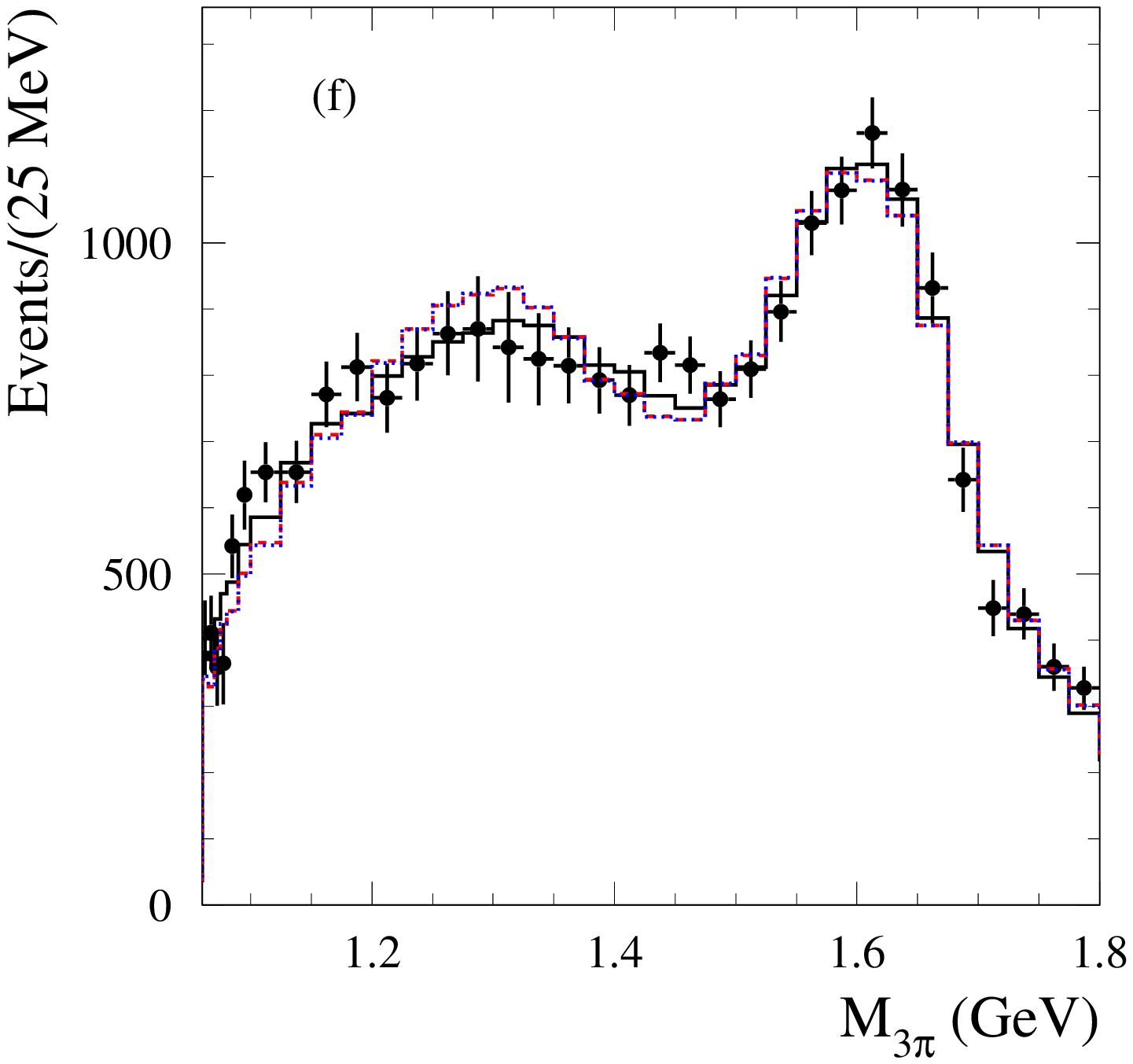}
\caption{The $3\pi$ mass spectrum in six regions:
(a) 0.62--0.70 GeV/$c^2$, (b) 0.70--0.74 GeV/$c^2$, (c) 0.74--0.82 GeV/$c^2$,
(d) 0.82--0.98 GeV/$c^2$, (e) 0.98--1.06 GeV/$c^2$, and (f) 1.06-1.80 GeV/$c^2$.
The solid, dashed, and dotted histograms represent the results of the fit 
in Models 1, 2, and 3 listed in Table~\ref{model40}, respectively.
\label{s40}}
\end{figure*}
We fit the $3\pi$ mass spectrum with a series of 4 models, allowing or not for
Lorentzian smearing and leaving the $\rho\to 3\pi$ branching fraction as a 
free parameter or forced to zero, as shown in Table~\ref{model40}.
The results of the fit in Models 1,2,3 are shown in Fig.~\ref{s40}. 
The smearing Gaussian standard deviation obtained in Model 1 is $1.5\pm 0.2$
GeV/$c^2$ at 
the $\omega$ and $1.5\pm 0.4$ GeV/$c^2$ at the $\phi$. Therefore, in what
follows we use a mass-independent smearing standard deviation.
The fitted parameters of 
the smearing Lorentzian function are the following: $\epsilon=0.007\pm0.002$
and $\gamma=63\pm 35$ GeV/$c^2$. The physical fit parameters
will be discussed below.

It is seen from Fig.~\ref{s40} that the fit in Model 2 
(${\cal B}(\rho\to 3\pi)=0$, no smearing Lorentzian) cannot describe
data well below 0.73 GeV/$c^2$ and in the region 0.82--0.9 GeV/$c^2$.
Including the smearing Lorentzian function (Model 3) improves the fit in
the energy region below the $\omega$. The region 0.82--0.9 GeV/$c^2$ cannot be
described reasonably well without the 
$\rho\to 3\pi$ decay. Models 2 and 3 also have a worse fit quality in the
mass range 1.05--1.8 GeV/$c^2$. This is because the fit tries to
compensate the absence of the $\rho\to 3\pi$ decay by increasing the
contribution from the tails of the $\omega(1420)$ and  $\omega(1650)$
resonances.

\begin{figure*}
\centering
\includegraphics[width=0.47\textwidth]{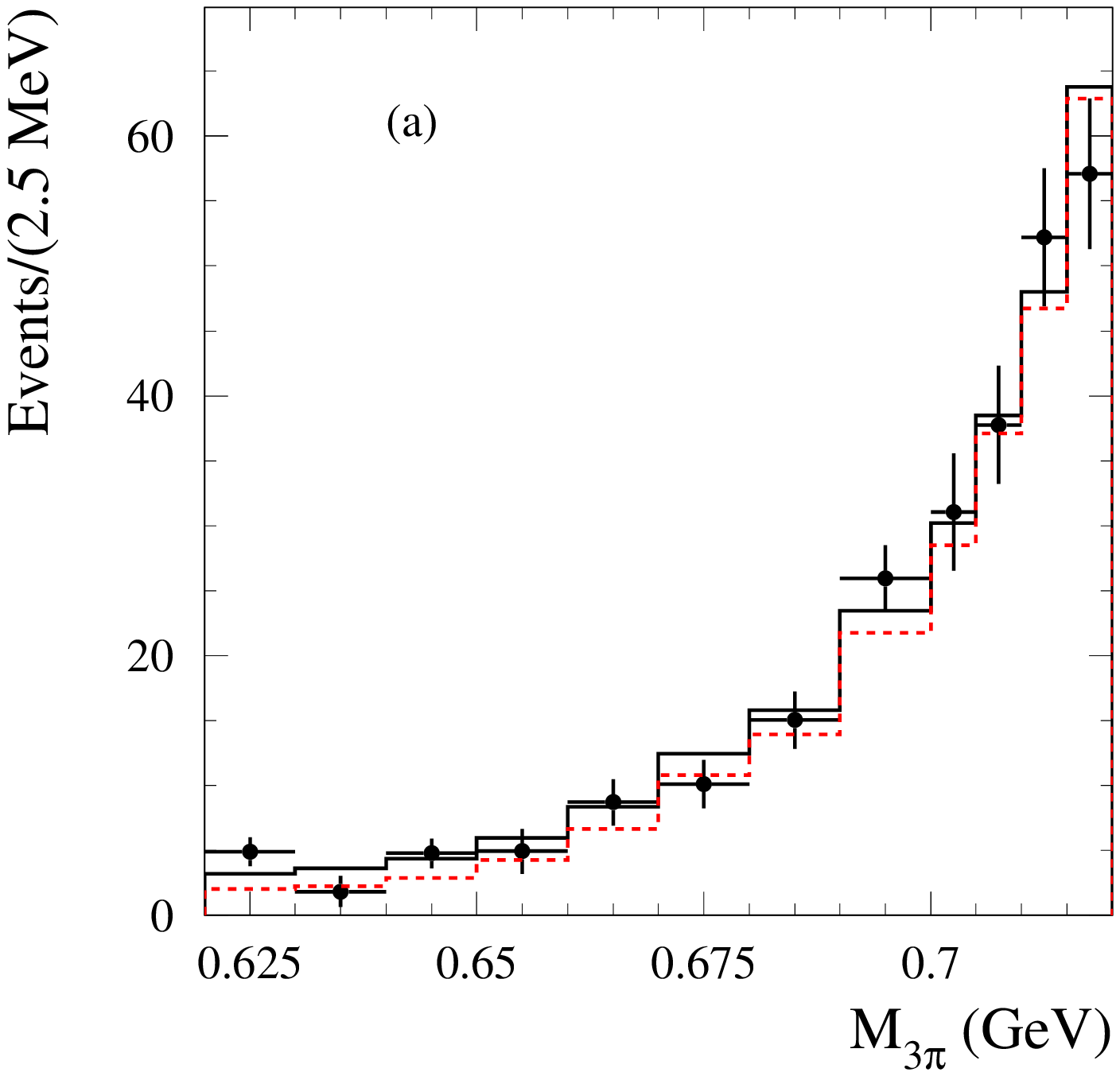}\hfill
\includegraphics[width=0.47\textwidth]{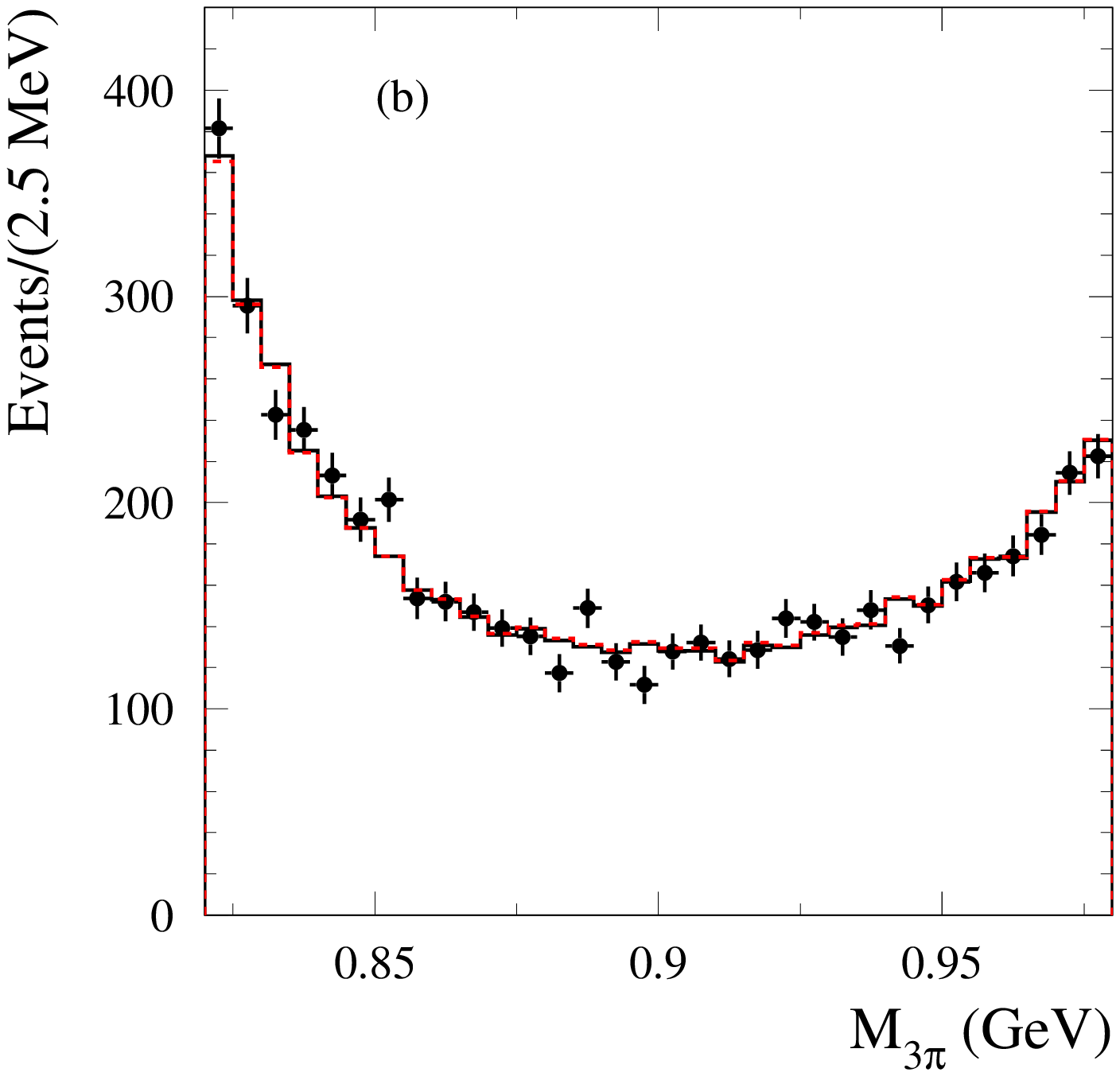}
\caption{The $3\pi$ mass spectrum in the regions
(a) 0.62--0.72 GeV/$c^2$ and (b) 0.82--0.98 GeV/$c^2$.
The solid and dashed histograms represent the results of the fit 
in Models 1 and 4 listed in Table~\ref{model40}, respectively.
\label{s40bw}}
\end{figure*}

The difference between Model 1 and Model 4 (free ${\cal B}(\rho\to 3\pi)$,
no Lorentzian smearing) is maximal in the $3\pi$ mass region 0.62--0.72 GeV/$c^2$
(see Fig.~\ref{s40bw}). To decrease the difference between
the measured and predicted spectrum below the $\omega$ peak in absence of
the Lorentzian smearing, the fit in Model 4 increases ${\cal B}(\rho\to 3\pi)$
by 14\%.

To reduce the influence on the fitted parameters of the data-simulation
difference in resolution, 
we tighten the condition on $\chi^2_{3\pi\gamma}$ 
from 40 to 20. In the last four rows of Table~\ref{model40}, we compare
the quality of the fit with Models 1--4 to the mass spectrum obtained with 
the tighter $\chi^2_{3\pi\gamma}$ requirement. It is seen that inclusion of the
Lorentzian smearing in this case improves the fit quality insignificantly.  
The obtained parameters of the Lorentzian smearing function are 
$\epsilon=0.0022\pm0.0016$ and $\gamma=59_{-31}^{+54}$ MeV/$c^2$. 
Therefore, below, we quote the fit parameters for Model 4. The standard
deviation of the smearing Gaussian function and the mass shifts for the
$\omega$ and $\phi$ mesons are found to be $\sigma_s=1.4\pm0.2$ MeV/$c^2$, 
$m_\omega-m_\omega^{\rm PDG}=0.04\pm 0.06$ MeV/$c^2$,  and
$m_\phi-m_\phi^{\rm PDG}=0.08\pm 0.08$ MeV/$c^2$. The latter two parameters
are consistent with zero.

Since the mass resolution (full width at half-maximum is about 13 MeV/$c^2$ at 
the $\omega$ and 15 MeV/$c^2$ at the $\phi$) is larger than the  $\omega$
and $\phi$ widths, the parameter with least sensitivity to resolution
effects is the area under the resonance curve, i.e.,
$P_V\equiv\Gamma(V\to e^+e^-){\cal B}(V\to \pi^+\pi^-\pi^0)$.  
From the fit we obtain
\begin{eqnarray}
P_\omega&=&
(0.5698\pm0.0031\pm0.0082) \mbox{ keV},\nonumber\\ 
P_\phi&=&
(0.1841\pm0.0021\pm0.0080) \mbox{ keV}.\label{fpar1} 
\end{eqnarray}
For the $\rho$ meson, we determine 
\begin{eqnarray}
{\cal B}(\rho\to 3\pi)&=&(0.88\pm 0.23\pm 0.30)\times 10^{-4},
\nonumber\\
\varphi_\rho&=&-(99\pm 9 \pm 15)^\circ.\label{fpar2} 
\end{eqnarray}
The significance of the $\rho\to 3\pi$ decay estimated from the difference
between the $\chi^2$ values for Models 4 and 2 is greater than $6\sigma$. 

The first uncertainties in Eqs.~(\ref{fpar1}) and (\ref{fpar2}) are
statistical, while the second are systematic.
The latter include the systematic uncertainties in the ISR luminosity,
radiative correction, and detection efficiency. The uncertainty due to the
data/MC difference in the mass resolution line shape is estimated as 
a difference between results of the fits with Models 4 and 1. 
We also vary within the uncertainties the values of the 
parameters $\varphi_\phi$, $\Gamma_\omega$, and $\Gamma_\phi$, and
the scale factors for the background processes. These contributions to
the systematic uncertainties are listed in Table~\ref{parsys}.
\begin{table*}
\caption{Contributions to the systematic uncertainties in fit parameters from 
different effects ($P_1=\Gamma(\omega\to e^+e^-){\cal B}(\omega\to
\pi^+\pi^-\pi^0)$, $P_2=\Gamma(\phi\to e^+e^-){\cal B}(\phi\to
\pi^+\pi^-\pi^0)$, $P_3={\cal B}(\rho\to \pi^+\pi^-\pi^0)$, and
$P_4=\varphi_\rho$).\label{parsys}}
\begin{ruledtabular}
\begin{tabular}{lcccc}
Effect & $P_1$ (\%) & $P_2$ (\%) & $P_3$ (\%) & $P_4$ (deg)\\
\hline
Luminosity                     & 0.4 & 0.4  & 0.4 & --  \\
Radiative correction           & 0.5 & 0.5  & 0.5 & --  \\
Detection efficiency           & 1.1 & 1.1  & 1.1 & --  \\
MC statistics                  & 0.1 & 0.2  & 0.2 & --  \\
Lorentzian smearing            & 0.3 & 0.4  & 4.7 & 12  \\
$\Gamma_\omega$                & 0.4 & 0.2  & 13.0& 8   \\
$\Gamma_\phi$                  & 0.0 & 0.0  & 0.3 & 0   \\
$\varphi_\phi$                    & 0.2 & 3.1  & 6.1 & 1   \\
Background subtraction         & 0.1 & 0.2  & 7.3 & 2   \\
$\omega(1680)\to\rho(1450)\pi$ & 0.4 & 2.7  & 30.0& 0 \\
\hline
total                          & 1.4 & 4.3  & 34.5& 15 \\
\end{tabular}
\end{ruledtabular}
\end{table*}

Our fitting model given by Eq.~(\ref{born}) assumes that the 
$e^+e^-\to \pi^+\pi^-\pi^0$ process proceeds via the $\rho(770)\pi$ 
intermediate state. Actually, due to a sizable 
$\omega(1650)\to \rho(1450)\pi\to 3\pi$ transition~\cite{snd3} and the 
existence of the $e^+e^-\to \omega\pi^0 \to \pi^+\pi^-\pi^0$ process, 
the $e^+e^-\to \pi^+\pi^-\pi^0$ amplitude above the $\phi$ cannot be presented
as a simple coherent sum of the $\omega(1420)$ and $\omega(1650)$ amplitudes.
To study the effect of non-$\rho(770)\pi$ mechanisms, we substitute the \babar\
data above 1.1 GeV/$c^2$ by the SND data on the $\rho(770)\pi$ and
$\rho(1450)\pi$ cross sections~\cite{snd3}. The data on the phase difference
between the $\rho(770)\pi$ and $\rho(1450)\pi$ amplitudes measured in
Ref.~\cite{snd3} is also included in the fit. The new fitting function
takes into account $\omega(1650)$ transitions to $\rho(770)\pi$ and 
$\rho(1450)\pi$, and interference between $\rho(770)\pi$ and $\rho(1450)\pi$
amplitudes. This new approach
modifies the contribution of the $\omega(1420)$ and $\omega(1650)$ resonances
in the $3\pi$ mass region below 1.1 GeV/$c^2$ and shifts the parameters of the
$\rho$, $\omega$ and $\phi$ resonances. In particular, for the $\rho$ decay we
obtain
\begin{eqnarray}
{\cal B}(\rho\to 3\pi)&=&(1.14_{-0.28}^{+0.32})\times 10^{-4}.
\end{eqnarray}
The difference between the results of this new fit and our nominal fit is used
as an estimate of systematic uncertainty due to the
$\omega(1650)\to\rho(1450)\pi$ decay (see Table~\ref{parsys}.)
\begin{figure*}[p]
\centering
\includegraphics[width=0.47\textwidth]{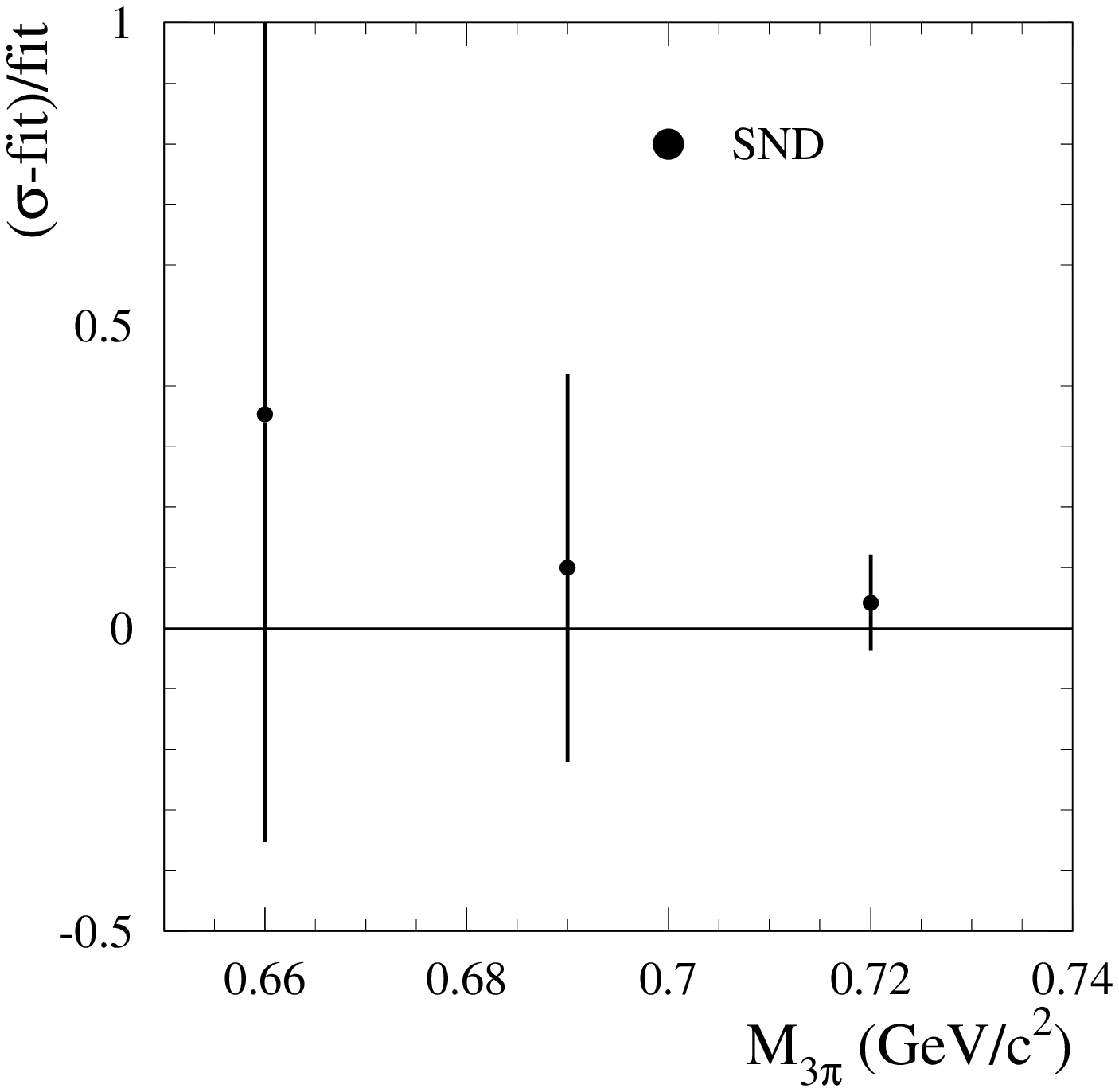}\hfill
\includegraphics[width=0.47\textwidth]{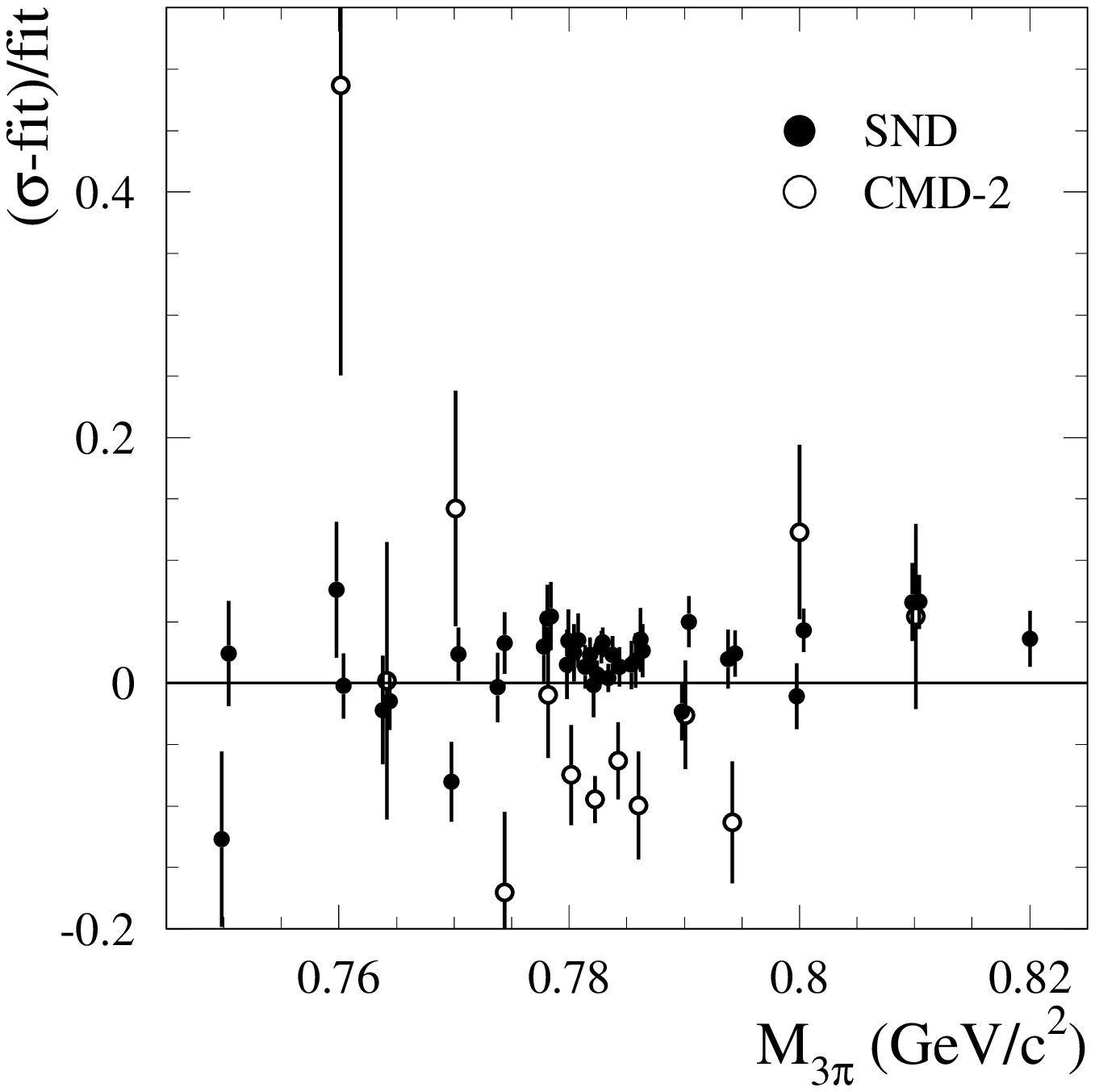}
\includegraphics[width=0.47\textwidth]{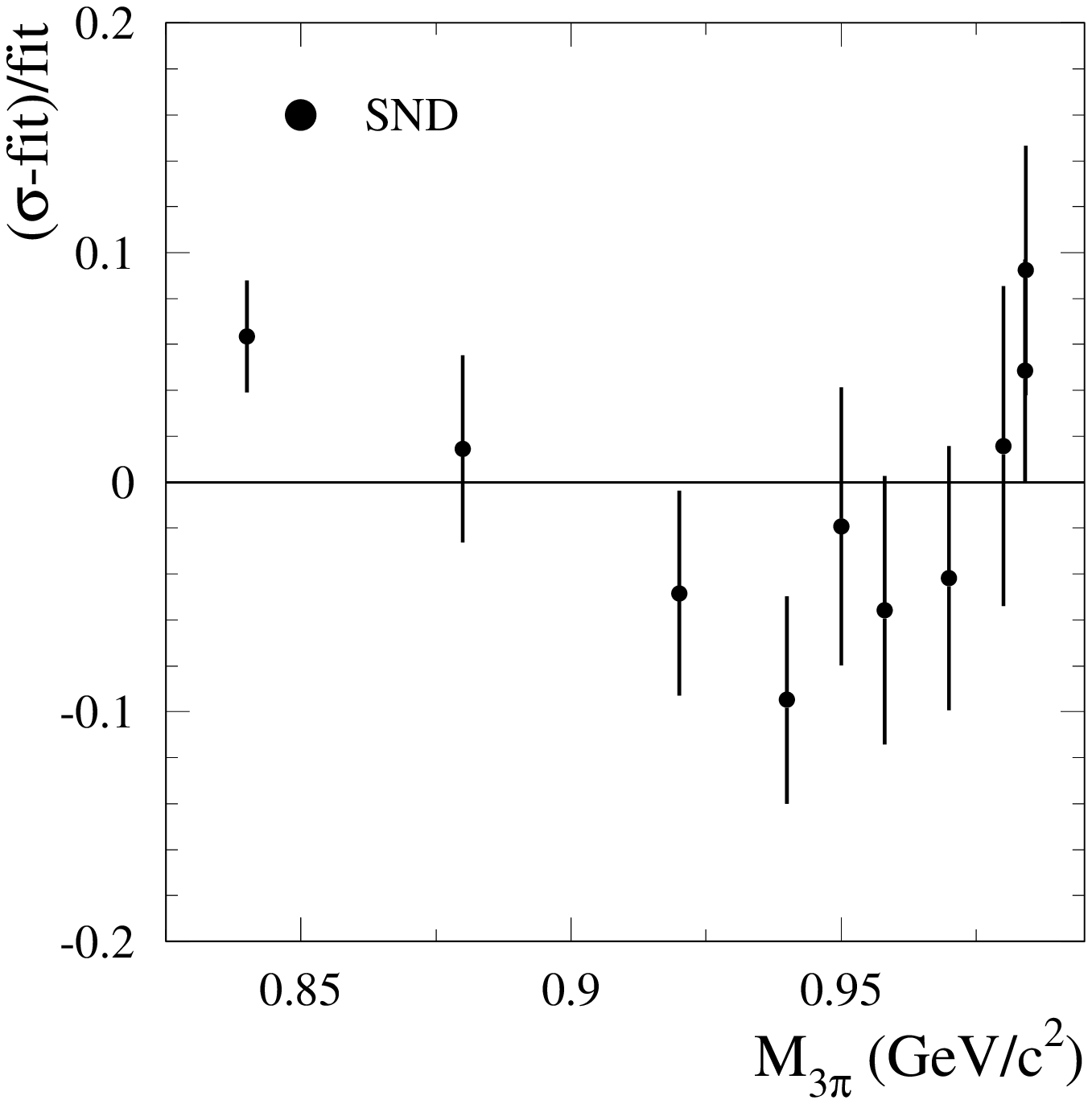}\hfill
\includegraphics[width=0.47\textwidth]{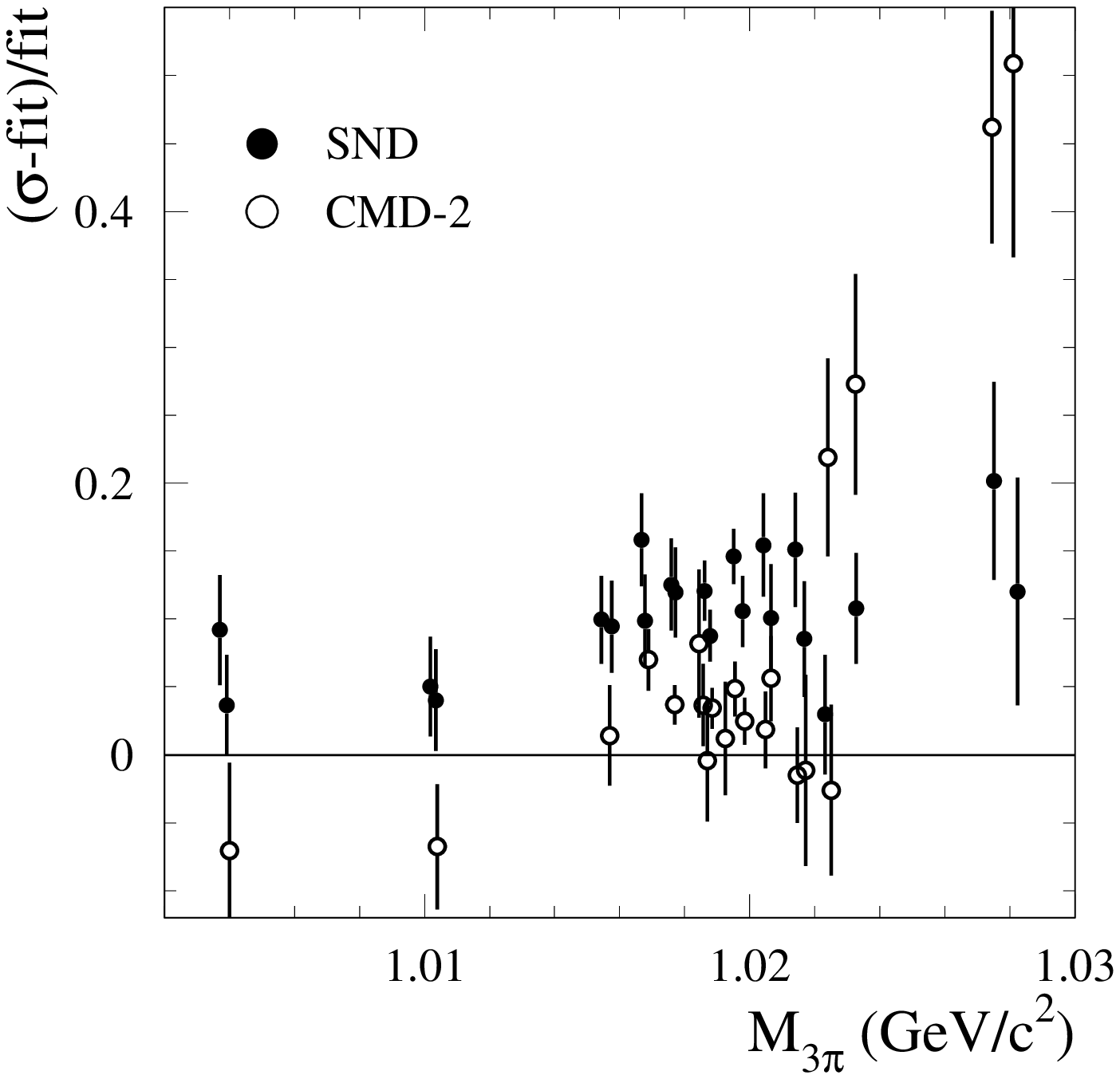}
\caption{The relative difference between SND~\cite{snd1,snd2} and 
CMD-2~\cite{cmd1,cmd2} data on the $e^+e^-\to \pi^+\pi^-\pi^0$ cross section
and the cross section calculated using Eq.~(\ref{born}) with parameters
obtained from the fit to the $3\pi$ mass spectrum. The uncertainties shown for
the SND and CMD2-2 data are statistical. The systematic uncertainty is
3.4\% for the SND data at the $\omega$~\cite{snd2}, 1.3\% for the CMD-2 data at
the $\omega$~\cite{cmd1}, 5\% for the SND data at the $\phi$~\cite{snd2},
and 2.5\% for the CMD-2 data at the $\phi$~\cite{cmd2}.
The systematic uncertainty in the \babar\ cross section is about 1.5\%.  
\label{comp}}
\end{figure*}

The fitted values of $\Gamma(V\to e^+e^-){\cal B}(V\to 3\pi)$ for the $\omega$ 
and $\phi$ mesons given by Eq.~(\ref{fpar1}) are in reasonable agreement with
the corresponding world average values~\cite{pdg}:
$0.557\pm 0.011$ keV and $0.1925\pm0.0043$ keV, respectively.
For the $\omega$ meson the accuracy of our result is comparable with
the accuracy of the PDG value. For the $\phi$ meson, we have a large systematic
uncertainty related to the interference between $\phi$-meson amplitude
and amplitudes of the resonances of the $\omega$ family. The fitted values of
${\cal B}(\rho\to 3\pi)$ and $\varphi_\rho$ given by Eq.~(\ref{fpar2})
are in agreement with the SND results: 
$(1.01^{+0.54}_{-0.36}\pm 0.34)\times 10^{-4}$ and 
$-(135^{+17}_{-13}\pm 9)^\circ$~\cite{snd2}. 

It is instructive to compare the cross section calculated using 
Eq.~(\ref{born}) with the SND and CMD-2 
data~\cite{snd1,snd2,cmd1,cmd2}. Such a comparison is presented in 
Fig.~\ref{comp}, where the difference between SND and CMD-2 data and the
\babar\
fit is shown in the energy region of the $\omega$ and $\phi$ resonances.
A shift in the energy ($3\pi$ mass) scale between different sets of  data
leads to the appearance of wiggles in the relative difference between them near
the resonance maximum. To eliminate these wiggles we shift the SND (CMD-2)
data by $-0.18$ (0.09) MeV at the $\omega$ region, and  0.09 (0.13) MeV at
the $\phi$ region. It is seen that the \babar\ cross section is
in reasonable agreement with the SND data below the $\phi$. At the $\omega$
the difference between the SND and \babar\ cross sections is about 2\%, well
below the systematic uncertainty (3.4\% for SND and 1.3\% for \babar). The
CMD-2 data in the vicinity of the $\omega$ lie about 7\% below zero. 
With the CMD-2 statistical and systematic uncertainties of 
of 1.8\% and 1.3\%, respectively, the difference between 
CMD-2 and \babar\ is $2.7\sigma$. Near the maximum 
of the $\phi$-meson resonance the CMD-2 and SND data with systematic
uncertainties of 2.5\% and 5\%, respectively, lie about 4\% and 11\% higher 
than the \babar\ cross section.

\section{
Measurement of the $e^+e^-\to \pi^+\pi^-\pi^0$ 
cross section below 1.1 GeV/$c^2$\label{cs_low}}
\begin{figure*}
\includegraphics[width=0.9\textwidth]{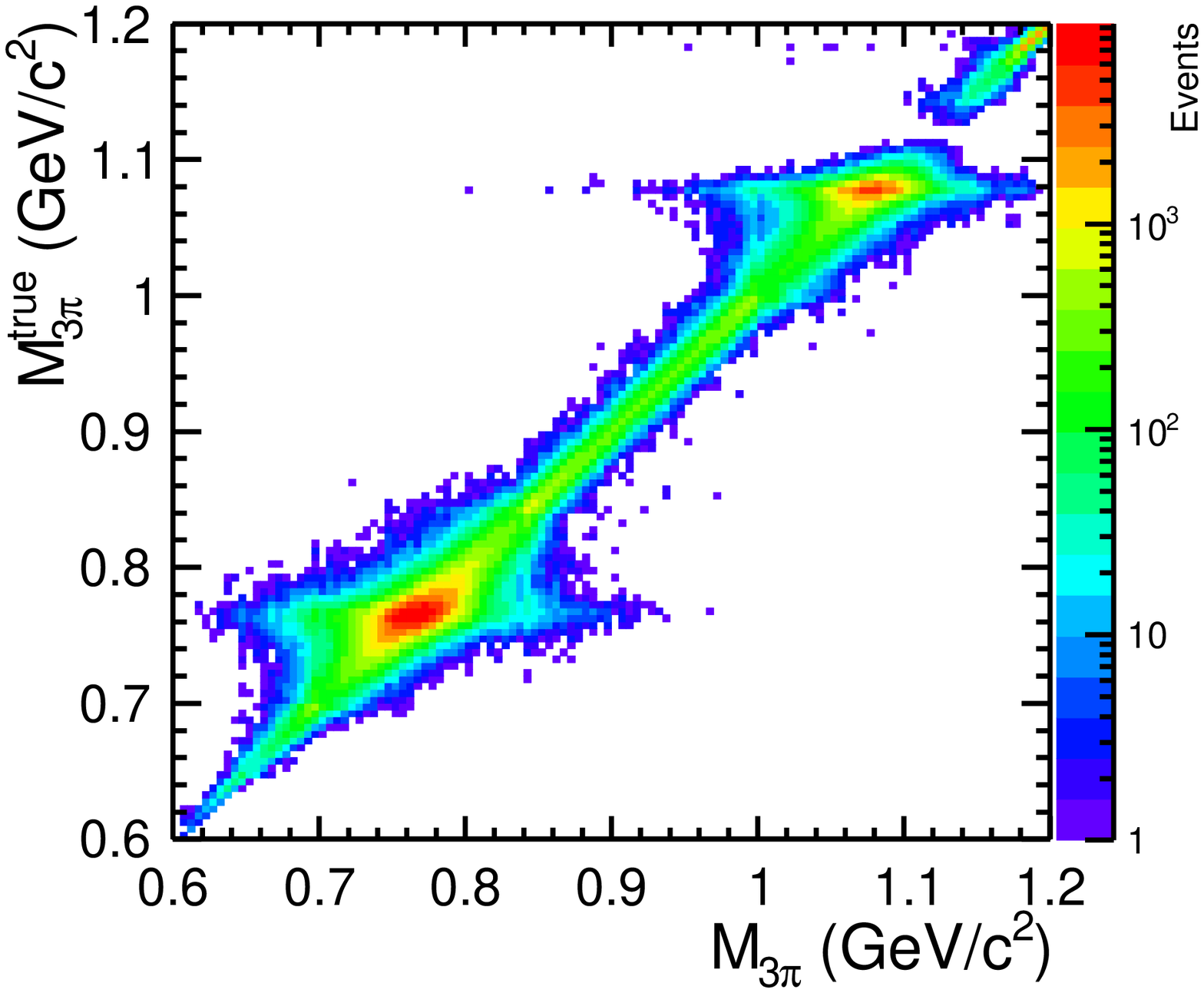}
\caption{The mass-transfer matrix giving the number of events generated
with a (true) mass $M_{3\pi}^{\rm true}$ in a bin $j$ and reconstructed with 
a (measured) mass $M_{3\pi}$ in a bin $i$.
\label{fig:fold}}
\end{figure*}
In the $M_{3\pi}$ region below 1.1 GeV/$c^2$, the detector resolution
strongly distorts the $3\pi$ mass spectrum as shown in 
Fig.~\ref{mcspec}~(left). To obtain the true mass ($M_{3\pi}^{\rm true}$) 
spectrum, unfolding must be applied to the measured $M_{3\pi}$ spectrum.  
Similar to the previous \babar\ analyses~\cite{Babar2pi,Babar2K}, we use a 
simplified version of the iterative unfolding method developed in 
Ref.~\cite{MALAESCU}. 

In Sec.~\ref{masssp} we reweight the signal MC simulation using the
results of the fit to the measured $M_{3\pi}$ spectrum and obtain the folding
matrix $P_{ij}$. This matrix must be corrected to take into account the data-MC
difference in the mass resolution. This difference is described by the smearing
Gaussian ($G$) and Lorentzian ($L$) functions, the parameters of which are
determined from the fit described in Sec.~\ref{masssp}. The corrected folding
matrix is calculated as
\begin{equation}
P_{ij}^\ast=(1-\epsilon)\sum_k P_{ik}G_{kj}+\epsilon L_{ij},
\end{equation}
where the matrices $G_{ij}$ and $L_{ij}$ are obtained
using the fitted theoretical mass spectrum $({dN}/{dm})^{\rm true}_{\rm FIT}$
and its convolution with the smearing functions $G$ and $L$, respectively.
In the unfolding procedure 
described below we use $P_{ij}^\ast$ with $\epsilon=0$. Unfolding with non-zero
$\epsilon=0.002$ (Model 1 in Table~\ref{model40}) is performed to 
estimate a systematic uncertainty due to possible unaccounted Lorentzian
smearing.

Figure~\ref{fig:fold} represents the transfer matrix 
$A_{ij}=P_{ij}^\ast T_j$, where the vector $T_i$ is obtained by
integration of the theoretical mass spectrum $({dN}/{dm})^{\rm
true}_{\rm FIT}$ 
over bin $j$. The unfolding matrix can also be obtained as
$\tilde{P}_{ij}=A_{ij}/M_i$, where $M_i=\sum_{j}A_{ij}=\sum_{j}P_{ij}^\ast T_j$ is
the reconstructed spectrum corresponding to the true spectrum $T_i$.
For small bin size the folding matrix describes detector resolution
and FSR effects and does not depend on the true spectrum $T_i$, while
$A_{ij}$ and $\tilde{P}_{ij}$ depend on it. The unfolding method used
is based on the idea that if $T_i$ is close to the true spectrum
and the folding matrix describes resolution and FSR effects well, the
matrix $\tilde{P}_{ij}$ can be applied to the measured spectrum to obtain 
the true spectrum.

The unfolding process consists of several iteration steps. At each step,
differences between the unfolded data spectrum and $T_i$ are used
to correct $T_i$ and the unfolding matrix $\tilde{P}_{ij}$, keeping
the folding probabilities unchanged. A regularization function is
used to suppress unfolding large statistical fluctuations in the data
and guarantee the stability of the method.

Since the transfer matrix shown in Fig.~\ref{fig:fold} is non-diagonal,
the values obtained for the true data spectrum are correlated. The 
covariance matrix containing the statistical uncertainties and their
bin-to-bin correlations is obtained from pseudo-experiments (toy MC), where
both the spectrum and the transfer matrix are statistically fluctuated.
In this analysis, we generate 1000 toy-MC samples.

\begin{figure*}
\centering
\includegraphics[width=0.8\textwidth]{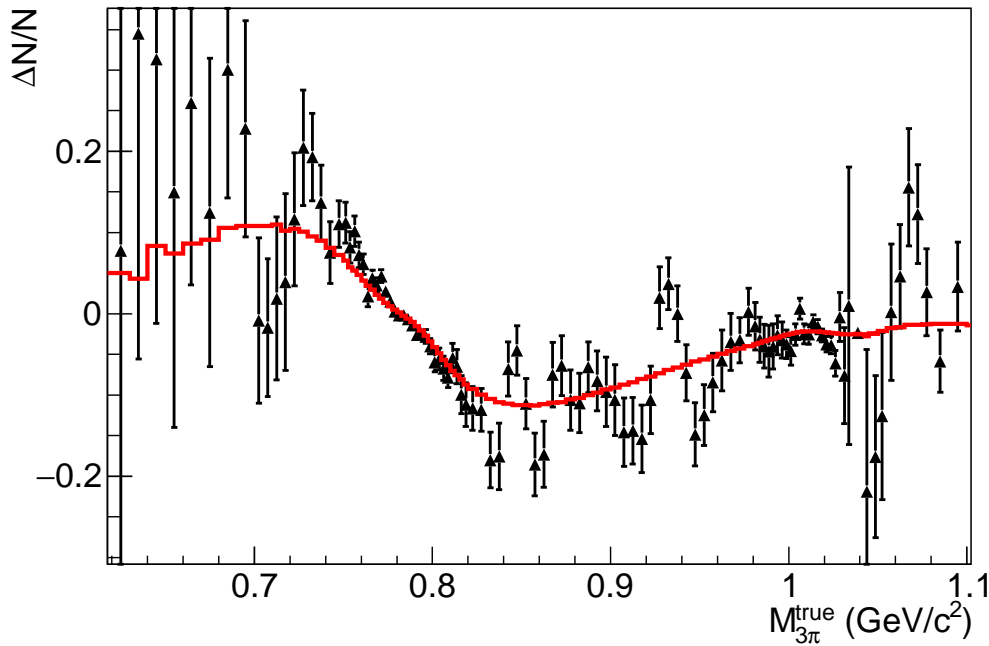}
\caption{The relative difference between the two model spectra
described in the text (curve) and the corresponding unfolded spectra
(points with error bars).
\label{relative}}
\end{figure*}
To test the unfolding procedure and choose parameters of
the regularization functions, we examine two model
spectra, each with the number of events equal to the number of events in data.
The first spectrum is the true MC spectrum $T_i$,
while the second is based on the fit with zero $\rho(770) \to 3\pi$ amplitude
(Model 2 in Table~\ref{model40}). Both spectra are convolved with the folding 
matrix $P_{ij}^\ast$, statistically fluctuated, and then subjected to
the same unfolding procedure. In Fig.~\ref{relative} the relative difference
between the two unfolded spectra is compared with the same difference in the
true spectra. The regularization parameter is chosen to minimize the 
difference between the points and the curve. The unfolded data shown in
Fig.~\ref{relative} are obtained after the first iteration step, and further
iterations do not improve the result.
\begin{figure*}
\centering
\includegraphics[width=0.9\textwidth]{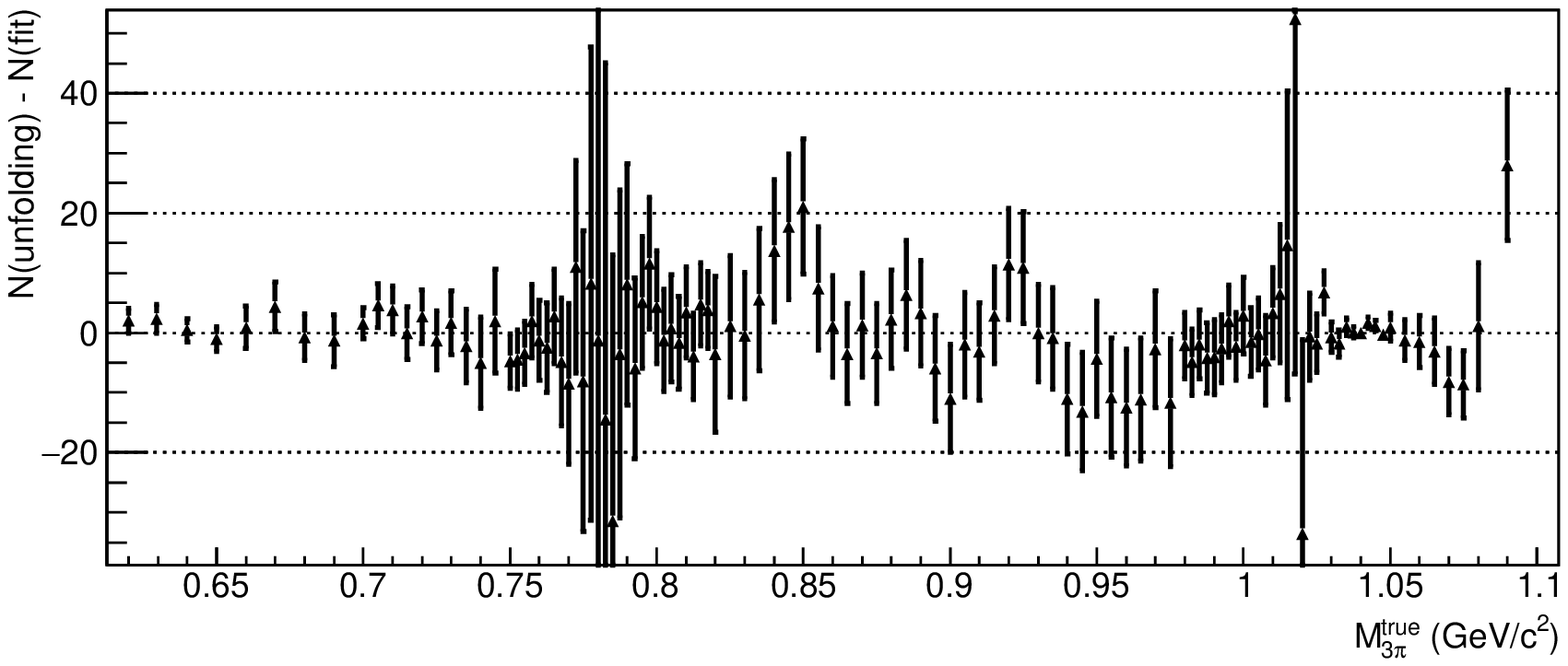}
\caption{The comparison of the unfolded data spectrum with the result of the
fit to the 3$\pi$ mass spectrum discussed in Sec.~\ref{masssp}. The error bars
correspond to the diagonal elements of the covariance matrix for the unfolded
spectrum.\label{fig:compare}}
\end{figure*}
Another test is performed to assess a systematic uncertainty in the unfolding
method. A set of 100 spectra are generated as described above,
using the true spectrum $T_i$. They are unfolded and averaged. The deviation
of the average unfolded spectrum from $T_i$ is taken as a measure of the
systematic uncertainty in the unfolding method. 

Figure~\ref{fig:compare} shows the difference between the unfolded data 
spectrum and the result of fit to the measured $3\pi$ mass spectrum (vector 
$T_i$). The error bars correspond to the diagonal elements of the covariance
matrix for the unfolded spectrum given in Ref.~\cite{cov_stat}.
The comparison demonstrates good agreement of fit results and unfolding
and establishes the adequacy of the model used in the fit.
\begin{figure*}
\centering
\includegraphics[width=0.9\textwidth]{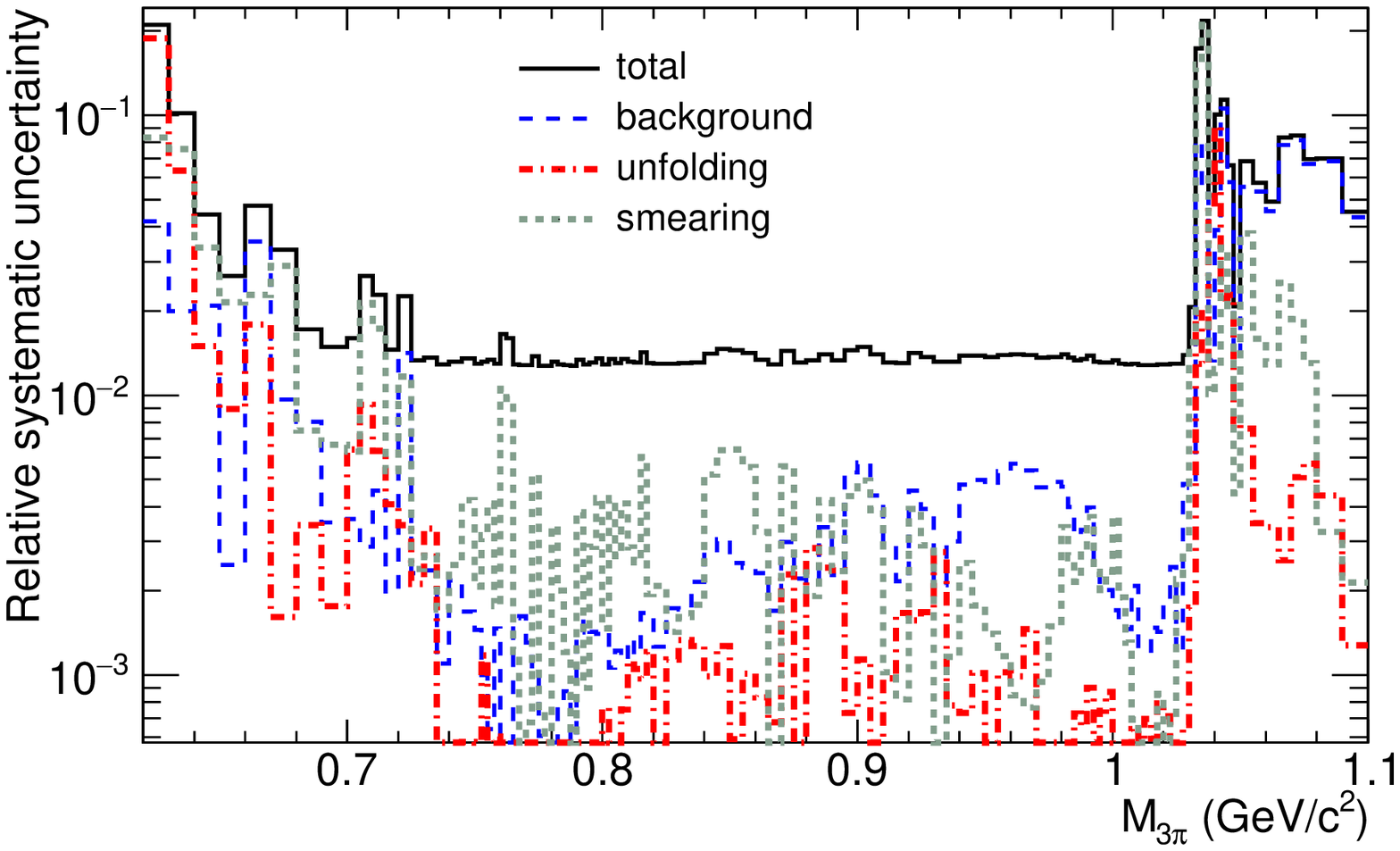}
\caption{The relative systematic uncertainty in the cross section
as a function of the $3\pi$ mass. The dashed, dash-dotted, and dotted 
histograms represent the contributions to the uncertainty due to background 
subtraction, the unfolding procedure, and Lorentzian plus Gaussian smearing.
\label{fig:bkg_contr}}
\end{figure*}

Using the unfolded $3\pi$ mass spectrum and Eq.~(\ref{thmspt}), we calculate
the Born cross section listed in Table~\ref{table_cross}, where the first
uncertainty is the square root of the diagonal element of the statistical
covariance matrix~\cite{cov_stat}. Systematic uncertainty is divided into 
two parts. The second error in Table~\ref{table_cross} represents
a correlated uncertainty that includes the uncertainties in the
luminosity, radiative correction, detection efficiency, and the uncertainty
due to the unfolding procedure. For the remaining part of the systematic
uncertainty associated with background subtraction and data-simulation 
difference in the mass resolution, we provide the covariance 
matrix~\cite{cov_sys}. The square root of the diagonal element of this 
matrix is listed in Table~\ref{table_cross} as the third error.

The mass dependence of the total systematic uncertainty is compared
with the uncertainties from the different sources in
Fig.~\ref{fig:bkg_contr}. It is seen that in the mass region between 0.73 and
1.03 GeV/$c^2$ the systematic uncertainty is dominated by the uncertainties
in the luminosity, radiative correction, and detection efficiency,
whose total contribution (1.3\%) is independent of mass.
\begin{table*}
\caption{Measured $e^+e^-\to \pi^+\pi^-\pi^0$
cross section below 1.1 GeV/$c^2$. The first uncertainty is the square root
of the diagonal element of the statistical covariance matrix, the second is
correlated systematic, and the third is the square root of the diagonal 
element of the systematic covariance matrix.
\label{table_cross}}
\begin{ruledtabular}
\begin{tabular}{lclc}
$M_{3\pi}$ (GeV/$c^2$) & $\sigma_{3\pi}$ (nb) & $M_{3\pi}$ (GeV/$c^2$)  &
$\sigma_{3\pi}$ (nb) \\
\hline
0.6200 -- 0.6300 &   0.192 $\pm$ 0.086 $\pm$ 0.036 $\pm$ 0.018 & 0.8750 -- 0.8800 &  11.051 $\pm$ 0.428 $\pm$ 0.148 $\pm$ 0.036 \\
0.6300 -- 0.6400 &   0.247 $\pm$ 0.098 $\pm$ 0.016 $\pm$ 0.019 & 0.8800 -- 0.8850 &  11.048 $\pm$ 0.421 $\pm$ 0.143 $\pm$ 0.032 \\
0.6400 -- 0.6500 &   0.223 $\pm$ 0.077 $\pm$ 0.004 $\pm$ 0.009 & 0.8850 -- 0.8900 &  11.024 $\pm$ 0.457 $\pm$ 0.141 $\pm$ 0.060 \\
0.6500 -- 0.6600 &   0.252 $\pm$ 0.081 $\pm$ 0.004 $\pm$ 0.005 & 0.8900 -- 0.8950 &  10.689 $\pm$ 0.443 $\pm$ 0.137 $\pm$ 0.035 \\
0.6600 -- 0.6700 &   0.446 $\pm$ 0.137 $\pm$ 0.010 $\pm$ 0.019 & 0.8950 -- 0.9000 &  10.101 $\pm$ 0.437 $\pm$ 0.130 $\pm$ 0.069 \\
0.6700 -- 0.6800 &   0.747 $\pm$ 0.154 $\pm$ 0.010 $\pm$ 0.023 & 0.9000 -- 0.9050 &   9.779 $\pm$ 0.446 $\pm$ 0.125 $\pm$ 0.074 \\
0.6800 -- 0.6900 &   0.818 $\pm$ 0.139 $\pm$ 0.011 $\pm$ 0.009 & 0.9050 -- 0.9100 &  10.180 $\pm$ 0.427 $\pm$ 0.131 $\pm$ 0.060 \\
0.6900 -- 0.7000 &   1.191 $\pm$ 0.154 $\pm$ 0.015 $\pm$ 0.009 & 0.9100 -- 0.9150 &  10.116 $\pm$ 0.397 $\pm$ 0.130 $\pm$ 0.031 \\
0.7000 -- 0.7050 &   1.775 $\pm$ 0.184 $\pm$ 0.025 $\pm$ 0.013 & 0.9150 -- 0.9200 &  10.443 $\pm$ 0.391 $\pm$ 0.134 $\pm$ 0.024 \\
0.7050 -- 0.7100 &   2.367 $\pm$ 0.254 $\pm$ 0.037 $\pm$ 0.051 & 0.9200 -- 0.9250 &  10.917 $\pm$ 0.443 $\pm$ 0.141 $\pm$ 0.066 \\
0.7100 -- 0.7150 &   2.801 $\pm$ 0.277 $\pm$ 0.040 $\pm$ 0.050 & 0.9250 -- 0.9300 &  10.987 $\pm$ 0.447 $\pm$ 0.141 $\pm$ 0.054 \\
0.7150 -- 0.7200 &   3.167 $\pm$ 0.302 $\pm$ 0.043 $\pm$ 0.018 & 0.9300 -- 0.9350 &  10.600 $\pm$ 0.387 $\pm$ 0.136 $\pm$ 0.022 \\
0.7200 -- 0.7250 &   4.193 $\pm$ 0.302 $\pm$ 0.056 $\pm$ 0.077 & 0.9350 -- 0.9400 &  10.733 $\pm$ 0.400 $\pm$ 0.138 $\pm$ 0.034 \\
0.7250 -- 0.7300 &   5.040 $\pm$ 0.330 $\pm$ 0.065 $\pm$ 0.017 & 0.9400 -- 0.9450 &  10.475 $\pm$ 0.427 $\pm$ 0.134 $\pm$ 0.057 \\
0.7300 -- 0.7350 &   6.754 $\pm$ 0.353 $\pm$ 0.089 $\pm$ 0.024 & 0.9450 -- 0.9500 &  10.650 $\pm$ 0.458 $\pm$ 0.137 $\pm$ 0.056 \\
0.7350 -- 0.7400 &   8.622 $\pm$ 0.402 $\pm$ 0.111 $\pm$ 0.018 & 0.9500 -- 0.9550 &  11.375 $\pm$ 0.441 $\pm$ 0.146 $\pm$ 0.057 \\
0.7400 -- 0.7450 &  11.494 $\pm$ 0.493 $\pm$ 0.147 $\pm$ 0.038 & 0.9550 -- 0.9600 &  11.471 $\pm$ 0.452 $\pm$ 0.147 $\pm$ 0.063 \\
0.7450 -- 0.7500 &  16.521 $\pm$ 0.557 $\pm$ 0.212 $\pm$ 0.075 & 0.9600 -- 0.9650 &  11.879 $\pm$ 0.443 $\pm$ 0.152 $\pm$ 0.068 \\
0.7500 -- 0.7525 &  20.723 $\pm$ 0.583 $\pm$ 0.266 $\pm$ 0.052 & 0.9650 -- 0.9700 &  12.538 $\pm$ 0.464 $\pm$ 0.161 $\pm$ 0.068 \\
0.7525 -- 0.7550 &  25.304 $\pm$ 0.624 $\pm$ 0.326 $\pm$ 0.084 & 0.9700 -- 0.9750 &  13.666 $\pm$ 0.435 $\pm$ 0.175 $\pm$ 0.065 \\
0.7550 -- 0.7575 &  31.395 $\pm$ 0.669 $\pm$ 0.403 $\pm$ 0.130 & 0.9750 -- 0.9800 &  14.232 $\pm$ 0.475 $\pm$ 0.182 $\pm$ 0.073 \\
0.7575 -- 0.7600 &  40.016 $\pm$ 0.765 $\pm$ 0.513 $\pm$ 0.092 & 0.9800 -- 0.9825 &  15.480 $\pm$ 0.488 $\pm$ 0.199 $\pm$ 0.090 \\
0.7600 -- 0.7625 &  50.575 $\pm$ 0.837 $\pm$ 0.648 $\pm$ 0.534 & 0.9825 -- 0.9850 &  15.978 $\pm$ 0.487 $\pm$ 0.205 $\pm$ 0.076 \\
0.7625 -- 0.7650 &  66.011 $\pm$ 0.927 $\pm$ 0.847 $\pm$ 0.641 & 0.9850 -- 0.9875 &  17.120 $\pm$ 0.504 $\pm$ 0.221 $\pm$ 0.065 \\
0.7650 -- 0.7675 &  89.749 $\pm$ 0.983 $\pm$ 1.150 $\pm$ 0.180 & 0.9875 -- 0.9900 &  17.981 $\pm$ 0.513 $\pm$ 0.231 $\pm$ 0.078 \\
0.7675 -- 0.7700 & 124.780 $\pm$ 1.318 $\pm$ 1.599 $\pm$ 0.215 & 0.9900 -- 0.9925 &  19.265 $\pm$ 0.547 $\pm$ 0.247 $\pm$ 0.104 \\
0.7700 -- 0.7725 & 183.933 $\pm$ 1.649 $\pm$ 2.358 $\pm$ 0.271 & 0.9925 -- 0.9950 &  20.984 $\pm$ 0.490 $\pm$ 0.274 $\pm$ 0.071 \\
0.7725 -- 0.7750 & 292.772 $\pm$ 2.168 $\pm$ 3.753 $\pm$ 1.548 & 0.9950 -- 0.9975 &  23.398 $\pm$ 0.522 $\pm$ 0.303 $\pm$ 0.068 \\
0.7750 -- 0.7775 & 495.904 $\pm$ 3.061 $\pm$ 6.358 $\pm$ 0.477 & 0.9975 -- 1.0000 &  25.669 $\pm$ 0.489 $\pm$ 0.349 $\pm$ 0.079 \\
0.7775 -- 0.7800 & 897.772 $\pm$ 4.793 $\pm$11.509 $\pm$ 0.583 & 1.0000 -- 1.0025 &  29.652 $\pm$ 0.553 $\pm$ 0.429 $\pm$ 0.120 \\
0.7800 -- 0.7825 &1430.557 $\pm$ 7.081 $\pm$18.334 $\pm$ 4.882 & 1.0025 -- 1.0050 &  34.276 $\pm$ 0.495 $\pm$ 0.447 $\pm$ 0.084 \\
0.7825 -- 0.7850 &1431.103 $\pm$ 7.153 $\pm$18.342 $\pm$ 3.081 & 1.0050 -- 1.0075 &  41.893 $\pm$ 0.516 $\pm$ 0.551 $\pm$ 0.082 \\
0.7850 -- 0.7875 & 919.240 $\pm$ 5.303 $\pm$11.787 $\pm$ 1.374 & 1.0075 -- 1.0100 &  53.705 $\pm$ 0.639 $\pm$ 0.736 $\pm$ 0.118 \\
0.7875 -- 0.7900 & 539.134 $\pm$ 3.262 $\pm$ 6.910 $\pm$ 0.343 & 1.0100 -- 1.0125 &  76.741 $\pm$ 0.645 $\pm$ 0.985 $\pm$ 0.097 \\
0.7900 -- 0.7925 & 336.835 $\pm$ 2.387 $\pm$ 4.318 $\pm$ 1.373 & 1.0125 -- 1.0150 & 126.211 $\pm$ 0.980 $\pm$ 1.619 $\pm$ 0.164 \\
0.7925 -- 0.7950 & 225.520 $\pm$ 1.780 $\pm$ 2.891 $\pm$ 0.405 & 1.0150 -- 1.0175 & 267.173 $\pm$ 2.181 $\pm$ 3.424 $\pm$ 0.417 \\
0.7950 -- 0.7975 & 163.164 $\pm$ 1.295 $\pm$ 2.092 $\pm$ 0.525 & 1.0175 -- 1.0200 & 552.239 $\pm$ 4.996 $\pm$ 7.078 $\pm$ 0.919 \\
0.7975 -- 0.8000 & 124.207 $\pm$ 1.286 $\pm$ 1.592 $\pm$ 0.561 & 1.0200 -- 1.0225 & 290.907 $\pm$ 2.727 $\pm$ 3.728 $\pm$ 0.561 \\
0.8000 -- 0.8025 &  97.049 $\pm$ 1.094 $\pm$ 1.246 $\pm$ 0.199 & 1.0225 -- 1.0250 &  74.305 $\pm$ 0.609 $\pm$ 0.954 $\pm$ 0.188 \\
0.8025 -- 0.8050 &  78.118 $\pm$ 0.993 $\pm$ 1.002 $\pm$ 0.345 & 1.0250 -- 1.0275 &  23.270 $\pm$ 0.410 $\pm$ 0.300 $\pm$ 0.058 \\
0.8050 -- 0.8075 &  65.131 $\pm$ 1.029 $\pm$ 0.835 $\pm$ 0.191 & 1.0275 -- 1.0300 &   9.109 $\pm$ 0.299 $\pm$ 0.117 $\pm$ 0.051 \\
0.8075 -- 0.8100 &  55.006 $\pm$ 0.894 $\pm$ 0.706 $\pm$ 0.234 & 1.0300 -- 1.0325 &   3.138 $\pm$ 0.207 $\pm$ 0.050 $\pm$ 0.051 \\
0.8100 -- 0.8125 &  48.036 $\pm$ 0.870 $\pm$ 0.618 $\pm$ 0.152 & 1.0325 -- 1.0350 &   0.935 $\pm$ 0.188 $\pm$ 0.041 $\pm$ 0.161 \\
0.8125 -- 0.8150 &  41.283 $\pm$ 0.821 $\pm$ 0.531 $\pm$ 0.115 & 1.0350 -- 1.0375 &   0.356 $\pm$ 0.125 $\pm$ 0.032 $\pm$ 0.077 \\
0.8150 -- 0.8175 &  37.556 $\pm$ 0.783 $\pm$ 0.483 $\pm$ 0.235 & 1.0375 -- 1.0400 &   0.035 $\pm$ 0.076 $\pm$ 0.026 $\pm$ 0.001 \\
0.8175 -- 0.8200 &  33.619 $\pm$ 0.725 $\pm$ 0.433 $\pm$ 0.085 & 1.0400 -- 1.0425 &   0.057 $\pm$ 0.057 $\pm$ 0.030 $\pm$ 0.002 \\
0.8200 -- 0.8250 &  28.538 $\pm$ 0.733 $\pm$ 0.366 $\pm$ 0.070 & 1.0425 -- 1.0450 &   0.338 $\pm$ 0.077 $\pm$ 0.004 $\pm$ 0.037 \\
0.8250 -- 0.8300 &  24.364 $\pm$ 0.658 $\pm$ 0.313 $\pm$ 0.055 & 1.0450 -- 1.0475 &   0.512 $\pm$ 0.073 $\pm$ 0.041 $\pm$ 0.031 \\
0.8300 -- 0.8350 &  21.055 $\pm$ 0.585 $\pm$ 0.271 $\pm$ 0.050 & 1.0475 -- 1.0500 &   0.623 $\pm$ 0.066 $\pm$ 0.042 $\pm$ 0.009 \\
0.8350 -- 0.8400 &  18.979 $\pm$ 0.651 $\pm$ 0.244 $\pm$ 0.053 & 1.0500 -- 1.0550 &   1.038 $\pm$ 0.095 $\pm$ 0.022 $\pm$ 0.070 \\
0.8400 -- 0.8450 &  17.583 $\pm$ 0.646 $\pm$ 0.226 $\pm$ 0.104 & 1.0550 -- 1.0600 &   1.388 $\pm$ 0.138 $\pm$ 0.022 $\pm$ 0.078 \\
0.8450 -- 0.8500 &  16.361 $\pm$ 0.653 $\pm$ 0.211 $\pm$ 0.116 & 1.0600 -- 1.0650 &   1.771 $\pm$ 0.172 $\pm$ 0.027 $\pm$ 0.084 \\
0.8500 -- 0.8550 &  15.400 $\pm$ 0.603 $\pm$ 0.198 $\pm$ 0.106 & 1.0650 -- 1.0700 &   2.042 $\pm$ 0.221 $\pm$ 0.027 $\pm$ 0.169 \\
0.8550 -- 0.8600 &  13.772 $\pm$ 0.547 $\pm$ 0.177 $\pm$ 0.083 & 1.0700 -- 1.0750 &   2.140 $\pm$ 0.220 $\pm$ 0.029 $\pm$ 0.179 \\
0.8600 -- 0.8650 &  12.715 $\pm$ 0.446 $\pm$ 0.163 $\pm$ 0.051 & 1.0750 -- 1.0800 &   2.377 $\pm$ 0.219 $\pm$ 0.031 $\pm$ 0.162 \\
0.8650 -- 0.8700 &  11.890 $\pm$ 0.435 $\pm$ 0.153 $\pm$ 0.021 & 1.0800 -- 1.0900 &   3.059 $\pm$ 0.207 $\pm$ 0.039 $\pm$ 0.210 \\
0.8700 -- 0.8750 &  11.673 $\pm$ 0.448 $\pm$ 0.152 $\pm$ 0.074 & 1.0900 -- 1.1000 &   3.920 $\pm$ 0.243 $\pm$ 0.051 $\pm$ 0.170 \\   
\end{tabular}
\end{ruledtabular}
\end{table*}

\section{
Measurement of the $e^+e^-\to \pi^+\pi^-\pi^0$ 
cross section above 1.1 GeV/$c^2$\label{cs_high}}
\begin{table*}
\caption{Measured $e^+e^-\to \pi^+\pi^-\pi^0$
cross section above 1.1 GeV/$c^2$. The first uncertainty is statistical, the
second is systematic. In the $M_{3\pi}$ intervals 3.0-3.1 GeV/$c^2$ and 
3.1-3.2 GeV/$c^2$, the values of the nonresonant cross section are listed,
which are obtained by subtraction of the $J/\psi$ contribution 
(see Sec.~\ref{jpsi}).
\label{tab_cshigh}}
\begin{ruledtabular}
\begin{tabular}{lclc}
$M_{3\pi}$ (GeV/$c^2$) & $\sigma_{3\pi}$ (nb) &$M_{3\pi}$ (GeV/$c^2$) &
$\sigma_{3\pi}$ (nb)\\
\hline
1.100--1.125& $4.29\pm 0.23\pm 0.22$& 2.000--2.025& $0.42\pm 0.07\pm 0.05$ \\
1.125--1.150& $4.19\pm 0.22\pm 0.23$& 2.025--2.050& $0.41\pm 0.07\pm 0.05$ \\
1.150--1.175& $4.83\pm 0.23\pm 0.24$& 2.050--2.075& $0.43\pm 0.07\pm 0.05$ \\
1.175--1.200& $4.97\pm 0.23\pm 0.25$& 2.075--2.100& $0.47\pm 0.07\pm 0.05$ \\
1.200--1.225& $4.58\pm 0.22\pm 0.24$& 2.100--2.125& $0.54\pm 0.07\pm 0.06$ \\
1.225--1.250& $4.78\pm 0.22\pm 0.26$& 2.125--2.150& $0.45\pm 0.07\pm 0.05$ \\
1.250--1.275& $4.94\pm 0.22\pm 0.31$& 2.150--2.175& $0.38\pm 0.06\pm 0.04$ \\
1.275--1.300& $4.89\pm 0.22\pm 0.40$& 2.175--2.200& $0.31\pm 0.06\pm 0.03$ \\
1.300--1.325& $4.64\pm 0.22\pm 0.42$& 2.200--2.225& $0.40\pm 0.06\pm 0.04$ \\
1.325--1.350& $4.45\pm 0.21\pm 0.33$& 2.225--2.250& $0.33\pm 0.06\pm 0.04$ \\
1.350--1.375& $4.32\pm 0.20\pm 0.25$& 2.250--2.275& $0.43\pm 0.06\pm 0.05$ \\
1.375--1.400& $4.12\pm 0.19\pm 0.20$& 2.275--2.300& $0.31\pm 0.06\pm 0.03$ \\
1.400--1.425& $3.94\pm 0.19\pm 0.17$& 2.300--2.325& $0.25\pm 0.05\pm 0.03$ \\
1.425--1.450& $4.19\pm 0.19\pm 0.16$& 2.325--2.350& $0.23\pm 0.05\pm 0.03$ \\
1.450--1.475& $4.03\pm 0.18\pm 0.15$& 2.350--2.375& $0.23\pm 0.05\pm 0.02$ \\
1.475--1.500& $3.72\pm 0.18\pm 0.14$& 2.375--2.400& $0.21\pm 0.05\pm 0.02$ \\
1.500--1.525& $3.88\pm 0.18\pm 0.14$& 2.400--2.425& $0.25\pm 0.05\pm 0.03$ \\
1.525--1.550& $4.23\pm 0.19\pm 0.15$& 2.425--2.450& $0.24\pm 0.04\pm 0.03$ \\
1.550--1.575& $4.80\pm 0.20\pm 0.17$& 2.450--2.475& $0.23\pm 0.04\pm 0.03$ \\
1.575--1.600& $4.95\pm 0.20\pm 0.18$& 2.475--2.500& $0.16\pm 0.04\pm 0.02$ \\
1.600--1.625& $5.28\pm 0.20\pm 0.19$& 2.500--2.525& $0.18\pm 0.04\pm 0.02$ \\
1.625--1.650& $4.82\pm 0.19\pm 0.19$& 2.525--2.550& $0.19\pm 0.04\pm 0.02$ \\
1.650--1.675& $4.11\pm 0.18\pm 0.18$& 2.550--2.575& $0.20\pm 0.04\pm 0.02$ \\
1.675--1.700& $2.79\pm 0.16\pm 0.16$& 2.575--2.600& $0.20\pm 0.04\pm 0.02$ \\
1.700--1.725& $1.93\pm 0.14\pm 0.13$& 2.600--2.625& $0.15\pm 0.04\pm 0.02$ \\
1.725--1.750& $1.87\pm 0.13\pm 0.11$& 2.625--2.650& $0.15\pm 0.04\pm 0.02$ \\
1.750--1.775& $1.51\pm 0.12\pm 0.09$& 2.650--2.675& $0.14\pm 0.04\pm 0.02$ \\
1.775--1.800& $1.36\pm 0.12\pm 0.08$& 2.675--2.700& $0.13\pm 0.04\pm 0.01$ \\
1.800--1.825& $1.22\pm 0.11\pm 0.14$& 2.700--2.800& $0.040\pm 0.032\pm 0.005$ \\
1.825--1.850& $0.91\pm 0.10\pm 0.11$& 2.800--2.900& $0.053\pm 0.030\pm 0.006$ \\
1.850--1.875& $1.11\pm 0.10\pm 0.12$& 2.900--3.000& $0.061\pm 0.022\pm 0.007$ \\
1.875--1.900& $0.73\pm 0.09\pm 0.08$& 3.000--3.100& $0.027\pm 0.030\pm 0.003$ \\
1.900--1.925& $0.64\pm 0.09\pm 0.07$& 3.100--3.200& $0.036\pm 0.025\pm 0.004$ \\
1.925--1.950& $0.58\pm 0.09\pm 0.07$& 3.200--3.300& $0.015\pm 0.025\pm 0.002$ \\
1.950--1.975& $0.43\pm 0.08\pm 0.05$& 3.300--3.400& $0.017\pm 0.021\pm 0.002$ \\
1.975--2.000& $0.49\pm 0.08\pm 0.05$& 3.400--3.500& $0.031\pm 0.016\pm 0.003$ \\
\end{tabular}
\end{ruledtabular}
\end{table*}
Above 1.1 GeV/$c^2$, the resolution effects distort the $3\pi$ mass spectrum
insignificantly. We test this by a convolution of the theoretical mass 
spectrum~(\ref{thmspt}) in the mass range 1--2 GeV/$c^2$ with the resolution
function obtained using simulation. The observed difference between the true
and measured spectra does not exceed 1\%. Therefore, the  $e^+e^-\to
\pi^+\pi^-\pi^0$ in the mass region 1.1-3.5 GeV/$c^2$ is determined as
\begin{equation}
\sigma_{3\pi}(m)=\frac{(dN/dm)^{\rm meas}}{\varepsilon\, R\,
d{\cal L}/dm}.
\end{equation}
The cross section thus obtained is listed in Table~\ref{tab_cshigh}. 
The quoted uncertainties are statistical and systematic. The latter includes 
uncertainties in the integrated luminosity (0.4\%) and radiative correction
(0.5\%), the statistical (0.3--2.4\%), systematic (1.7--1.8\%), and model 
(1.5\%) uncertainties in the detection efficiency, and the uncertainty
associated with background subtraction (3--15\%).

\begin{figure*}
\centering
\includegraphics[width=0.47\textwidth]{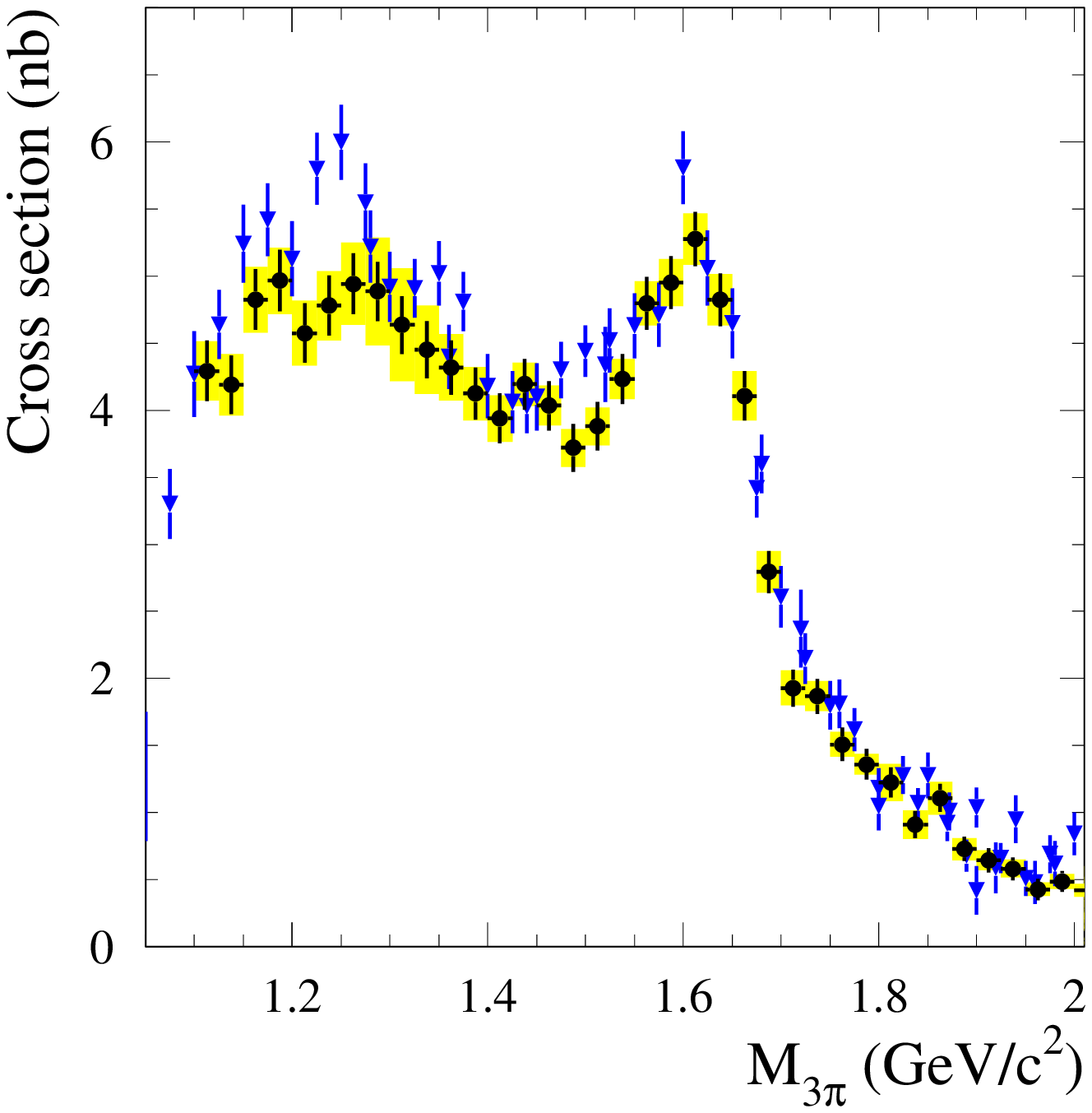}
\includegraphics[width=0.47\textwidth]{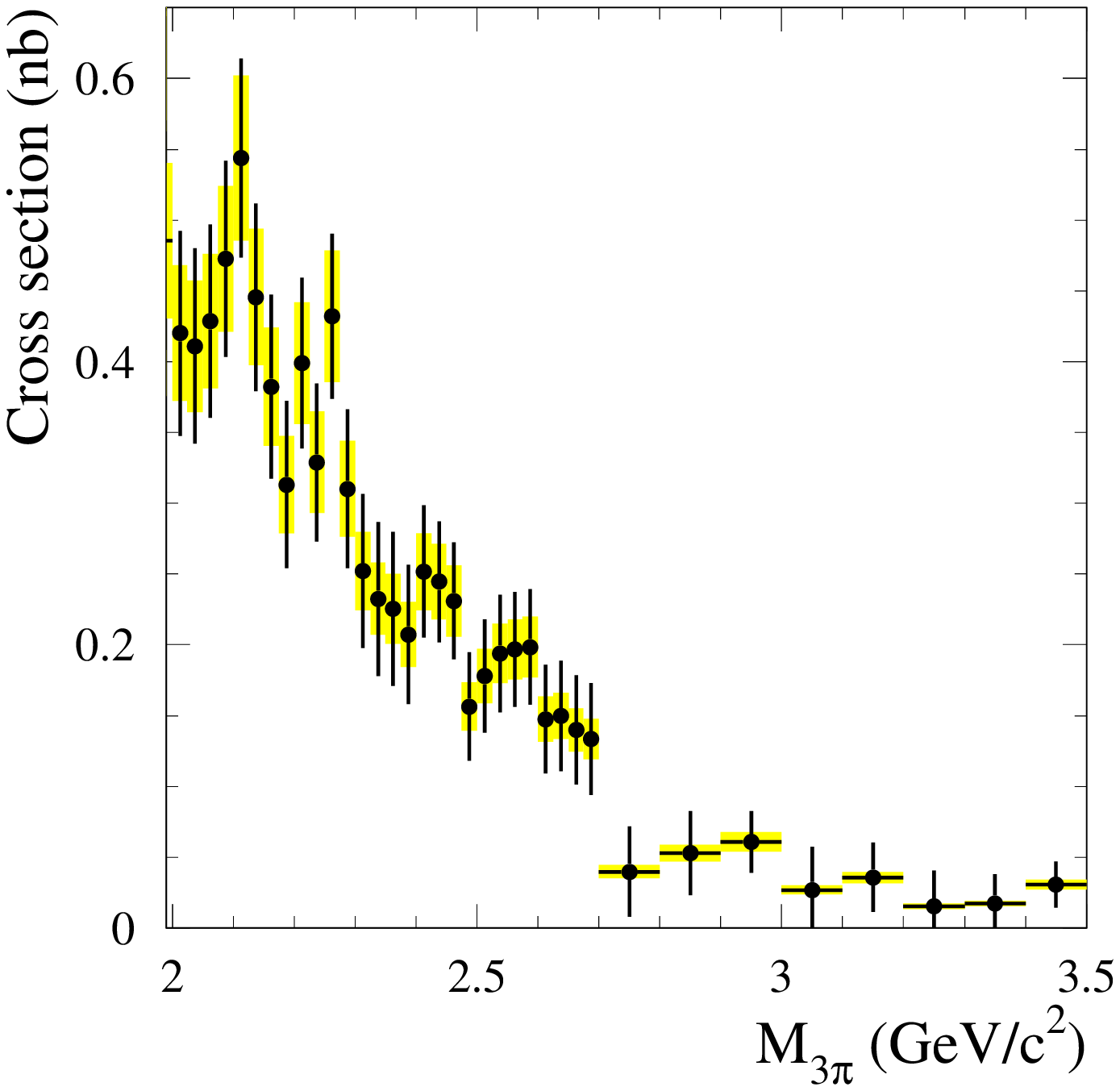}
\caption{The $e^+e^-\to \pi^+\pi^-\pi^0$ cross section measured in this work
(circles) in comparison with the SND result~\cite{snd3} (triangles). 
In the $M_{3\pi}$ interval 3.0-3.2 GeV/$c^2$, the nonresonant cross section 
obtained after subtraction of the $J/\psi$ contribution (see Sec.~\ref{jpsi})
is shown. For the
\babar\ data, the error bar represents the statistical uncertainty, while the
shaded
box shows the systematic uncertainty. For the SND data, only the statistical
uncertainty is shown; the systematic uncertainty is 4.4\%.
\label{cshigh}}
\end{figure*}

In Fig.~\ref{cshigh} (left) the measured cross section is compared with
the SND measurement~\cite{snd3} in the mass range 1.1--2 GeV/$c^2$.
A sizable difference between the two measurements is observed near 1.25 
GeV/$c^2$ and 1.5 GeV/$c^2$. The cross section above 2 GeV is shown
in Fig.~\ref{cshigh} (right). 

\section{The  $e^+e^-\to\pi^+\pi^-\pi^0$ contribution to $a_\mu$} 
The leading-order hadronic contribution to the muon anomalous magnetic
moment is calculated using the measured total hadronic cross section via
the dispersion integral (see, for example, Ref.~\cite{whp})
\begin{equation}
a_\mu=\frac{\alpha^2}{3\pi^2}\int_{m^2_\pi}^{\infty}
\frac{K(s)}{s}R(s)\,ds,
\label{amu}
\end{equation}
where the kernel function $K(s)$ can be found in Ref.~\cite{whp} and
\begin{equation}
R(s)=\frac{\sigma_0(e^+e^-\to{\rm hadrons})(s)}{4\pi\alpha^2/s}.
\label{amu1}
\end{equation}
Here $\sigma_0$ is the bare cross section, excluding effects from
vacuum polarization.

To calculate $a_\mu^{3\pi}$ we substitute $\sigma_0(e^+e^-\to{\rm
hadrons})(s)$ in Eq.~(\ref{amu1}) by 
\begin{equation}
\sigma_0(e^+e^-\to \pi^+\pi^-\pi^0)(s)= \sigma_{3\pi}(s)|1-\Pi(s)|^2,
\end{equation}
where $\sigma_{3\pi}(s)$ is the ``dressed'' cross section measured
in Secs.~\ref{cs_low} and \ref{cs_high}.
The vacuum polarization operator $\Pi(s)$ is tabulated in
Ref.~\cite{ignatov}. The integral~(\ref{amu}) is substituted by
a sum over mass bins with $s=M_{3\pi,i}^2$. In the sum, the
values of the functions $K$, $d{\cal L}/dm$, $\varepsilon$, 
and $|1-\Pi(s)|^2$ are
taken at the center of the bin. To estimate the uncertainty due to the
substitution of the integral by the sum, we
perform $a_\mu^{3\pi}$ calculations using the theoretical cross 
section~[Eq.~(\ref{born})] and mass spectrum~[Eq.~(\ref{thmspt})] with 
parameters~[Eq.~(\ref{fpar2})]. The difference between
the sum and integral is found to be 0.03\% for the mass range 0.62--1.1
GeV/$c^2$ and 0.007\% for the mass range 1.1-2.0 GeV/$c^2$. 
The main reason for the larger difference in the lower-mass region is 
the strong $s$ dependence of $|1-\Pi(s)|^2$ near the $\omega$ and $\phi$
resonances.

It should be noted that the exclusion/inclusion of the factor $|1-\Pi(s)|^2$
in the $a_\mu^{3\pi}$ calculation changes its value by about 3.5\%.
The theoretical cross section~[Eq.~(\ref{born})] is used to estimate the
uncertainty
in $a_\mu^{3\pi}$ associated with $\Pi(s)$ uncertainties ($\Delta\Pi(s)$)
given in Ref.~\cite{ignatov}. Assuming that the $\Delta\Pi(s)$ 
at different $s$ are fully correlated, we calculate  $a_\mu^{3\pi}$ with 
$\Pi(s)$ substituted by $\Pi(s)\pm\Delta\Pi(s)$. The resulting uncertainty
in $a_\mu^{3\pi}$ is found to be 0.06\% for the mass range 0.62--1.1 GeV/$c^2$
and 0.03\% for the mass range 1.1-2.0 GeV/$c^2$. 

The above-mentioned strong $s$ dependence of the $|1-\Pi(s)|^2$ term near the 
resonances leads to the $a_\mu^{3\pi}$ systematic uncertainty associated with
the ${3\pi}$ mass scale calibration. The $e^+e^-\to \rm{hadrons}$ cross section
near the $\omega$ and $\phi$ resonances used for the $\Pi(s)$ calculation in 
Ref.~\cite{ignatov} is based mainly on data obtained in the SND and CMD-2 
experiments at the VEPP-2M collider. In Sec.~\ref{masssp}, where 
the SND and CMD-2 measurements are compared with the \babar\ fit, we observe
0.1--0.2 MeV/$c^2$ shifts between the energy/mass scales of the \babar\ and 
VEPP-2M experiments. To estimate the associated systematic uncertainty we
introduce a mass shift $\Delta M=\pm0.2$ MeV/$c^2$ in the  theoretical cross
section~[Eq.~(\ref{born})] and calculate $a_\mu^{3\pi}$. The relative difference with 
the zero $\Delta M$ value is found to be 0.2\% for the mass range 0.62--1.1
GeV/$c^2$ and 0.03\% for the mass range 1.1-2.0 GeV/$c^2$.

Combining the three systematic uncertainties described above in quadrature,
we find the systematic uncertainty in $a_\mu^{3\pi}$ associated with the
vacuum polarization factor to be 0.21\% at $M_{3\pi}=0.62$--1.1 GeV/$c^2$ and 
0.04\% at $M_{3\pi}=1.1$--2.0 GeV/$c^2$.

The $a_\mu^{3\pi}$ values for different mass intervals obtained using
the $e^+e^-\to \pi^+\pi^-\pi^0$ cross section measured in this work are
listed in Table~\ref{amu_tab}.
\begin{table}
\caption{Values of $a_\mu^{3\pi}$ for different mass intervals. The first 
three rows represent the \babar\ result, while the last three are the 
calculations~\cite{dhmz,KNT,FJ,HHK} based on previous $e^+e^-\to
\pi^+\pi^-\pi^0$ measurements.
\label{amu_tab}}
\begin{ruledtabular}
\begin{tabular}{lccc}
$M_{3\pi}$ GeV/$c^2$ & $a_\mu^{3\pi}\times 10^{10}$ \\
\hline
0.62--1.10 & $42.91\pm0.14\pm 0.55 \pm 0.09$ \\
1.10--2.00 & $2.95\pm 0.03\pm 0.16$ \\
$<2.00$    & $45.86\pm0.14\pm 0.58$ \\
\hline
$<1.8$\cite{dhmz} & $46.21\pm 0.40\pm 1.40$\\
$<1.97$\cite{KNT} & $46.74\pm 0.94$\\
$<2$\cite{FJ} & $44.32\pm 1.48$ \\
$<1.8$\cite{HHK} & $46.2\pm 0.6 \pm 0.6$ \\
\end{tabular}
\end{ruledtabular}
\end{table}
For the mass range 0.62--1.10  GeV/$c^2$ the quoted uncertainties are
statistical, systematic due to the cross section measurement, and systematic 
due to the vacuum polarization. The statistical uncertainty in $a_\mu^{3\pi}$
is calculated using the toy MC simulation as described in Sec.~\ref{cs_low}.
\begin{table}
\caption{Contributions to the systematic uncertainty in
$a_\mu^{3\pi}(0.62<M_{3\pi}<1.1\mbox{ GeV}/c^2)$ from 
different effects.\label{amusys}}
\begin{ruledtabular}
\begin{tabular}{lcccc}
Effect & Uncertainty (\%) \\
\hline
Luminosity                     & 0.4 \\
Radiative correction           & 0.5 \\
Detection efficiency           & 1.1 \\
MC statistics                  & 0.15 \\
Background subtraction         & 0.073 \\
Gaussian smearing              & 0.0007 \\
Lorentzian smearing            & 0.003 \\
Unfolding procedure            & 0.045 \\
\hline
Total                          & 1.3 \\
\end{tabular}
\end{ruledtabular}
\end{table}

The contributions to the systematic uncertainty in
$a_\mu^{3\pi}(0.62<M_{3\pi}<1.1\mbox{ GeV}/c^2)$ from different effects are
listed in Table~\ref{amusys}. The uncertainties in the detection efficiency,
luminosity, and radiative correction dominate. These
three contributions are common for the mass intervals below and above 1.1
GeV/$c^2$. However, in the mass range 1.10--2.00 GeV/$c^2$ 
the largest contribution to the systematic uncertainty comes from the FSR
background. In the $a_\mu^{3\pi}(1.1<M_{3\pi}<2.0\mbox{ GeV}/c^2)$
calculation, the systematic uncertainties listed in Table~\ref{tab_cshigh}
are conservatively taken to be 100\% correlated.

For $a_\mu^{3\pi}(M_{3\pi}<2\mbox{ GeV}/c^2)$ we also add the contribution
from the region below 0.62 GeV/$c^2$, which is estimated to be 
$5.7\times 10^{-13}$ using the theoretical cross section~[Eq.~(\ref{born})]
with parameters~[Eq.~(\ref{fpar2})].

In Table~\ref{amu_tab} our result is compared with the calculations
of $a_\mu^{3\pi}$~\cite{dhmz,KNT,FJ,HHK} based on previous $e^+e^-\to
\pi^+\pi^-\pi^0$ measurements. Since the calculations are performed in
different $3\pi$ mass regions, we also give our result for 
the mass interval 1.8-2.0 GeV/$c^2$: 
$(0.116 \pm0.005 \pm0.013)\times 10^{-10}$.
Our $a_\mu^{3\pi}$ value is in reasonable agreement with the previous 
calculations~\cite{dhmz,KNT,FJ,HHK} but has better accuracy.

\section{
Measurement of the $J/\psi\to \pi^+\pi^-\pi^0$ decay
\label{jpsi}}
\begin{figure}
\centering
\includegraphics[width=0.95\linewidth]{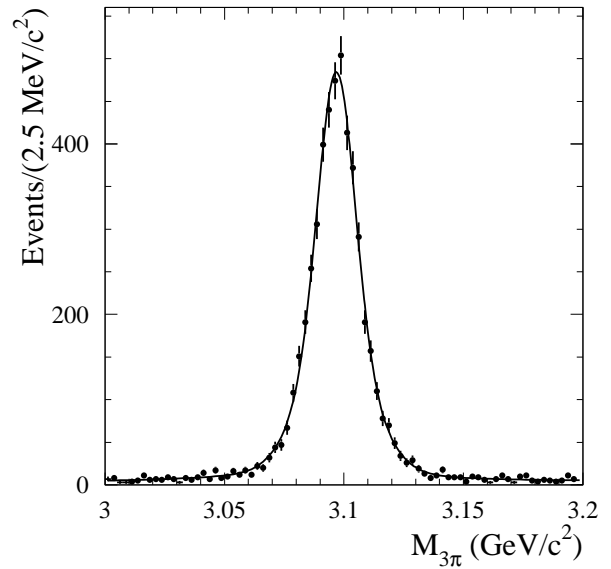}
\caption{The $3\pi$ mass spectrum for selected events in the $J/\psi$ mass
region. The curve is the result of the fit described in the text.}
\label{jpsisp}
\end{figure}
The $3\pi$ mass spectrum in the $J/\psi$ mass region for data events 
selected with the standard criteria 
is shown in Fig.~\ref{jpsisp}. The small width of the $J/\psi$ resonance
leads to negligible peaking background. In particular,
$e^+e^-\to J/\psi\gamma\to K^+K^-\pi^0\gamma$ events reconstructed
under the $3\pi\gamma$ hypothesis have the $3\pi$ invariant mass in the range
2.8 to 3.0 GeV/$c^2$. To determine the number of $J/\psi$ events, the spectrum
is fitted with a sum of a resonance distribution and a linear background. The
resonance line shape is a Breit-Wigner function convolved with a 
triple-Gaussian function describing detector resolution. The Breit-Wigner width
is fixed at its PDG value~\cite{pdg}. The parameters of the resolution
function are determined from simulation. To account for possible differences
in detector response between data and simulation, the simulated resolution
function is modified by adding a smearing variance $\sigma_s^2$ 
to each of the three variances of the triple-Gaussian function. The free 
parameters
in the fit are the number of resonance events ($N_{J/\psi}$), the number of
nonresonant background events, the slope of the background, $\sigma_s$, and 
the resonance mass.

The result of the fit is shown in Fig.~\ref{jpsisp}. The fitted resonance
parameters are the following: $N_{J/\psi}=4921\pm 74$, $\sigma_s^2=1.8\pm2.6$
MeV$^2$/$c^4$, and $M_{J/\psi}=3.0962\pm0.0002$ GeV/$c^2$. The latter differs
from the nominal $J/\psi$ mass ($3096.900\pm 0.006$ MeV/$c^2$) by
$-(0.7\pm0.2)$ MeV/$c^2$, while the $\sigma_s$ value is consistent with zero.

The differential cross section for ISR production of a narrow resonance, such
as $J/\psi$, can be calculated using~\cite{ivanch}
\begin{equation}
\frac{d\sigma(s,\theta_\gamma)}{d\cos{\theta_\gamma}} =
\frac{12\pi^2P_{J/\psi}}{m_{J/\psi}s}
W(s,x_{J/\psi},\theta_\gamma),
\label{eqpsi}
\end{equation}
where $P_{J/\psi}\equiv\Gamma(J/\psi\to e^+e^-){\cal B}(J/\psi\to 3\pi)$,
$m_{J/\psi}$ and $\Gamma(J/\psi\to e^+e^-)$ are the $J/\psi$ mass and
electronic width, $W(s,x_{J/\psi},\theta_\gamma)$ is the radiator
function from Eq.~(\ref{eq2}), $x_{J/\psi} = 1-m_{J/\psi}^2/s$, 
and ${\cal B}(J/\psi\to 3\pi)$ is the branching fraction of the decay
$J/\psi\to \pi^+\pi^-\pi^0$. Therefore, the measurement of the number 
of $J/\psi\to \pi^+\pi^-\pi^0$ decays in the $e^+e^-\to \pi^+\pi^-\pi^0\gamma$
reaction  determines the product of the electronic width and the branching
fraction $\Gamma(J/\psi\to e^+e^-){\cal B}(J/\psi\to 3\pi)$.

The cross section for
$e^+e^-\to J/\psi\gamma\to \pi^+\pi^-\pi^0\gamma$ for
$20^\circ<\theta_\gamma<160^\circ$ 
is calculated as
\begin{equation}
\sigma(20^\circ<\theta_\gamma<160^\circ)=\frac{N_{J/\psi}}
{\varepsilon\, R\, {\cal L}}=(114.7\pm1.7\pm2.4)\mbox{ fb}.
\end{equation}
Here ${\cal L}=468.6\pm2.0$ fb$^{-1}$, $R=1.0091\pm0.0050$, and the detection 
efficiency $\varepsilon$ corrected for the data-MC 
difference in the detector response and the decay model is $(9.07\pm0.18)\%$,
where the uncertainty includes the systematic uncertainty (1.8\%) and the
MC statistical uncertainty (0.8\%).
From the measured cross section and Eq.~(\ref{eqpsi}),
we determine:
\begin{equation}
P_{J/\psi}=
(0.1248\pm0.0019\pm0.0026)\mbox{ keV}.
\end{equation}
Using the PDG value $\Gamma(J/\psi\to e^+e^-)=(5.51\pm0.10)$ keV~\cite{pdg}
we obtain
\begin{equation}
{\cal B}(J/\psi\to 3\pi)=(2.265\pm0.034\pm0.062)\%,
\end{equation}
which is in reasonable agreement with the average PDG value
$(2.10\pm0.08)\%$~\cite{pdg} and the most precise measurement
$(2.137\pm0.064)\%$ by the BESIII Collaboration~\cite{bespsi}.

\section{Summary}
The cross section for the process $e^+e^-\to \pi^+\pi^-\pi^0$
has been measured by the \babar\ experiment in the c.m.~energy range 
from 0.62 to 3.5 GeV, using the ISR method. The cross section is dominated
by the $\omega$ and $\phi$ resonances. Near the maxima of these resonances
it is measured with a systematic uncertainty of 1.3\%. 
The leading-order hadronic contribution to the muon
magnetic anomaly, calculated using the measured $e^+e^-\to \pi^+\pi^-\pi^0$ 
cross section from threshold to 2.0 GeV, is
$(45.86 \pm 0.14 \pm 0.58)\times 10^{-10}$.
Our $a_\mu^{3\pi}$ value is in reasonable agreement with the  
calculations~\cite{dhmz,KNT,FJ,HHK} based on previous 
$e^+e^-\to \pi^+\pi^-\pi^0$
measurements but is more precise by about a factor of about 2.
From the fit to the measured
$3\pi$ mass spectrum in the process $e^+e^-\to \pi^+\pi^-\pi^0\gamma$
we have determined the resonance parameters 
\begin{eqnarray}
P_\omega&=&
(0.5698\pm0.0031\pm0.0082) \mbox{ keV},\nonumber\\ 
P_\phi&=&
(0.1841\pm0.0021\pm0.0080) \mbox{ keV}.\nonumber\\ 
{\cal B}(\rho\to 3\pi)&=&(0.88\pm 0.23\pm 0.30)\times 10^{-4},
\nonumber\\
\varphi_\rho&=&-(99\pm 9 \pm 15)^\circ,
\end{eqnarray}
where $P_V=\Gamma(V\to e^+e^-){\cal B}(V\to \pi^+\pi^-\pi^0)$.
The significance of the $\rho\to 3\pi$ decay is found to be greater than 
$6\sigma$. 
The measured values of $\Gamma(V\to e^+e^-){\cal B}(V\to 3\pi)$ for the 
$\omega$ and $\phi$ mesons are in agreement with
the world average values~\cite{pdg}.
For the $\omega$ meson, the accuracy of our result is comparable with
the accuracy of the PDG value. For the $\phi$ meson we have a large systematic
uncertainty related to the interference between the $\phi$-meson amplitude
and amplitudes of the resonances of the $\omega$ family. The measured values
of ${\cal B}(\rho\to 3\pi)$ and $\varphi_\rho$ 
are in agreement with the SND results~\cite{snd2}. 

For the $J/\psi$ resonance we have measured the product
\begin{eqnarray}
\Gamma(J/\psi\to e^+e^-){\cal B}(J/\psi\to 3\pi)&=&\nonumber\\
(0.1248\pm0.0019\pm0.0026)\mbox{ keV},
\end{eqnarray}
and the branching fraction ${\cal B}(J/\psi\to 3\pi)=(2.265\pm0.071)\%$. The
latter is in reasonable agreement with the average PDG value~\cite{pdg} and the
most precise measurement, which is by the BESIII Collaboration~\cite{bespsi}.

\section{Acknowledgments}
We thank V. L. Chernyak for useful discussions. We are grateful for the 
extraordinary contributions of our PEP-II colleagues in
achieving the excellent luminosity and machine conditions
that have made this work possible.
The success of this project also relies critically on the 
expertise and dedication of the computing organizations that 
support \babar.
The collaborating institutions wish to thank 
SLAC for its support and the kind hospitality extended to them. 
This work is supported by the
US Department of Energy
and National Science Foundation, the
Natural Sciences and Engineering Research Council (Canada),
the Commissariat \`a l'Energie Atomique and
Institut National de Physique Nucl\'eaire et de Physique des Particules
(France), the
Bundesministerium f\"ur Bildung und Forschung and
Deutsche Forschungsgemeinschaft
(Germany), the
Istituto Nazionale di Fisica Nucleare (Italy),
the Foundation for Fundamental Research on Matter (The Netherlands),
the Research Council of Norway, the
Ministry of Education and Science of the Russian Federation,
Ministerio de Economia y Competitividad (Spain), and the
Science and Technology Facilities Council (United Kingdom).
Individuals have received support
from the Russian Foundation for Basic Research (grant
No. 20-02-00060),
the Marie-Curie IEF program (European Union), the A. P. Sloan Foundation (USA) 
and the Binational Science Foundation (USA-Israel).

\end{document}

%% file: authors_sep2021_frozen.tex
\author{J.~P.~Lees}
\author{V.~Poireau}
\author{V.~Tisserand}
\affiliation{Laboratoire d'Annecy-le-Vieux de Physique des Particules (LAPP), Universit\'e de Savoie, CNRS/IN2P3,  F-74941 Annecy-Le-Vieux, France}
\author{E.~Grauges}
\affiliation{Universitat de Barcelona, Facultat de Fisica, Departament ECM, E-08028 Barcelona, Spain }
\author{A.~Palano}
\affiliation{INFN Sezione di Bari, I-70126 Bari, Italy}
\author{G.~Eigen}
\affiliation{University of Bergen, Institute of Physics, N-5007 Bergen, Norway }
\author{D.~N.~Brown}
\author{Yu.~G.~Kolomensky}
\affiliation{Lawrence Berkeley National Laboratory and University of California, Berkeley, California 94720, USA }
\author{M.~Fritsch}
\author{H.~Koch}
\author{T.~Schroeder}
\affiliation{Ruhr Universit\"at Bochum, Institut f\"ur Experimentalphysik 1, D-44780 Bochum, Germany }
% ...
\collaboration{The \babar\ Collaboration}
\noaffiliation

%% file: babar-pub-2104.bbl
\begin{thebibliography}{99}
\bibitem{bgf}
The background filters are the offline filters providing selection of events
before the full reconstruction. They use information about charged tracks
and clusters in the calorimeter to classify events as multihadron, 
$\tau^+\tau^-$, ISR, etc. Events that do not pass filters are rejected.
\end{thebibliography}
